\documentclass[12pt]{spieman}  
\usepackage{amsmath,amsfonts,amssymb}
\usepackage{graphicx}
\usepackage{setspace}
\usepackage{tocloft}
\usepackage{lineno}
\usepackage{multirow}

\usepackage{siunitx}
\usepackage{graphicx}
\graphicspath{{Figures/}}

\newif\iffigures
\figurestrue

\newcommand{\nor}[1]{\textcolor{black}{#1}}

\newcommand{\ctf}{{CT5TEA}}
\newcommand{\ctc}{{CTC}}

\title{Design and Performance of the Upgraded Prototype Schwarzschild-Couder Telescope Camera Module}

\author[0]{Giovanni~Ambrosi}
\author[1]{Carla~Aramo}
\author[0]{Mattia~Barbanera}
\author[2, 3]{Chiara~Bartolini}
\author[4]{Wystan~Benbow}
\author[0, 5]{Bruna~Bertucci}
\author[3, 6]{Elisabetta~Bissaldi}
\author[7]{Massimiliano~Bitossi}
\author[8]{\nor{James~H.~Buckley}}
\author[9, 10]{Massimo~Capasso}
\author[0]{Mirco~Caprai}
\author[3, 6]{Davide~Cerasole}
\author[11,*]{Zachary~Curtis-Ginsberg}
\author[3, 6]{Gaia~De~Palma}
\author[3,*]{Leonardo~Di~Venere}
\author[12]{Miguel~Escobar~Godoy}
\author[13]{Qi~Feng}
\author[0]{Emanuele~Fiandrini}
\author[14]{Lucy~Fortson}
\author[15]{Stefan~Funk}
\author[12, 16]{Amy~Furniss}
\author[17]{Alasdair~Gent}
\author[0, 5]{Stefano~Germani}
\author[3, 6]{Nicola~Giglietto}
\author[3, 6]{Francesco~Giordano}
\author[4]{William~Hanlon}
\author[11]{Sam~Heiman}
\author[12]{Olivier~Hervet}
\author[0]{Maria~Ionica}
\author[18]{Weidong~Jin}
\author[13]{David~Kieda}
\author[3]{Francesco~Licciulli}
\author[2, 3]{Pierpaolo~Loizzo}
\author[3, 6,*]{Serena~Loporchio}
\author[19, 20]{Giovanni~Marsella}
\author[9]{Reshmi~Mukherjee}
\author[17]{Nepomuk~Otte}
\author[3, 6]{Francesca~Romana~Pantaleo}
\author[7, 21]{Riccardo~Paoletti}
\author[14]{Deivid~Ribeiro}
\author[11,*]{Luca~Riitano}
\author[4]{Emmet~Roache}
\author[22]{Duncan~Ross}
\author[4]{Lab~Saha}
\author[12]{Heiko~Salzmann}
\author[15]{Benjamin~Schwab}
\author[9]{Ruo-Yu~Shang}
\author[0]{Gianluigi~Silvestre}
\author[7, 21]{Leonardo~Stiaccini}
\author[23]{Hiroyasu~Tajima}
\author[24]{Svanik~Tandon}
\author[19]{Giovanni~Tripodo}
\author[11]{Justin~Vandenbroucke}
\author[18]{Vladimir~V.~Vassiliev}
\author[25]{Richard~White}
\author[12]{David~A.~Williams}
\author[15]{Adrian~Zink}

\affil[0]{INFN Sezione di Perugia, 06123 Perugia, Italy}
\affil[1]{INFN Sezione di Napoli, 80126 Napoli, Italy}
\affil[2]{Dipartimento di Fisica, Università di Trento, 38123 Trento, Italy}
\affil[3]{INFN Sezione di Bari, 70125 Bari, Italy}
\affil[4]{Center for Astrophysics | Harvard \& Smithsonian, Cambridge, MA 02138, USA}
\affil[5]{Dipartimento di Fisica e Geologia dell’Universit\`a degli Studi di Perugia, 06123 Perugia, Italy}
\affil[6]{Dipartimento Interateneo di Fisica dell’Universit\`a e del Politecnico di Bari, 70126 Bari, Italy	INFN Sezione di Bari, 70125 Bari, Italy}
\affil[7]{INFN Sezione di Pisa, 56127 Pisa, Italy}
\affil[8]{\nor{Department of Physics, Washington University, 1 Brookings Drive, St. Louis, MO 63130, USA}}
\affil[9]{Department of Physics and Astronomy, Barnard College, Columbia University, NY 10027, USA}
\affil[10]{Broadcom Inc., \nor{Gertraud-Kaltenecker-Straße 1, 93049 Regensburg, Germany}}
\affil[11]{Department of Physics and Wisconsin IceCube Particle Astrophysics Center, University of Wisconsin, Madison, WI 53706, USA}
\affil[12]{Santa Cruz Institute for Particle Physics and Department of Physics, University of California, Santa Cruz, CA 95064, USA}
\affil[13]{Department of Physics and Astronomy, University of Utah, Salt Lake City, UT 84112, USA}
\affil[14]{School of Physics and Astronomy, University of Minnesota, Minneapolis, MN 55455, USA}
\affil[15]{Erlangen Centre for Astroparticle Physics (ECAP), Friedrich-Alexander-Universität Erlangen-Nürnberg, Nikolaus-Fiebiger-Str. 2, Erlangen, 91058, Germany}
\affil[16]{Department of Physics, California State University - East Bay, Hayward, CA 94542, USA}
\affil[17]{School of Physics \& Center for Relativistic Astrophysics, Georgia Institute of Technology, Atlanta, GA 30332-0430, USA}
\affil[18]{Department of Physics and Astronomy, University of California, Los Angeles, CA 90095, USA}
\affil[19]{Dipartimento di Fisica e Chimica ``E. Segr\`e'', Universit\`a degli Studi di Palermo, via delle Scienze, 90128 Palermo, Italy}
\affil[20]{INFN Sezione di Catania, 95123 Catania, Italy}
\affil[21]{Dipartimento di Scienze Fisiche, della Terra e dell'Ambiente, Universit\`a degli Studi di Siena, 53100 Siena, Italy}
\affil[22]{Space Park Leicester, 92 Corporation Road, Leicester, LE4 5SP, UK}
\affil[23]{Institute for Space--Earth Environmental Research and Kobayashi--Maskawa Institute for the Origin of Particles and the Universe, Nagoya University, Nagoya 464-8601, Japan}
\affil[24]{Department of Physics, Columbia University, New York, NY 10027, USA}
\affil[25]{Max-Planck-Institut für Kernphysik, P.O. Box 103980, 69029 Heidelberg, Germany}

\cftpagenumbersoff{figure}
\cftpagenumbersoff{table} 
\begin{document} 
\DeclareSIUnit{\sample}{Sa}
\maketitle

\begin{abstract}
The Cherenkov Telescope Array Observatory (CTAO) is a ground-based observatory that will improve upon the sensitivities of the current generation of very-high-energy gamma-ray instruments. The Schwarzschild-Couder Telescope (SCT) is a dual-mirror candidate design for a CTAO Medium-Sized Telescope (MST). The prototype Schwarzschild-Couder Telescope (pSCT) was inaugurated in 2019 at Fred Lawrence Whipple Observatory (FLWO) in Arizona and observed significant gamma-ray emission from the Crab Nebula with a partially populated camera. The pSCT camera is currently being upgraded to fully instrument the focal plane with 11,328 silicon photomultiplier (SiPM) pixels split between 177 camera modules. Additionally, the modules will feature upgraded electronics designed to reduce electronics crosstalk and noise. A module calibration procedure has been developed using a preproduction test module. Following this calibration procedure, performance testing shows that the upgrade module has low noise, minimal electronics crosstalk, and excellent charge resolution. After calibration and optimization, the 177 production modules will be installed in the pSCT camera for commissioning. This will be followed by observations of known VHE gamma-ray sources for camera performance validation.
\end{abstract}


{\noindent \footnotesize\textbf{*}Luca Riitano,  \linkable{riitano@wisc.edu}, Serena Loporchio,  \linkable{serena.loporchio@ba.infn.it}, Zachary Curtis-Ginsberg,  \linkable{curtisginsbe@wisc.edu}, Leonardo Di Venere,  \linkable{leonardo.divenere@ba.infn.it}}


\date{\today}

\begin{spacing}{2}   

\section{Introduction}
The Cherenkov Telescope Array Observatory (CTAO) is the next-generation ground-based observatory for very-high-energy (VHE \qty{\geq 100}{\GeV}) gamma rays. At such high energies, the gamma-ray fluxes are too small to be detected by satellite detectors, but it is possible to exploit their interaction with Earth's atmosphere by means of imaging atmospheric Cherenkov telescopes (IACTs). In extensive air showers produced by VHE gamma rays, the constituent particles emit Cherenkov radiation which is detected by IACTs.

The CTAO will consist of two multiple-telescope arrays, one for each hemisphere. The Northern CTAO site (CTAO-N) is located at the Roque de los Muchachos Observatory on the island of La Palma (Spain), while the Southern CTAO site (CTAO-S) will be constructed near the Paranal Observatory in the Atacama Desert (Chile). The so-called Alpha configuration foresees a total of 64 telescopes of three different designs to provide a wide energy range coverage from tens of GeV to hundreds of TeV. The Alpha Configuration includes four Large-Sized Telescopes (LSTs) and nine Medium-Sized Telescopes (MSTs) at CTAO-N, and 14 MSTs and 37 Small-Sized Telescopes (SSTs) at CTAO-S. As of the time of writing, this configuration does not consider LSTs in the CTAO-S array yet. However, the preparation of the foundation for four LSTs and additional SSTs are planned, paving the way in the future enhancement of the CTAO-S array. The construction of 2 LSTs and 5 SSTs is currently proposed as part of the Italian CTA+ project.

Along with the approved single-mirror Davies-Cotton MST design, the Schwarzschild-Couder telescope (SCT) is proposed as a dual-mirror alternative type of medium telescope that could be added to the CTAO-S array in the future. The dual-mirror design, also planned for the SSTs, has so far seen only very recent implementations in Cherenkov astronomy, such as the ASTRI-Horn telescope. The innovative dual-mirror design offers compensation of optical aberration and de-magnification of images, expecting to allow for a better gamma-ray and optical angular resolution than classical single-mirror telescopes as a result of a very large number of camera pixels over a wider field-of-view (FoV) \cite{Vassiliev_2007}. The SCT design includes a \qty{9.66}{\meter} diameter primary mirror and a \qty{5.4}{\meter} diameter secondary mirror, which will result in a focal surface plate scale of \qty{1.625}{\mm} per arcminute \cite{vassiliev_2017}, making the design compatible with silicon photomultipliers (SiPMs) to be employed in the place of photomultiplier tubes (PMTs) used by other IACTs. \nor{The SiPM technology, which will also be adopted in smaller telescopes such as the SSTs, has already demonstrated its imaging performance in operational instruments like the FACT \cite{2014JInst...9P0012B}.}

A prototype of the SCT (pSCT) is located at the Fred Lawrence Whipple Observatory in Arizona, United States of America. The aim of the prototype development is to prove the feasibility of the innovative design. The prototype has proven its imaging capability through the detection of the Crab Nebula \cite{ADAMS2021102562}. Currently, the telescope is undergoing major upgrades in its camera design and equipment, with the aim of improving its performance.
In this work we focus on the front-end electronics (FEE) developed to upgrade the photo-detection modules of the pSCT camera. In Section \ref{sect:design} we provide an overview of the module design, including the photo-sensors and the electronic read-out chain.
In Section \ref{Calibration} we describe the calibration procedure and tests performed to optimize the module for data-taking. Subsequently, in Section \ref{Performance} we describe the tests conducted to verify the module performance after the calibration procedure. In Section \ref{sect:gain} we describe a flat-fielding procedure to equalize the gain. Finally in Section \ref{sect:conclusion} we summarize how the upgraded pSCT camera module was calibrated and tested to meet design goals for its future use in SCTs at CTAO.

\section{Design Overview}
\label{sect:design}
The pSCT camera employs a modular design consisting of nine backplanes and 177 photo-detection modules, each with 64 channels, for a total of 11,328 6 $\times$ 6 \unit{\mm\squared} SiPM pixels, as can be seen in Figure \ref{fig:camera_drawing}. 

\iffigures
\begin{figure}[htp]
    \centering
    \includegraphics[width=0.9\textwidth]{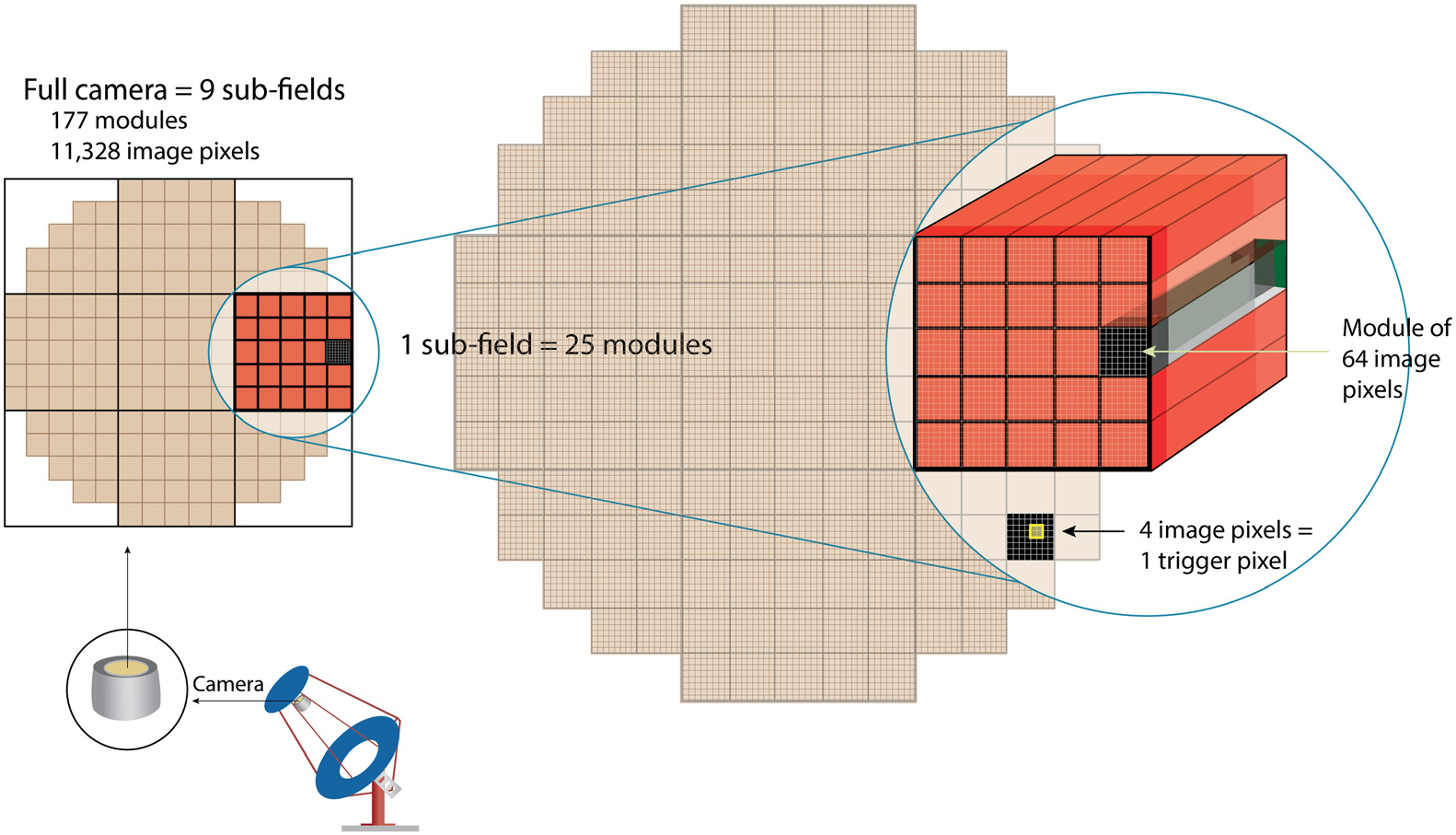}
    \caption{Design of the SCT camera. The full camera is composed of nine sectors, each of which can hold up to 25 photo-detection modules. The entire camera has 177 modules (corner sectors are equipped with fewer modules to replicate a circular shape). Each module has 64 pixels, for a total of 11,328 pixels in the fully-equipped camera \cite{camera_paper_jatis}.}
    \label{fig:camera_drawing}
\end{figure}
\fi

Each module integrates an FEE and a focal-plane module (FPM). The SiPMs are arranged in 16-channel square tiles with groups of four tiles connected to a single FEE module. The shape of the camera is approximately circular with a diameter of \qty{0.8}{\meter} or $8.04$ degrees, resulting in a 50.8 square degree FoV (assuming a circular focal plane).

The current pSCT camera has only the central sector populated, comprising 1,600 pixels equipped with both the Hamamatsu S12642-0404PA-50(X) SiPMs and the Fondazione Bruno Kessler (FBK) Near Ultra-Violet High-Definition (NUV-HD) SiPMs \cite{AMBROSI2023168023,AMBROSI2022167359}.
The camera's FEE manages signal amplification, shaping, digitization, SiPM bias voltage control, low-level trigger generation, temperature monitoring, and data packaging for storage. These functionalities are currently supported by the TARGET (TeV Array Readout with GSa/s sampling and Event Trigger) ASIC, specifically the TARGET-7 \cite{TARGET7}.

A detailed description of the pSCT camera's FPM and FEE with TARGET-7, including their mechanics and performance, is available in Ref.~\citenum{camera_paper_jatis}. Upgrades to the pSCT camera are currently underway to increase the number of pixels from 1,600 to 11,328, thereby fully populating the camera. After the upgrade, the entire camera will use newer FBK NUV-HD-MT SiPMs, which feature low crosstalk probability and enhanced photon detection efficiency \cite{Merzi_2023, LOIZZO2024169751}. 

\begin{figure}[htp]
    \centering
    \includegraphics[width=0.9\textwidth]{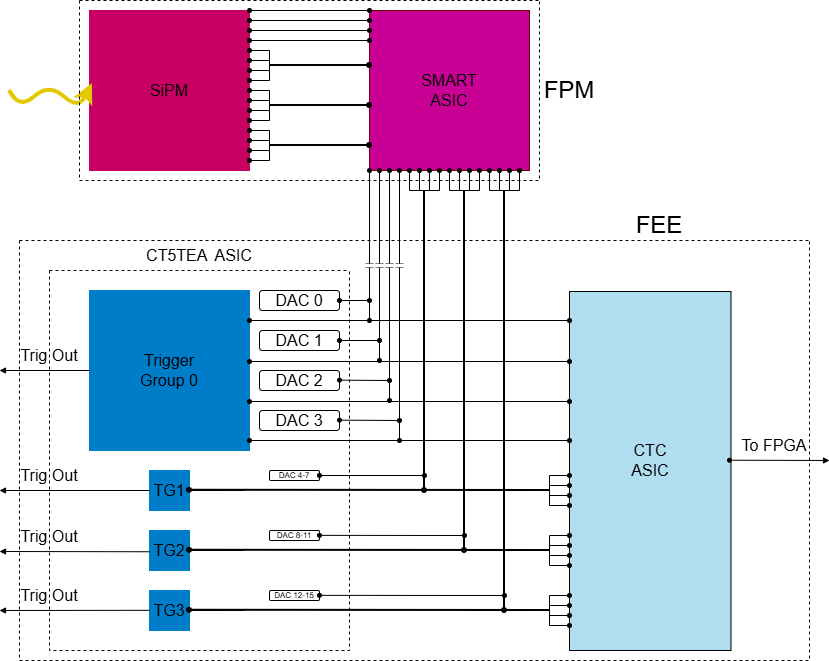}
    \caption{Block diagram of the module signal path. The thicker lines indicate bundles of four channels that are not drawn out separately. The signal is first detected by the SiPMs before traveling to the SMART ASIC and then the FEE. The Vped DAC in the CT5TEA provides a DC bias that is AC coupled to the input signal. The signal from each channel then splits to travel to its respective trigger group (formed by four signal channels) for triggering and to the CTC ASIC for digitization. This diagram represents only one quadrant which contains 1/4 of the channels in the module.}
    \label{fig:pSCT_signal_diagram}
\end{figure}

Additionally, the upgraded FEE will employ the latest TARGET ASICs: the Cherenkov TARGET-C (\ctc) for sampling and digitization and the Cherenkov TARGET-5 Trigger Extension ASIC (\ctf) for trigger generation \cite{2024NIMPA106969841S}. A third ASIC, the SiPM Multichannel ASIC for High-Resolution Cherenkov Telescopes (SMART), has also been developed. This ASIC integrates the signal shaping and pre-amplification stage — currently implemented with discrete components on the TARGET-7 board — directly at the SiPM sensors, improving SiPM bias voltage control, pulse shaping, and amplification. The analog signal will then be transferred to the \ctc\, and \ctf\, ASICs on the FEE. A block diagram of the signal path from photon detection to digitization is shown in Figure \ref{fig:pSCT_signal_diagram}.

\begin{figure}[htp]
    \centering
    \includegraphics[width=0.9\textwidth]{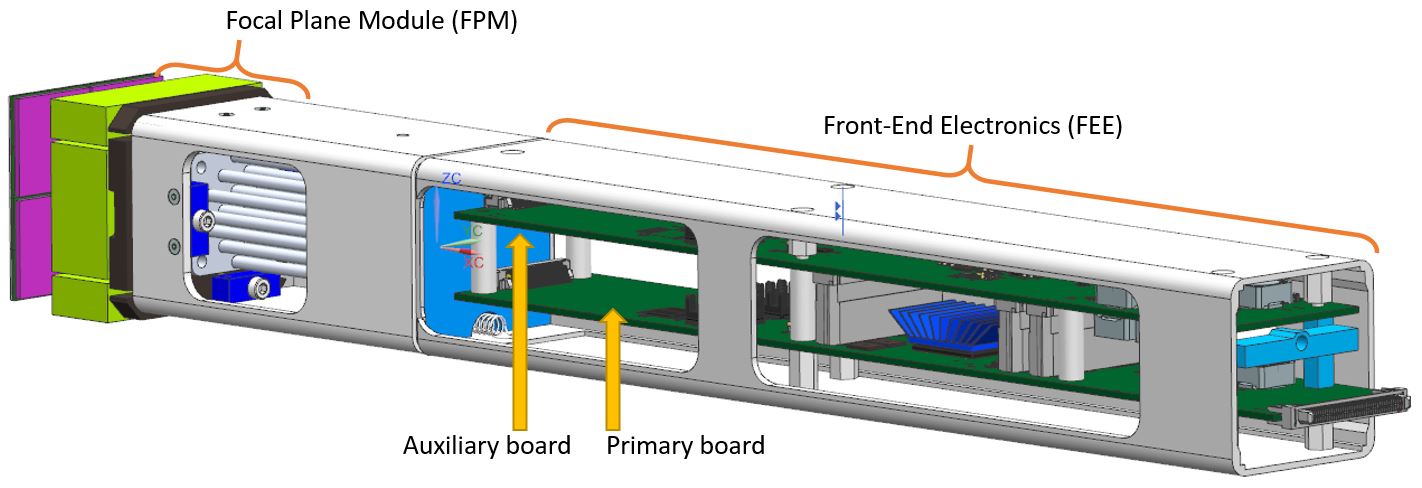}
    \caption{Drawing of the pSCT camera upgraded module. The module consists of two main parts: the FPM (front) and the FEE (rear) parts. The FEE module is composed of an Auxiliary and a Primary board. Between the FPM and the FEE, a 3D-printed cage houses a Peltier cooling system to temperature stabilize the SiPMs \cite{10.1117/12.2530431}. \nor{The length from the rear of the FEE cage (on the right), to the rear of the yellow foam block (on the left), is \qty{33.3}{cm} (\qty{13.1}{in}).}}
    \label{fig:module_drawing}
\end{figure}

\begin{figure}[htp]
    \centering
    \includegraphics[width=0.9\textwidth]{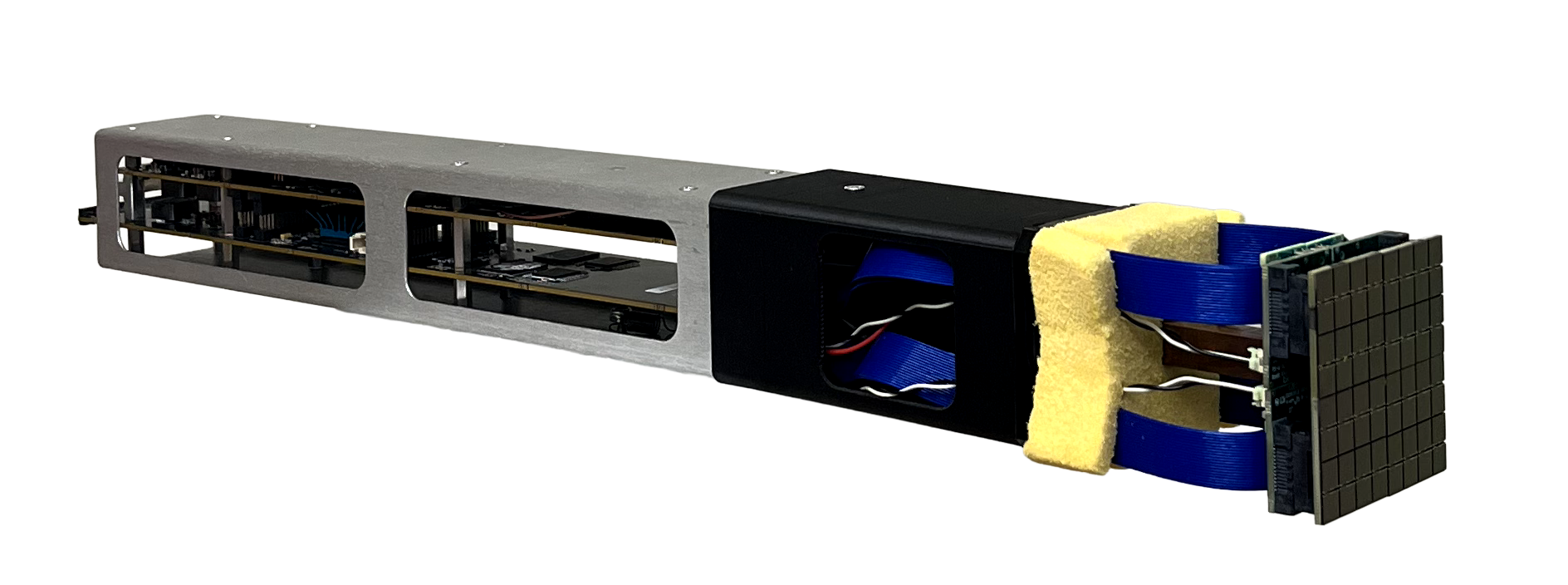}
    \caption{Photograph of the upgraded pSCT camera module. \nor{The length from the rear of the FEE cage (on the left), to the rear of the yellow foam block (on the right), is \qty{33.3}{cm} (\qty{13.1}{in}).}}
    \label{fig:module_photo}
\end{figure}

To stabilize the SiPM temperature, a thermoelectric element managed by a microcontroller on the FEE is thermally coupled to the SiPMs. The thermoelectric element transfers heat to a large heat sink that sits between the focal plane and FEE. The open sides of the FPM cage expose the heat sink to the ambient air so that excess heat can be drawn out of the module by the camera fans and cooling system. A drawing of the upgraded camera module is shown in Figure \ref{fig:module_drawing} and a photograph is shown in Figure \ref{fig:module_photo}. Further details on the FPM and FEE are provided in subsequent sections.

Simulations indicate that the ideal pSCT operating trigger threshold, defined by the rate at which the trigger rates from \nor{night-sky background (NSB)} and protons are matched, is \qty{\sim 1}{\kHz} \nor{(full camera trigger rate)}. Triggering and saving data at a higher rate would require a very high data volume due to the camera's 11,328 channels. Therefore, we have an existing goal of operating at \qty{\sim 2}{\kHz}, with the possibility of achieving up to \qty{10}{\kHz} \nor{for a future SCT}, with sufficient hardware and software improvements. \nor{Initial estimates of the camera temperature predicted an operating temperature of \qty{35}{\degreeCelsius}; therefore, most calibrations and tests were performed at \qty{35}{\degreeCelsius}.} Calculations of the expected camera temperature \nor{were later refined to} show that active cooling and insulation will keep the camera between \qtyrange{27}{30}{\degreeCelsius} and that the internal camera temperature will not vary more than one degree Celsius over the course of the night, despite the presence of significant ambient temperature changes. \nor{The effect of temperature was evaluated for all calibrations to determine whether the temperature differences would affect the camera performance. Any calibrations with significant temperature dependence, such as pedestal subtraction (see Section \ref{TARGET DC Transfer Function}), will be performed in situ on the telescope.}

\subsection{Focal Plane Module}
\label{sect:fpm}

\begin{figure}
\centering
\begin{tabular}{cc}
\includegraphics[width=0.45\columnwidth]{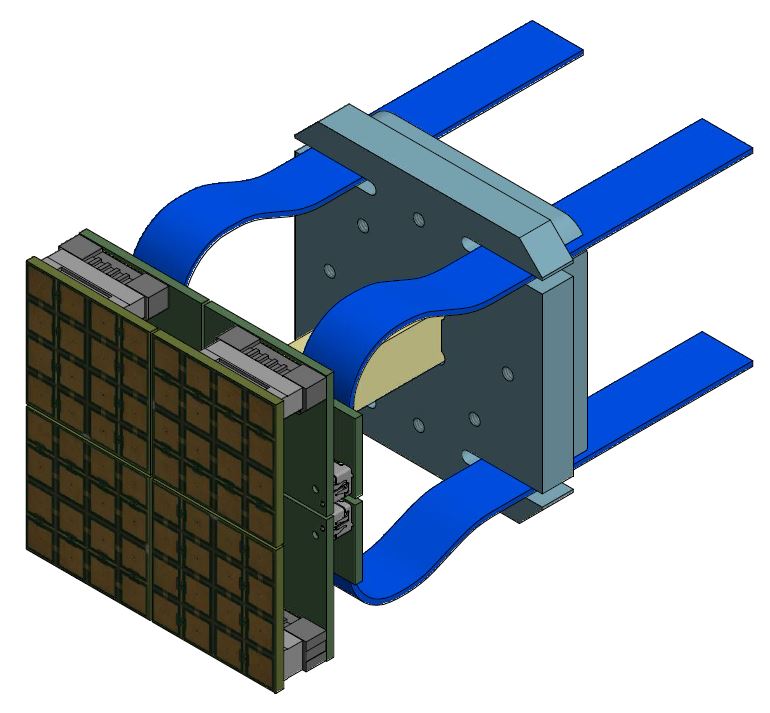}\,
\includegraphics[width=0.45\columnwidth]{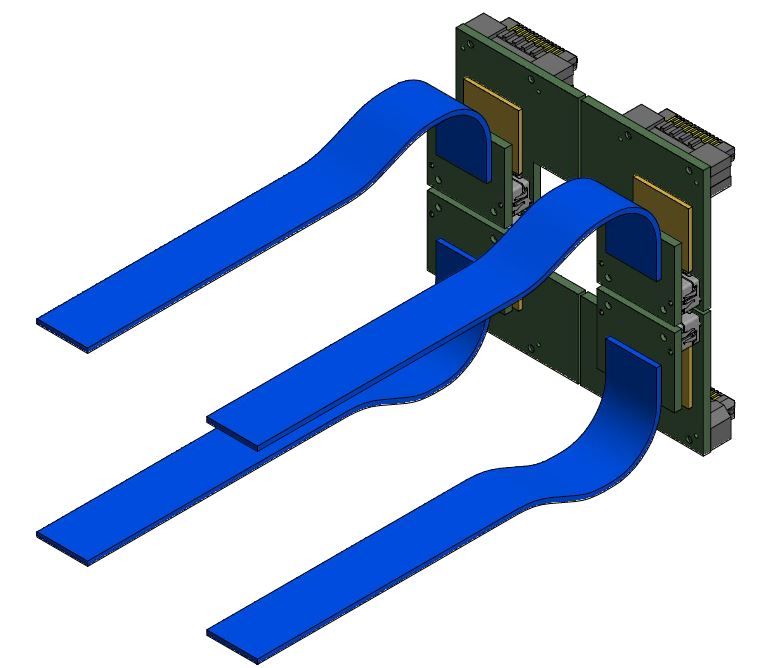} 
\end{tabular}
\caption{Drawing of the FPM as seen from the front (left) and from the back (right). On the left side, the 4 SiPMs tiles are visible, connected to the SMART ASIC boards. The blue ribbon cable connects the SMART ASICs to the FEE. The system is supported by a base-plate (left drawing, on the right end) to avoid movements \cite{10.1117/12.2530431}.} 
\label{fig:fpm_front_and_back}
\end{figure}

Each FPM contains 64 SiPM pixels, arranged in four tiles and connected to the SMART ASIC boards for pre-amplification, as illustrated in Figure \ref{fig:fpm_front_and_back}. From this point on, we will refer to the whole chain composed by SiPM tile, SMART ASIC and corresponding front-end electronics as a \emph{quadrant}. Facing the FPM, we will refer to quadrants 0, 1, 2, and 3, with the quadrants 0 and 1 being the bottom-left and the bottom-right ones, respectively, and quadrants 2 and 3 being the top-left and top-right ones, respectively.

The SiPMs are mounted to a copper post. A curved focal plane that is isochronous is established by varying the length of the copper post between modules. The copper posts are mounted to base-plates which slot into the camera mechanical lattice to create a singular plane. Additionally, the base-plate houses the thermoelectric element and routes the signal and thermistor cables. Small foam blocks (visible in Figure \ref{fig:module_photo}) are attached to the base-plates to create a layer of insulation that thermally isolates the focal plane from the camera electronics.
        
The upgraded pSCT camera will feature a new generation of SiPMs optimized for the Cherenkov emission spectrum, achieving a photon detection efficiency of up to \qty{\sim 50}{\percent} at a wavelength of around \qty{410}{\nm}, while minimizing crosstalk to reduce intrinsic device noise \cite{Merzi_2023, LOIZZO2024169751}. Each 16-channel SiPM tile will be directly connected to a SMART ASIC v2 specific printed circuit board (PCB), enabling immediate amplification of SiPM signals. Only the amplified signals will be transmitted over micro-coaxial ribbon cables to the FEE module.

This design addresses a key limitation of the previous camera, where non-amplified SiPM signals were transmitted via ribbon cables to the FEE, along with connections for ground, SiPM bias supply, and thermistors. This earlier configuration was susceptible to a form of electromagnetic interference, called crosstalk, which could distort the SiPM signals.

\begin{figure}
\centering
\includegraphics[width=0.9\textwidth, bb=120 100 920 500,clip]{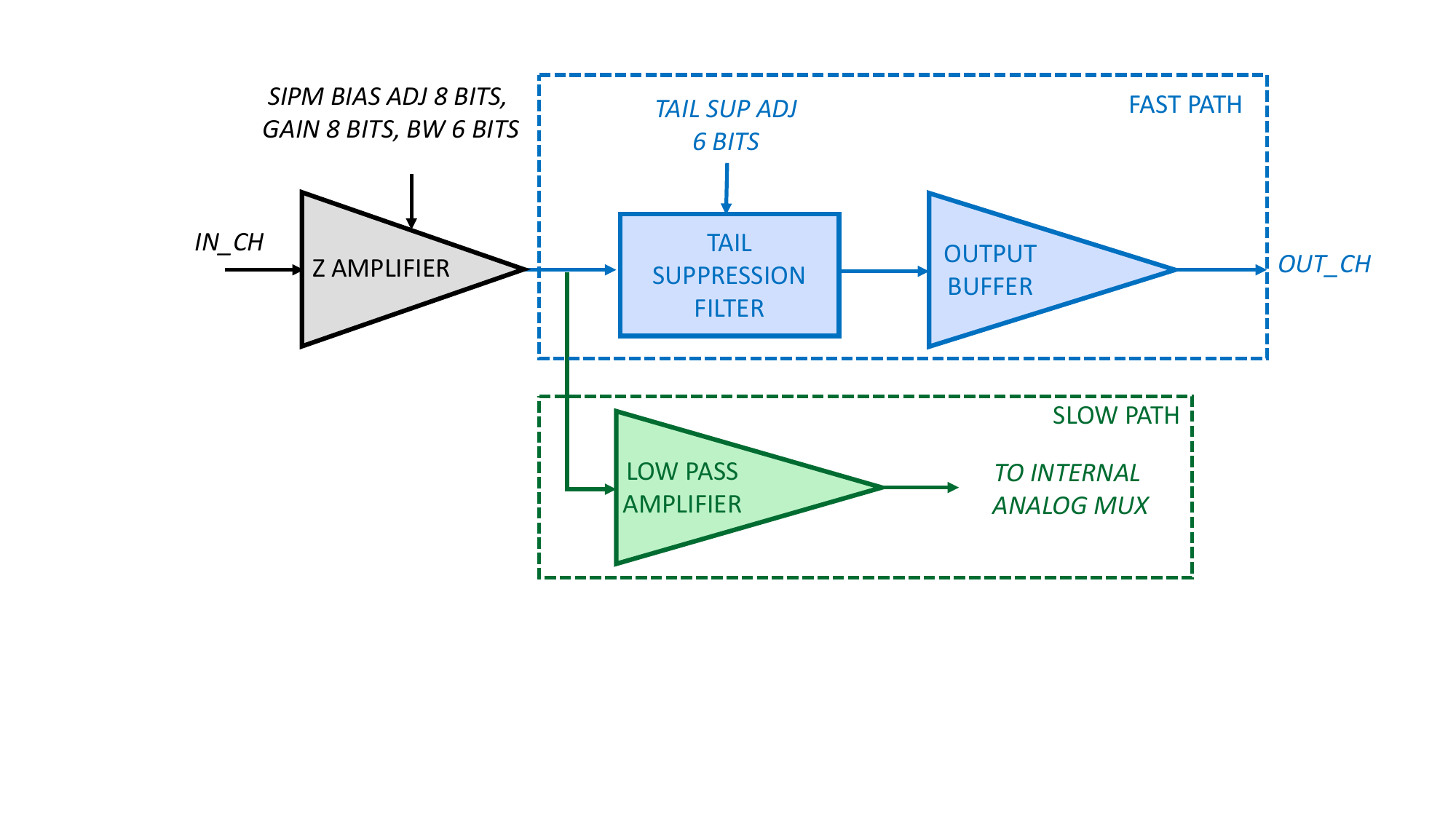}
\caption{Internal architecture of one of the 16 channels of the SMART ASIC. Starting from the left: the trans-impedance amplifier with adjustable bias gain and bandwidth, followed by the two paths. The fast path includes a tail suppression filter and an output buffer, while the slow path incorporates an additional gain stage.}
\label{fig:smart_ch}
\end{figure}

The SMART ASIC is a 16-channel pre-amplifier, developed in a \qty{0.35}{\um} Si-Ge technology, by the electronics CAD service of INFN Bari.
Each of the 16 input channels consists of a trans-impedance amplifier with two paths: a \emph{fast} path optimized for low-noise charge integration and photon counting, and a \emph{slow} path designed for SiPM mean current measurements.
The SMART chip includes a 20-bit register for global adjustment of the gain resistance (8 bits), the bandwidth (6 bits) and the pole-zero network filter for tail suppression (6 bits). Additionally, each channel is equipped with an 8-bit digital-to-analog converter (DAC) for fine-tuning the SiPM bias voltage. The fast path features a signal-shaping filter followed by a single-ended output buffer, while the slow path exploits a low-pass amplifier to measure the SiPM mean current with a resolution of \qty{20}{\nA}. This output is then routed through an analog multiplexer to an internal 10-bit successive-approximation-register (SAR) analog-to-digital converter (ADC). The architecture of the single channel is shown in Figure \ref{fig:smart_ch}.

The bias of each channel can be adjusted using the 8-bit DAC, ensuring that each pixel is properly biased based on its breakdown voltage. This way, the SiPM bias voltage is given by $V_\textrm{bias,SiPM} = V_\textrm{external} - V_\textrm{DAC,channel}$. When the DAC is set to 0, the external voltage is decreased by \qty{0.75}{\V} due to the the base-emitter voltage shift of the heterojunction bipolar transistor employed in the amplification stage. The DAC voltages range between \qty{0.75}{\V} and \qty{1.9}{\V}, allowing individual pixel biases to be adjusted within a \qty{1.15}{\V} range. As will be discussed in Section \ref{sect:gain}, this range is sufficient to adjust the SiPM bias and achieve uniform gain on all channels in a module. In later developments of the SMART, this range has been extended to \qty{1.7}{\V}.

Additionally, the SMART registers can enable or disable individual input channels. The channel power consumption is \qty{20}{\mW}. The SMART ASIC is programmed via a \qty{1}{\MHz} Serial Peripheral Interface (SPI) protocol. The SMART ASIC version used in the camera will have \qty{0.7}{\V} of input dynamic range, although a newer version with a dynamic range of up to \qty{1.3}{\V} may be implemented in some modules.
Details on the SMART performance when paired with FBK SiPMs are provided in Refs.~\citenum{ARAMO2023167839, 10164617}, while the quality control tests performed during SMART ASIC production are described in Ref.~\citenum{ARAMO2023167605}.

\subsection{Front-End Electronics}

In the first version of the pSCT camera, the FEE was distributed across two interconnected circuit boards: the primary and auxiliary boards. The primary board connected directly to the backplane of the camera and hosted the digital control of the FEE along with four TARGET-7 chips, which handled the readout and triggering of the SiPM signals. The auxiliary board was dedicated to the analog processing of the 64 incoming SiPM signals. The analog processing included a pulse-shaping circuit for each channel, which shortened the SiPM pulses using a high-pass filter, as well as 16 current sensors to monitor the combined current from groups of four SiPM pixels. In the upgraded camera, analog processing will be integrated into the SMART ASIC.
To reduce the electronics noise to the lowest possible level, the FEE modules have been completely redesigned compared to the initial development of the pSCT camera. 

\begin{figure}
\centering
\includegraphics[width=0.9\textwidth]{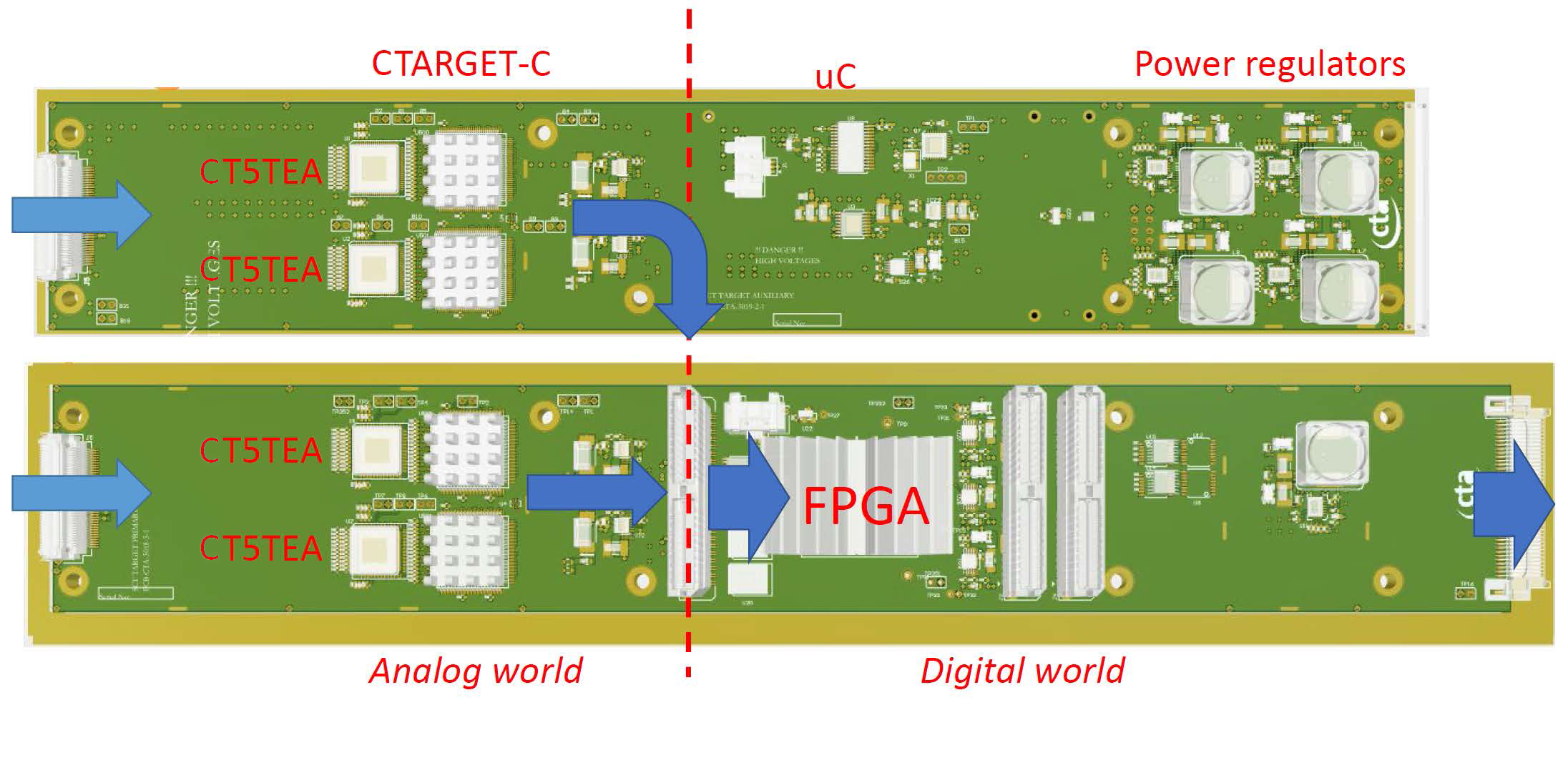}
\caption{Sketch of the FEE boards, the auxiliary (top) and the primary (bottom). A separation between analog and digital/power section can be seen on both boards \cite{10.1117/12.2530431}. }
\label{fig:fee_sketch}
\end{figure}

The new design features two boards, the primary and the auxiliary, with separate areas for analog and digital/power circuitry \cite{10.1117/12.2530431}. This design provides isolation between analog and digital noise sources, preventing any noise return path in the analog chain. The separation of the trigger and sampling paths into two different ASICs reduces crosstalk between the two paths, resulting in a lower trigger threshold and a larger dynamic range. The previous camera \nor{operated at} a trigger threshold of \num{22} \nor{photo-electrons (p.e.)} at \qty{1}{\kHz} and a dynamic range of several p.e. to \num{\sim 330} p.e. The upgraded camera will \nor{operate at} a trigger threshold of \num{\sim 10} p.e. (according to simulations) and a dynamic range of \num{\sim 1} p.e. to \num{> 350} p.e. \cite{2017AIPC.1792h0012F}. Each board houses two CTC and two CT5TEA chips, which collect 32 analog pre-amplified signals. The trigger and digitization tasks are now evenly distributed between the two boards, as shown in Figure \ref{fig:fee_sketch}.

The FEE digitizes SiPM signals via CTC ASICs, while CT5TEA ASICs provide fast low-level triggering. As can be seen in the figure, both boards host two CTC and two CT5TEA ASICs, for a total of 32 digitized channels per board. The SMART ASICs' ribbon cable connectors are shown on the left of the picture. The primary board hosts Xilinx Artix-7 field-programmable gate array (FPGA, XC7A100T-FGG484), which connects to the SMART ASIC amplifier for SiPM bias and signal-shaping adjustments, configures the CTC and the CT5TEA ASICs on both boards, retrieves the digitized data, builds UDP data packets, and sends the packets to a computer via Gbit ethernet. In addition, the FPGA communicates with a microcontroller placed on the auxiliary board that is controlling the thermoelectric element located between the FPM and FEE, to maintain the SiPM sensors at a stable temperature \nor{and prevent variations in the gain over the course of an observing night.}





A single pair of \ctc{} and \ctf, combined with an FPGA, is sufficient to set up a \qty{1}{\giga\sample}/s waveform digitizer with self-triggering capabilities for 16 channels and a 16,384 sample storage depth.

\nor{In the trigger system inside the CT5TEA ASICs, groups of four adjacent channels are summed together to form a single trigger pixel (see Figure \ref{fig:camera_drawing}). If this sum exceeds an adjustable threshold a low-level trigger signal (module trigger) is generated as a low-voltage differential signaling (LVDS) transmission. The threshold can be adjusted by the parameters \texttt{Thresh} and \texttt{PMTRef4}, which sets the threshold applied to the comparator and changes the baseline of the summed signal, respectively. The module trigger signal is sent directly to the backplane. Module triggers are then evaluated by the backplane to produce the so-called backplane trigger, which is produced when three adjacent (i.e. orthogonally or diagonally adjacent) trigger pixels produce a module trigger at the same time. 
In case of a backplane trigger, data for all channels are sent to the FPGA for transfer to storage.}

Each SiPM signal is sent simultaneously to both \ctc\, and \ctf. Each channel has two capacitor arrays, the \emph{sampling buffer} and the \emph{storage buffer}. The sampling array, consisting of two blocks of 32 capacitors (referred to as cells), is used to sample the analog SiPM signal. The cells track the signal until they are sequentially disconnected after a \qty{1}{ns} delay, resulting in \qty{1}{\giga\sample}/s sampling speed, and retain a charge proportional to the amplitude of the SiPM signal during that \qty{1}{ns} interval. The $2\times32$ sampling capacitors operate continuously in ping-pong mode, allowing signal sampling in one block while the other block transfers the sampled voltages to the storage buffer, with the roles reversed in the subsequent half sampling cycle. Ping-pong operation of the sampling array decreases the dead time due to digitization and readout, while ensuring continuous sampling and large trigger latency of up to \qty{16}{\us} at \qty{1}{\giga\sample}/s. The storage buffer consists of a 16,384 sample-deep capacitor array, arranged in 512 blocks of 32 cells. The samples in the storage buffer are continuously overwritten until a high-level \nor{backplane} trigger occurs. Then, the blocks where the \nor{backplane} trigger occurred will no longer be overwritten while the sampling continues for the rest of the storage buffer. \nor{Up to 14} blocks of 32 are randomly accessible for digitization. \nor{One additional storage block is digitized, except in the case the readout window falls exactly in line with storage array block boundaries. For example, for 2-block readout, if the trigger does not occur exactly at the boundary of a storage block, 96 samples (3 blocks) are digitized before the FPGA trims the data down to 64 samples (2 blocks).} In the case of a trigger signal, for each channel, all 32 cells \nor{of a block} are digitized in parallel by Wilkinson ADCs in \qty{\sim 27}{\us} (see Section \ref{Throughput and Power Consumption} for more details) \nor{before the subsequent block is digitized}. 
In the \ctf{} ASIC, the DACs generate an adjustable DC pedestal voltage for each channel, known as the operational pedestal voltage. The pedestal voltage places the signal into the ideal range for readout by the Wilkinson ADCs and allows negative signals to be digitized. 


\subsection{SiPM bias control system}

\begin{figure}[htp]
    \centering
    \includegraphics[width=0.9\textwidth]{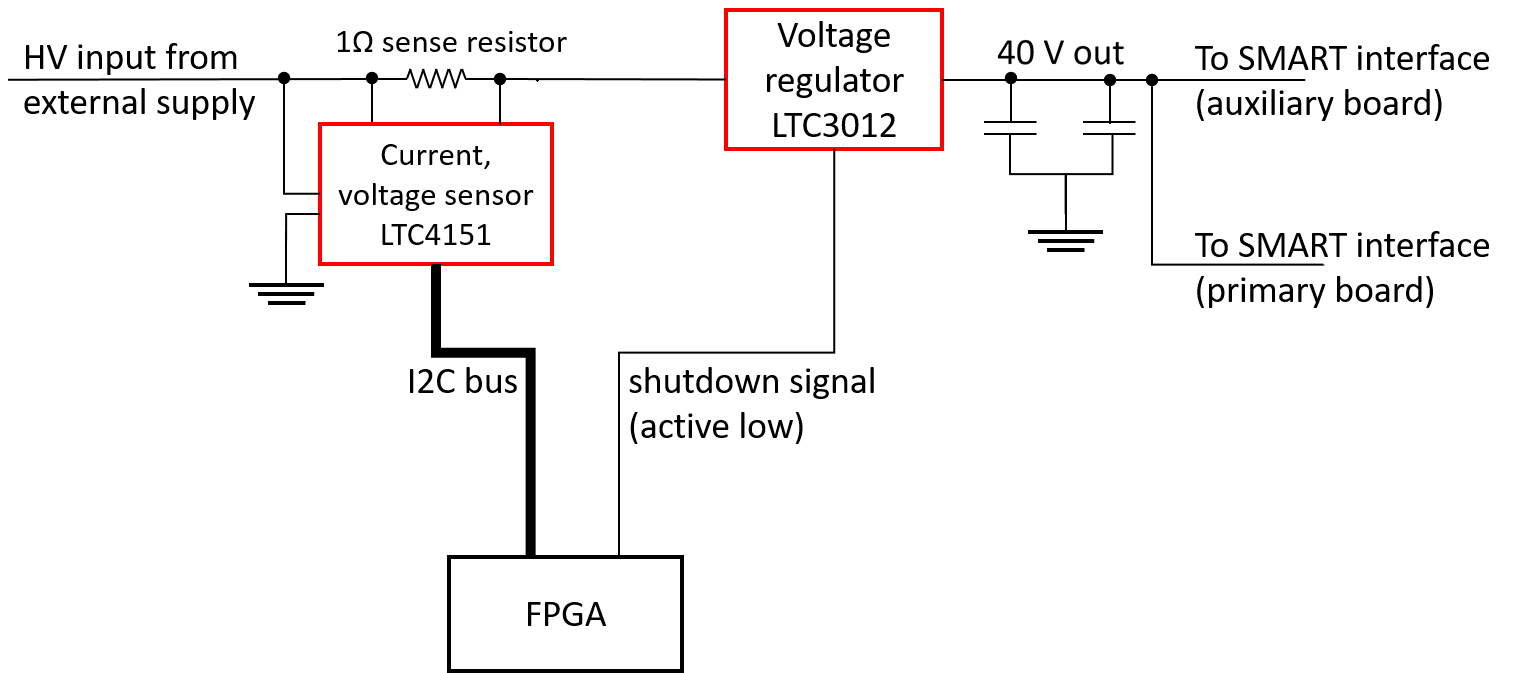}
    \caption{High level block diagram of main FEE components that deliver HV to the FPM.}
    \label{fig:hv_control_system}
\end{figure}

The FEE includes a system to regulate the high side of the bias voltage which gets applied to the common cathode of all SiPMs in one camera module, denoted as Vbias-out on the SMART interface connectors. Figure \ref{fig:hv_control_system} shows a high level block diagram of the HV system with the main components and controls. The bias voltage is provided by the LTC3012 linear regulator, configured to bias the SiPMs at their operating voltage. It has the capability to deliver a constant voltage at a maximum current of \qty{128}{\mA}. The noise on the output voltage is quoted to be \qty{100}{\uV} RMS with appropriate filter capacitors. The FPGA can control the output of the voltage regulator through the shutdown pin. 
In addition to this regulator, a LTC4151 sensor is used to measure current and voltage delivered to the system from the outside, through an I2C bus from the FPGA.

\subsection{SiPM Temperature Monitoring and Control} \label{SiPM Temperature Monitoring and Control}

\begin{figure}[htp]
    \centering
    \includegraphics[width=0.9\textwidth]{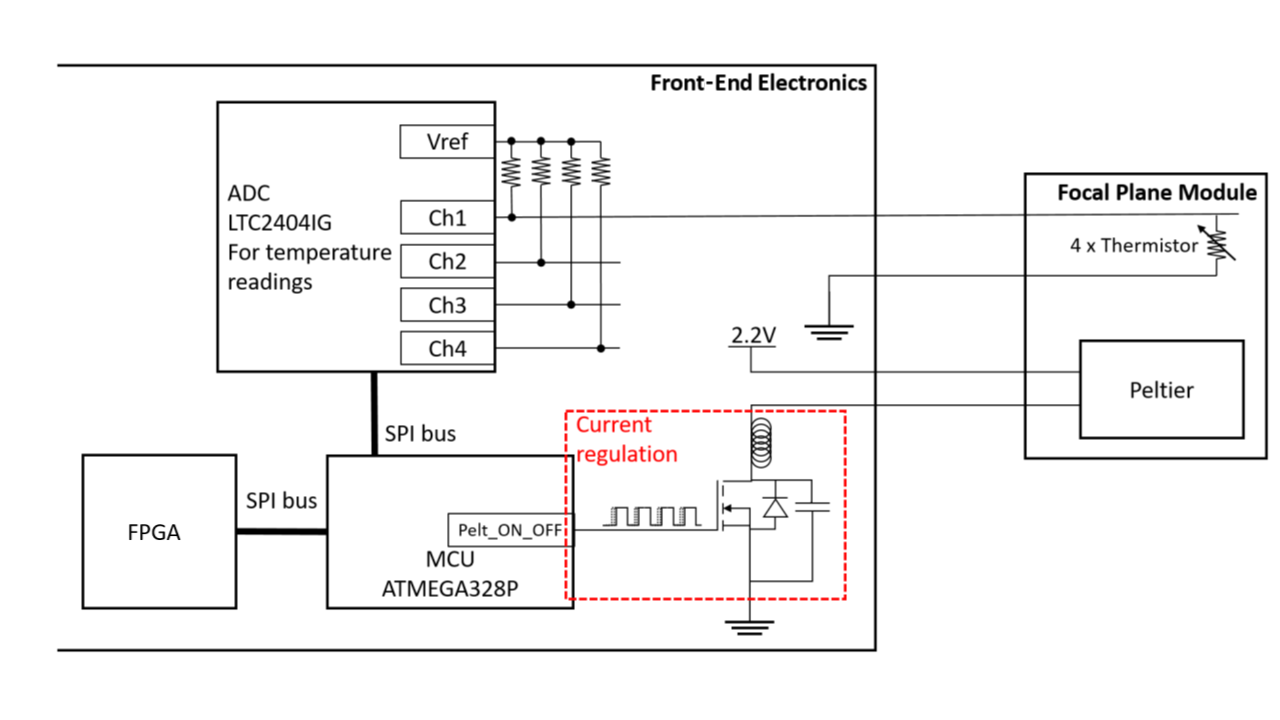}
    \caption{Block diagram of the main FEE and FPM components that are involved in the temperature monitoring and control of the thermoelectric element.}
    \label{fig:Peltier_control}
\end{figure}

The temperature of the SiPMs is measured with four thermistors, one mounted on the back side of each of the four SiPM matrices. The thermistor values are digitized with a 24-bit ADC LTC2404 on the FEE and then transferred into an \nor{ATmega328-P} microcontroller. The microcontroller implements a proportional–integral–derivative (PID) controller that calculates the average of the four thermistor values, compares it to the set temperature, and adjusts the duty cycle of the pulse-width modulated bias of the thermoelectric element based on the temperature difference. The set temperature is forwarded to the microcontroller via an 8-bit SPI interface from the FEE FPGA. Figure \ref{fig:Peltier_control} shows a block diagram of the SiPM temperature control.

\section{Calibration} \label{Calibration}

The tests described in this section were performed on a single preproduction upgrade module for testing purposes. In the course of developing the module production process, a single \ctc{} ASIC was damaged as well as a couple of other channels. Due to these damaged channels, the number of channels evaluated in each calibration varies. \nor{Despite simulations of the upgrade camera temperature later being refined to \qtyrange{27}{30}{\degreeCelsius}, initial estimates predicted a camera temperature of \qty{35}{\degreeCelsius}}. Therefore, unless specified otherwise, all calibrations were performed at \qty{35}{\degreeCelsius}.

Several of the ASIC settings must be tested and calibrated before the module performance is optimized for taking data. During calibration and testing, it was decided that \qty{850}{\mV} would be the operating pedestal voltage due to the low baseline noise at that voltage. Additionally, the SMART ASIC that will be used in the upgraded camera has a dynamic range of up to \qty{1.3}{V}; the maximum voltage to be digitized will therefore be limited to \qty{2.15}{\V} (\qty{850}{\mV} plus \qty{1.3}{\V}), within the range of the \texttt{Vped} DAC (see Section \ref{Vped DAC}). Since the \texttt{Vped} DAC is used to calibrate the full range of the Wilkinson ramp ADC (see Sections \ref{Wilkinson Ramp}, \ref{TARGET DC Transfer Function}), it is important that it can produce voltages of up to \qty{2.15}{\V}. More details on calibration of the \ctc{} and \ctf{} ASICs are available in Ref.~\citenum{2024NIMPA106969841S}.

\subsection{\texttt{Vped} DAC} \label{Vped DAC}

Each module channel is provided with a pedestal voltage by the \texttt{Vped} DAC on the CT5TEA. A 12-bit R-2R ladder supplies over \qty{2.3}{\V} of dynamic range. More details of the \texttt{Vped} DAC can be found in Ref.~\citenum{2024NIMPA106969841S}. The \texttt{Vped} voltage can be probed by measuring test points on the module board. The test points were probed using a Keysight DAQ970A multimeter to measure the \texttt{Vped} voltage at the 512 DAC values that are before or after a fourth or higher bit flip (e.g. 0, 7, 8, 15, 16...), while the remaining \texttt{Vped} DAC values can be linearly interpolated with sufficient accuracy.

\begin{figure}[htp]
    \centering
    \includegraphics[width=0.9\textwidth]{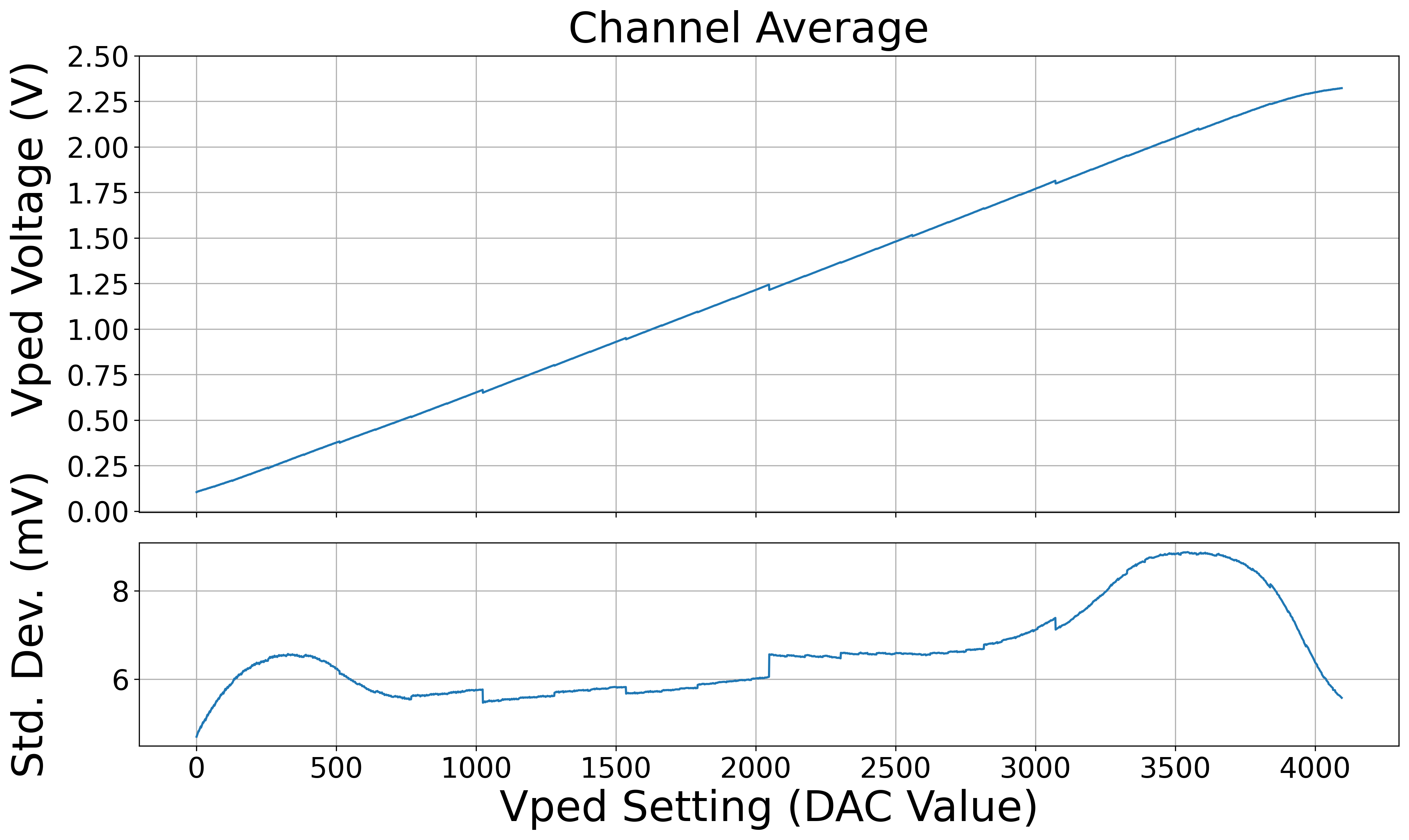}
    \caption{Top: Average transfer function at \qty{35}{\degreeCelsius} relating the \texttt{Vped} DAC values to the \texttt{Vped} voltage for all 62 good channels. Bottom: Standard deviation of the transfer function voltages across all 62 good channels.}
    \label{fig:Vped_DAC_Ch0}
\end{figure}

The average and standard deviation across all channels of the \texttt{Vped} voltage values at \qty{35}{\degreeCelsius} are plotted as a function of DAC value in Figure \ref{fig:Vped_DAC_Ch0}. The R-2R DAC seams, \nor{caused by resistor tolerances,} are visible at the largest bit flips \nor{(all multiples of 512, e.g. 511 to 512, 1023 to 1024, 1535 to 1536, 2047 to 2048, 2559 to 2560, 3071 to 3072, and 3583 to 3584). The resistors are tuned to cause downward shifts in the voltage at seams so that no large gaps in the voltage values available to the DAC can occur.}

The slope of the \texttt{Vped} transfer function can be adjusted through the \texttt{VpedBias} ASIC setting. \texttt{VpedBias} was set to a DAC value of 1000 for these tests but will be updated to a DAC value of 900 during mass calibration to maximize the dynamic range and linearity of the \texttt{Vped} transfer function. \nor{A known flattening of the transfer function can be seen above \num{\sim 3800} DAC.}

\begin{figure}[htp]
    \centering
    \includegraphics[width=0.7\textwidth]{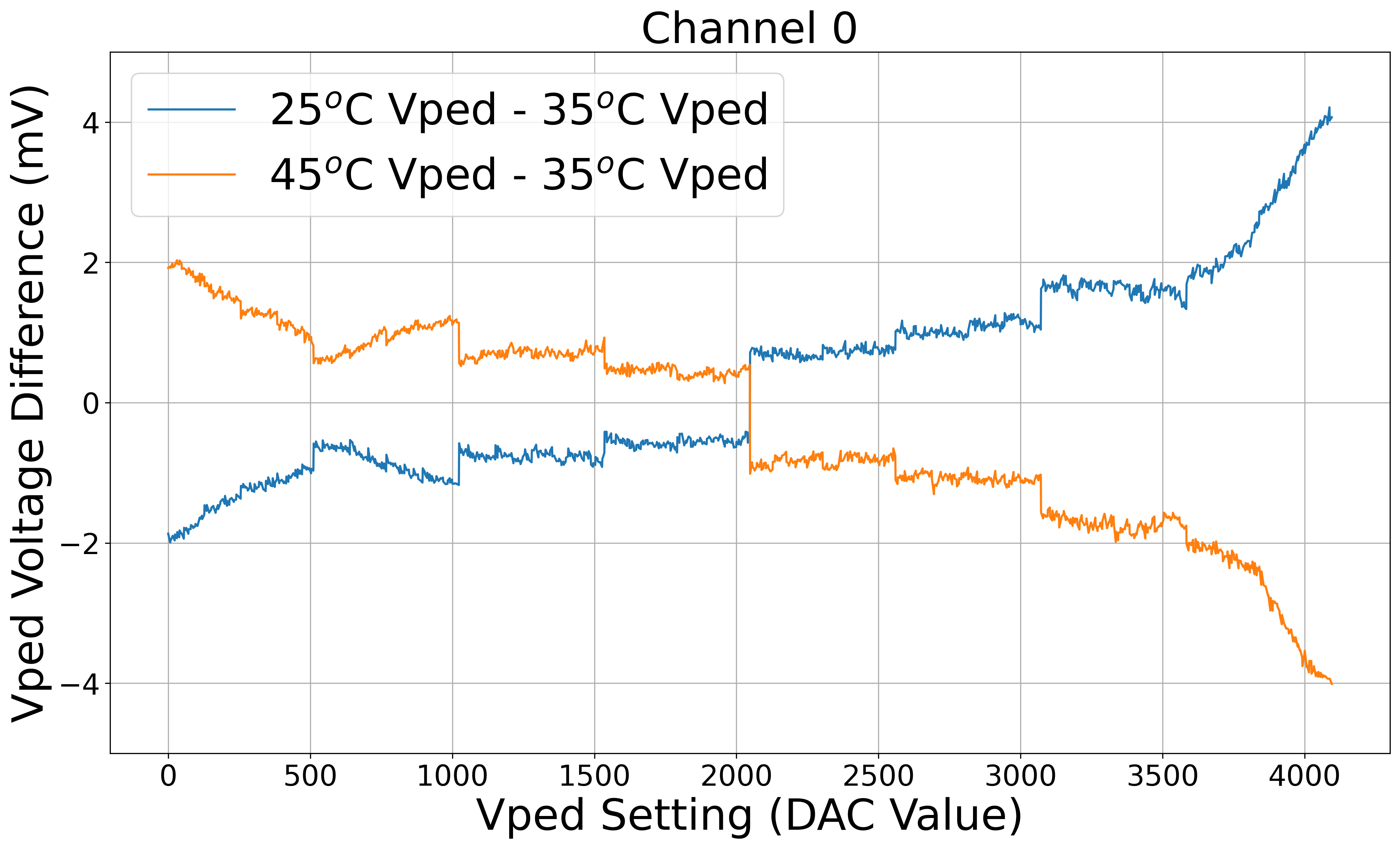}
    \caption{The difference in \texttt{Vped} voltage when comparing channel 0 data taken at \qty{25}{\degreeCelsius} and \qty{45}{\degreeCelsius} to data taken at a nominal \qty{35}{\degreeCelsius} for all \texttt{Vped} DAC count values. \nor{There are large jumps in the voltage difference at the R-2R DAC seams, particularly 2047-2048.}}
    \label{fig:Vped_Temp_Dep_Ch0}
\end{figure}

\begin{figure}[htp]
    \centering
    \includegraphics[width=0.9\textwidth]{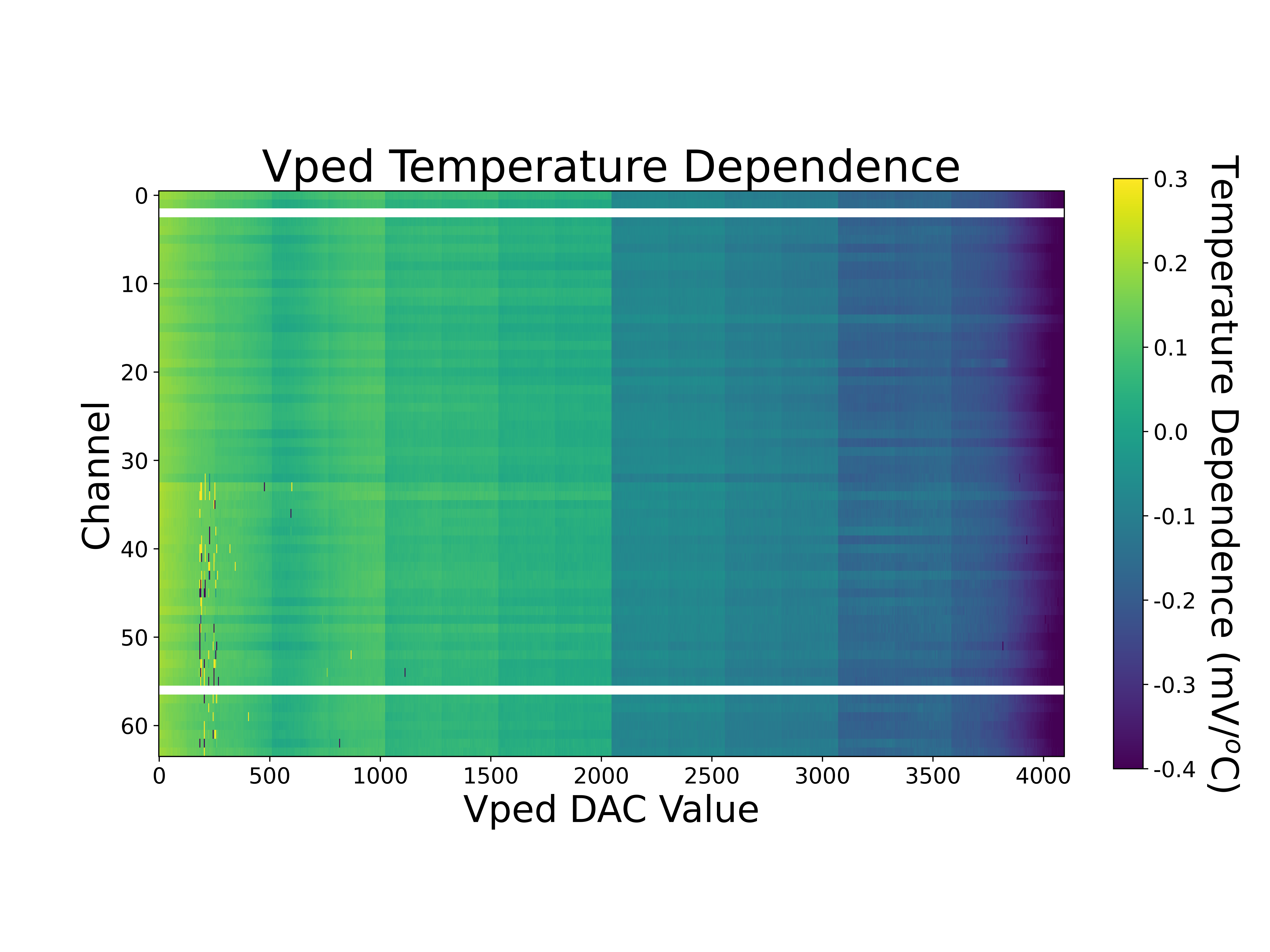}
    \caption{\texttt{Vped} temperature dependence in \unit[per-mode = symbol]{\mV\per\degreeCelsius} for 62 channels across all \texttt{Vped} DAC count values with the the remaining two channels that have measurement issues plotted white. There are some artifacts visible due to poor measurements but they are below the operating voltage of \qty{850}{\mV} (approximately a DAC value of 1300 for the selected \texttt{VpedBias} DAC value of 1000).}
    \label{fig:Vped_Temp_Dep_All}
\end{figure}

The \texttt{Vped} voltage measurements were taken at \qty{25}{\degreeCelsius}, \qty{35}{\degreeCelsius}, and \qty{45}{\degreeCelsius} to characterize the temperature dependence of the DAC. The difference of the \texttt{Vped} transfer functions when comparing channel 0 data taken at \qty{25}{\degreeCelsius} and \qty{45}{\degreeCelsius} to data taken at a nominal \qty{35}{\degreeCelsius} is plotted in Figure \ref{fig:Vped_Temp_Dep_Ch0}. At low DAC values, higher temperatures produce higher voltages and at high DAC values the opposite is true. The absolute voltage difference due to a \qty{10}{\degreeCelsius} temperature difference is as large as \qty{4}{\mV} at the most extreme values, although it is within \qty{1}{\mV} for moderate values. Large jumps in voltage difference exist at the R-2R seam DAC values \nor{(particularly at the 2047-2048 seam where the sign of the temperature difference changes)}, indicating that the R-2R seams in the \texttt{Vped} transfer function are temperature dependent. The temperature dependence of \texttt{Vped} was measured for 62 channels and plotted in Figure \ref{fig:Vped_Temp_Dep_All}.

\subsection{Wilkinson Ramp} \label{Wilkinson Ramp}

The \ctc{} ASIC uses a Wilkinson ADC to digitize the voltage value recorded by the storage capacitors. The Wilkinson ramp of the ADC can be tuned using the \texttt{Vdischarge} and \texttt{Isel} settings, which control the null level and slope of the ramp, respectively. \texttt{Vdischarge} and \texttt{Isel} are set globally for each ASIC and should therefore be optimized at the ASIC level.

To understand the Wilkinson ramp settings indirectly, pedestal data can be taken with a range of set \texttt{Vped} voltage values to evaluate the ADC value the Wilkinson ramp records for a given voltage. The ADC range should be maximized between the operating voltage and maximum voltage (as described in Section \ref{Calibration}) to give the maximum resolution. If the ramp slope is too large, the ADC counts will overflow and the voltage will be recorded as lower than the true value.

Tests of the \texttt{Vdischarge} parameter indicate that the default DAC value of 300 produces sufficiently optimized Wilkinson ramp behavior across ASICs. The \texttt{Isel} setting was optimized for 3 ASICs by measuring the average pedestal ADC value at the maximum \texttt{Vped} voltage for a range of \texttt{Isel} settings and selecting the \texttt{Isel} which maximizes the average ADC counts without overflowing. For ASICs 0, 2, and 3, the ideal \texttt{Isel} DAC settings were 2565, 2525, and 2565, respectively.

\begin{figure}[htp]
    \centering
    \includegraphics[width=0.7\textwidth]{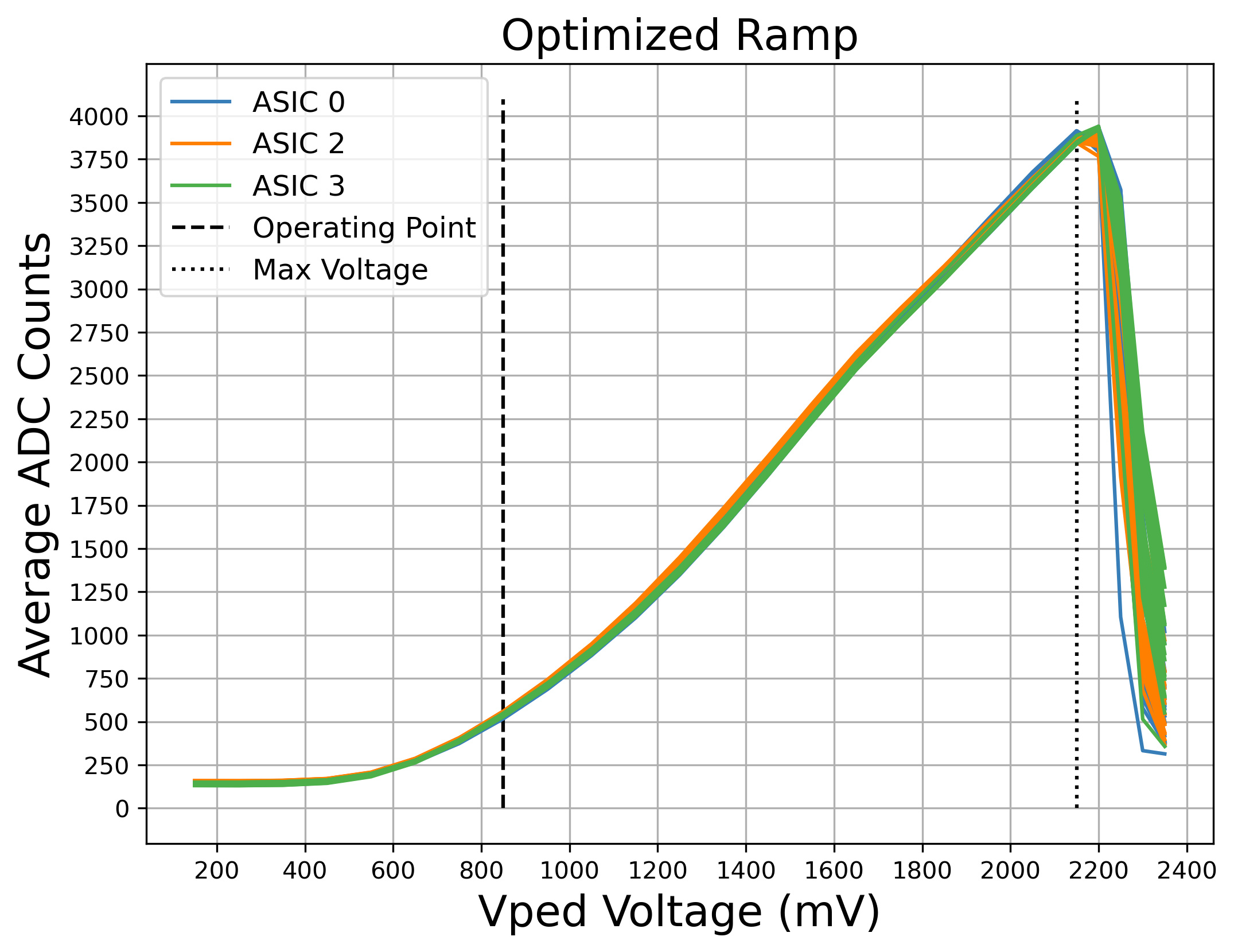}
    \caption{Plot of average ADC counts across all 128 samples of each of the approximately 30,000 waveforms as a function of \texttt{Vped} voltage for all 47 measured channels. The operating voltage of \qty{850}{\mV} and the maximum input voltage of \qty{2150}{\mV} are marked to indicate the useable range.}
    \label{fig:Ramp_Tuning}
\end{figure}

A form of the DC transfer function (discussed in Section \ref{TARGET DC Transfer Function}) after ramp optimization was plotted in Figure \ref{fig:Ramp_Tuning} by taking pedestal data at 25 \texttt{Vped} voltages for 47 channels and averaging the digitized pedestal voltage across approximately 30,000 waveforms for each voltage. The average value across channels at the maximum voltage is 3860 $\pm$ 22 ADC counts, which is within \qty{\approx 5}{\percent} of the maximum value.

\subsection{\texttt{VTrimT}}

\begin{figure}[htp]
    \centering
    \includegraphics[width=0.9\textwidth]{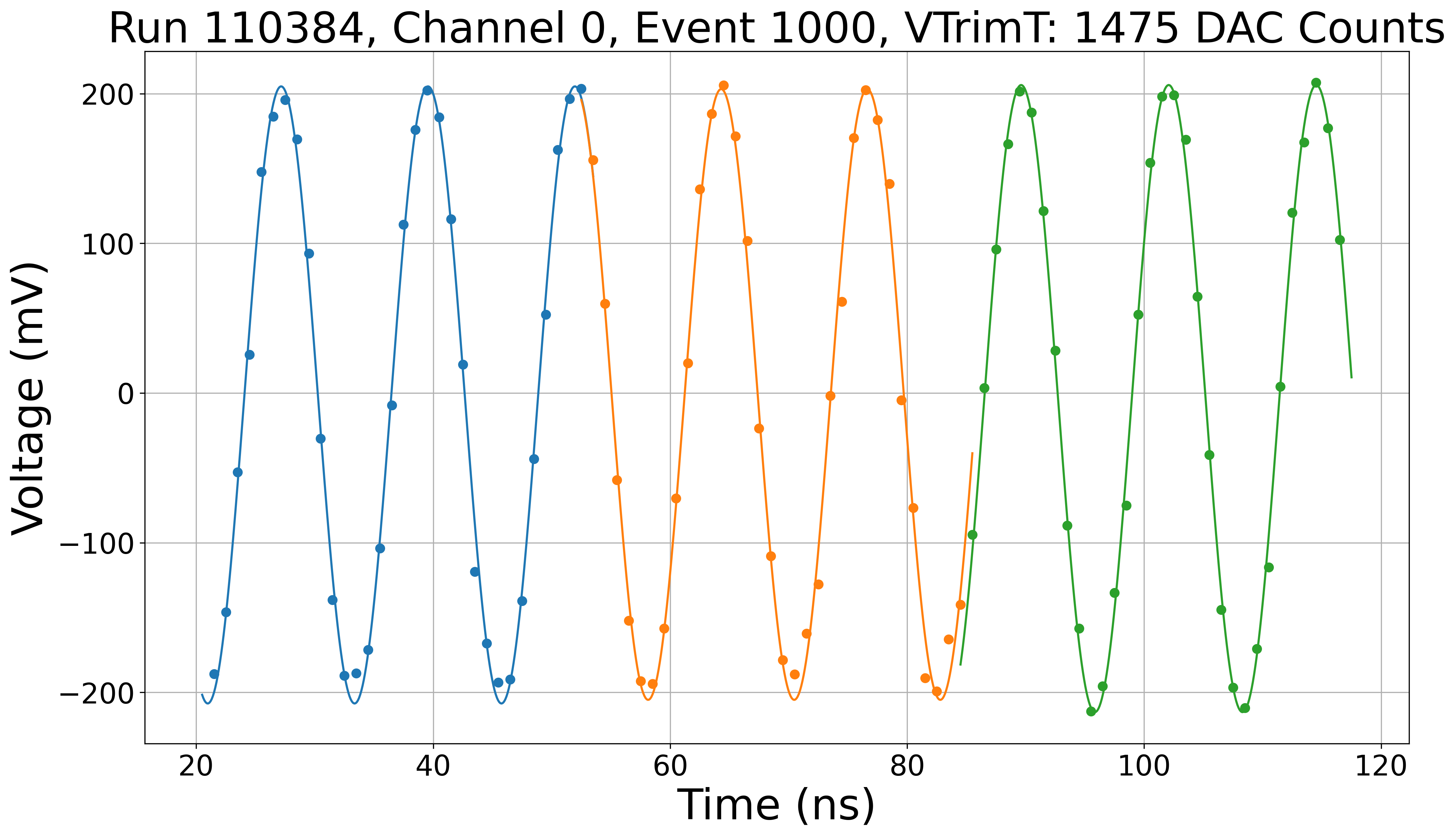}
    \caption{Fit of sine wave electrically injected by the function generator into the FEE through the breakout board. The data (dots) are fit by a sine function (solid lines). The fits extend slightly beyond the data in either direction to show the phase match or mismatch. The first transition (blue to orange) has well-matched phases, while the second transition (orange to green) has a clear phase mismatch.}
    \label{fig:VTrimT_Fit}
\end{figure}

The \texttt{VTrimT} \ctc{} ASIC setting is optimized to obtain a sampling array width of \qty{64}{\ns}. \texttt{VTrimT} is a fine control setting in a feedback loop that maintains the sampling array length. The optimization of \texttt{VTrimT} can be tested by injecting a known electrical signal such as a sine wave and analyzing the module output. This procedure was done using a Keysight 33612A function generator and a custom breakout board that routes the electrical pulses into the module channels via a ribbon cable. Each block of \qty{32}{\ns} is independently fit to a sine wave and the phase is extracted as shown in Figure \ref{fig:VTrimT_Fit}.

\begin{figure}[htp]
    \centering
    \includegraphics[width=0.9\textwidth]{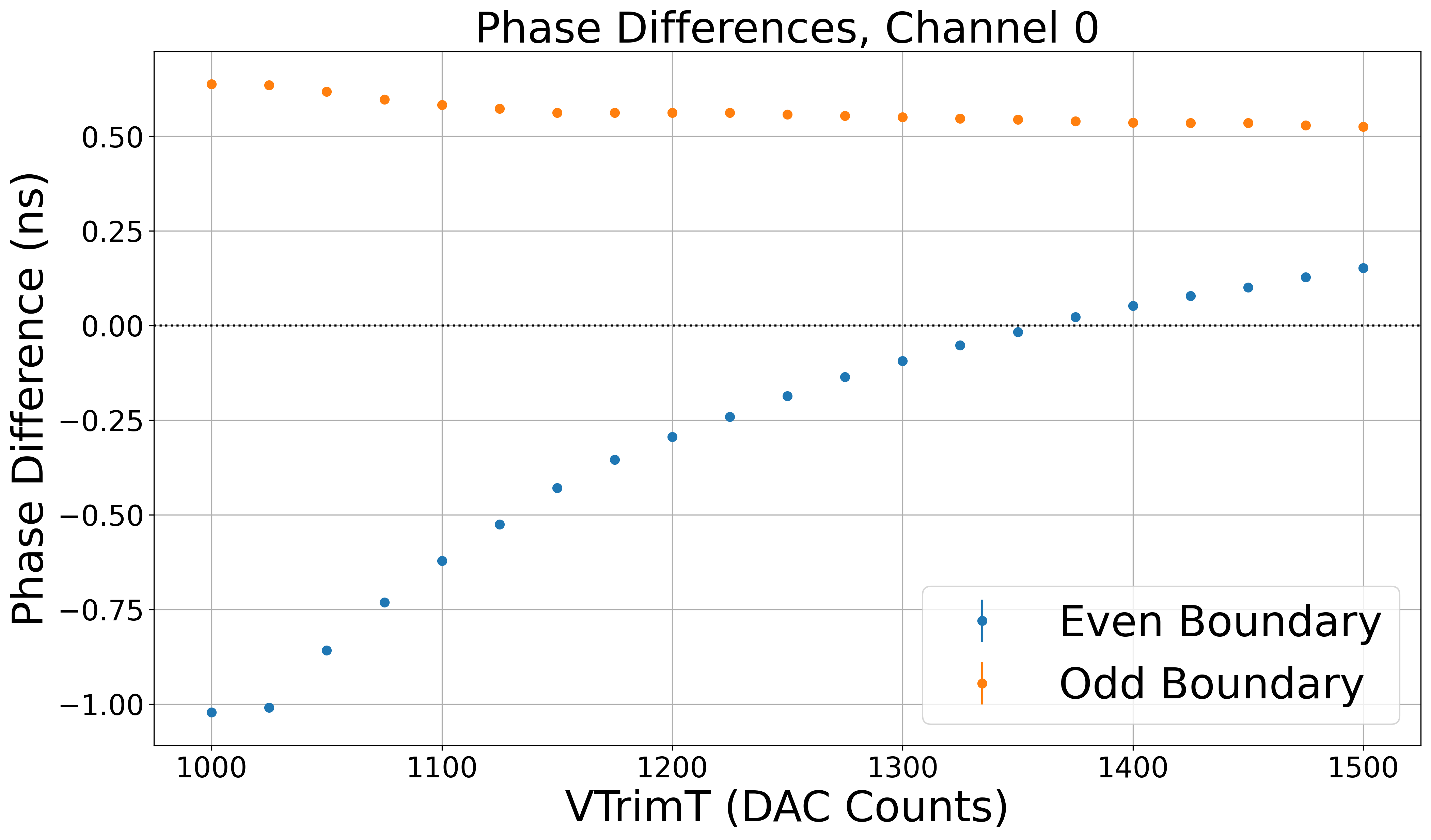}
    \caption{The value of the average phase shift between blocks across all events as a function of \texttt{VTrimT} DAC value. The phase shift is separated into boundaries that precede odd blocks and boundaries that precede even blocks.}
    \label{fig:VTrimT_Summary}
\end{figure}

\texttt{VTrimT} can be tuned by analyzing the difference of fit sine wave phases between blocks, which is minimized when the sampling array width is exactly \qty{64}{\ns}. The phase differences at the block boundaries with odd indexed blocks preceding them are different from those with even blocks preceding them. Since the sampling array is two blocks, the phase difference only occurs before even blocks. A constant phase difference of less than \qty{1}{ns} is expected before odd blocks due to a known clock timing issue. \nor{This phase difference is caused by the cumulative effect of the timebase variation across the 64 cells of the sampling array (see Ref. \citenum{2024NIMPA106969841S} for more details). The mismatch is not planned to be corrected for in the prototype SCT but is not expected to have a significant effect.} The phase differences as a function of \texttt{VTrimT} are plotted for channel 0 in Figure \ref{fig:VTrimT_Summary}. The points were obtained by fitting 10,000 events and averaging the phase differences. For channel 0, the optimized \texttt{VTrimT} value is the value at which the phase difference equals zero. After linearly interpolating between DAC values of 1350 and 1375, the optimum value of 1362 was selected. 

\subsection{TARGET DC Transfer Function} \label{TARGET DC Transfer Function}

\begin{figure}[htp]
    \centering
    \includegraphics[width=0.9\textwidth]{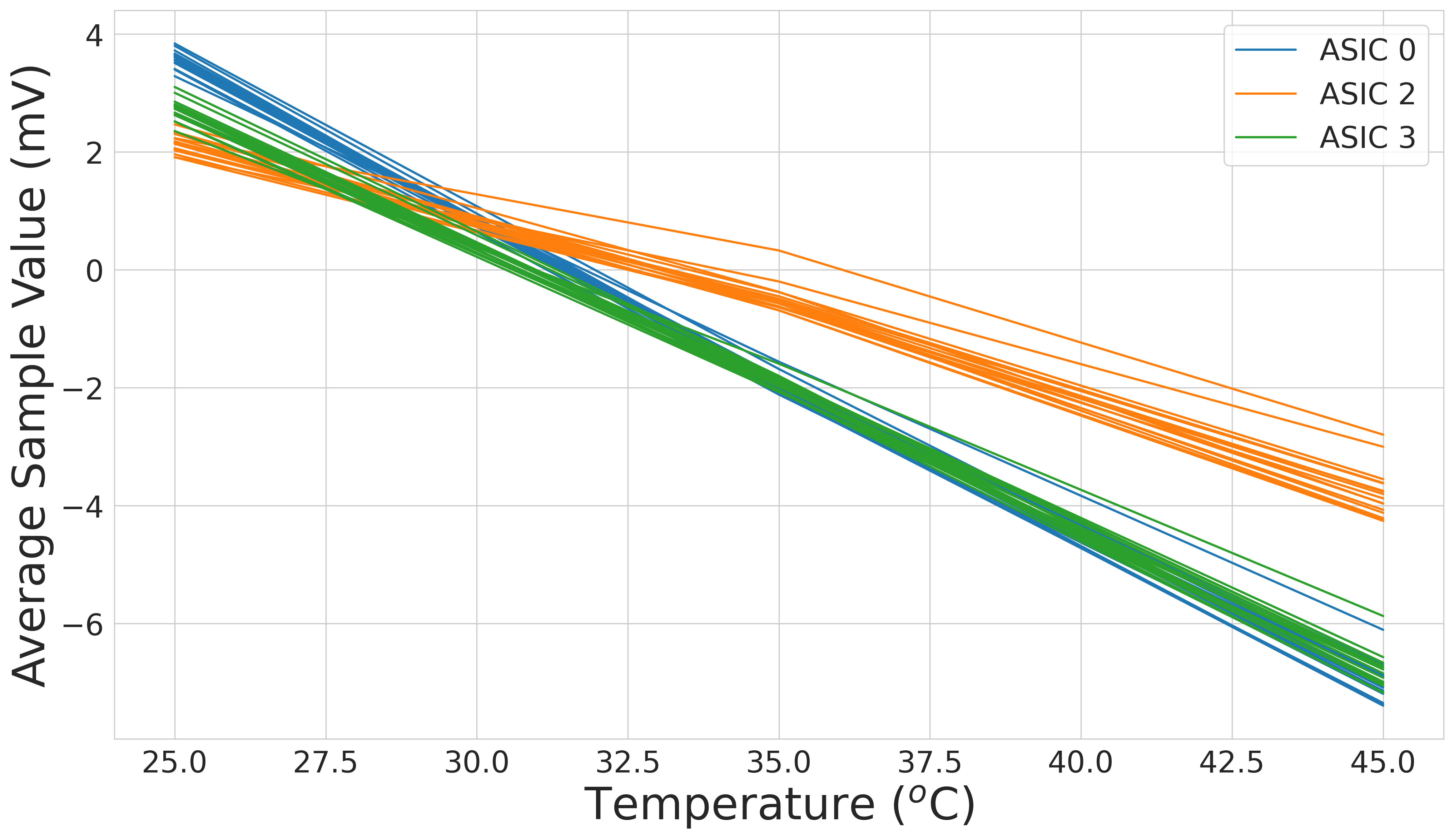}
    \caption{The average sample voltage after pedestal subtraction at \qty{35}{\degreeCelsius} and transfer function calibration as a function of temperature for 47 channels. The channels are color coded by ASIC. The absolute value of each point is not important since the offset can be fixed by subtracting a more suitable pedestal file.}
    \label{fig:pedestal_temp_dep}
\end{figure}

The input signal from the module FPM is added to the baseline supplied by the pedestal voltage after AC coupling. A number of factors contribute to the pedestal voltage that need to be controlled for, including storage array position, sampling array position, and temperature. By taking sufficient data without any input signal, pedestal files can be calculated and later subtracted from signal files to bring the signal baseline to zero. New pedestal files should be created when the module's environmental variables, especially temperature, vary. The module baseline can shift by up to \qty[per-mode = symbol]{0.5}{\mV\per\degreeCelsius} for some channels (Figure \ref{fig:pedestal_temp_dep}). During pSCT operations, we plan to take a new pedestal file before or after every data run. Updating the pedestal runs to use a hardsync trigger will reduce the time of a run to less than a minute. \nor{Nominal data runs will last approximately 30 minutes and with the possibility of introducing a temperature and condition dependent pedestal lookup table, pedestal datataking is not expected to require a significant amount of observing time.}

The output voltage in ADC counts varies by channel and storage cell due to the differing properties of the Wilkinson ADCs and storage capacitors. The accuracy of the recorded voltages can be improved by implementing DC transfer functions to convert the recorded ADC counts to physical voltage values. The transfer function can be constructed by stepping through the voltage pedestal values using the \ctf{} \texttt{Vped} DAC and recording the ADC counts seen by each channel and storage cell at each step.

\begin{figure}[htp]
    \centering
    \includegraphics[width=0.9\textwidth]{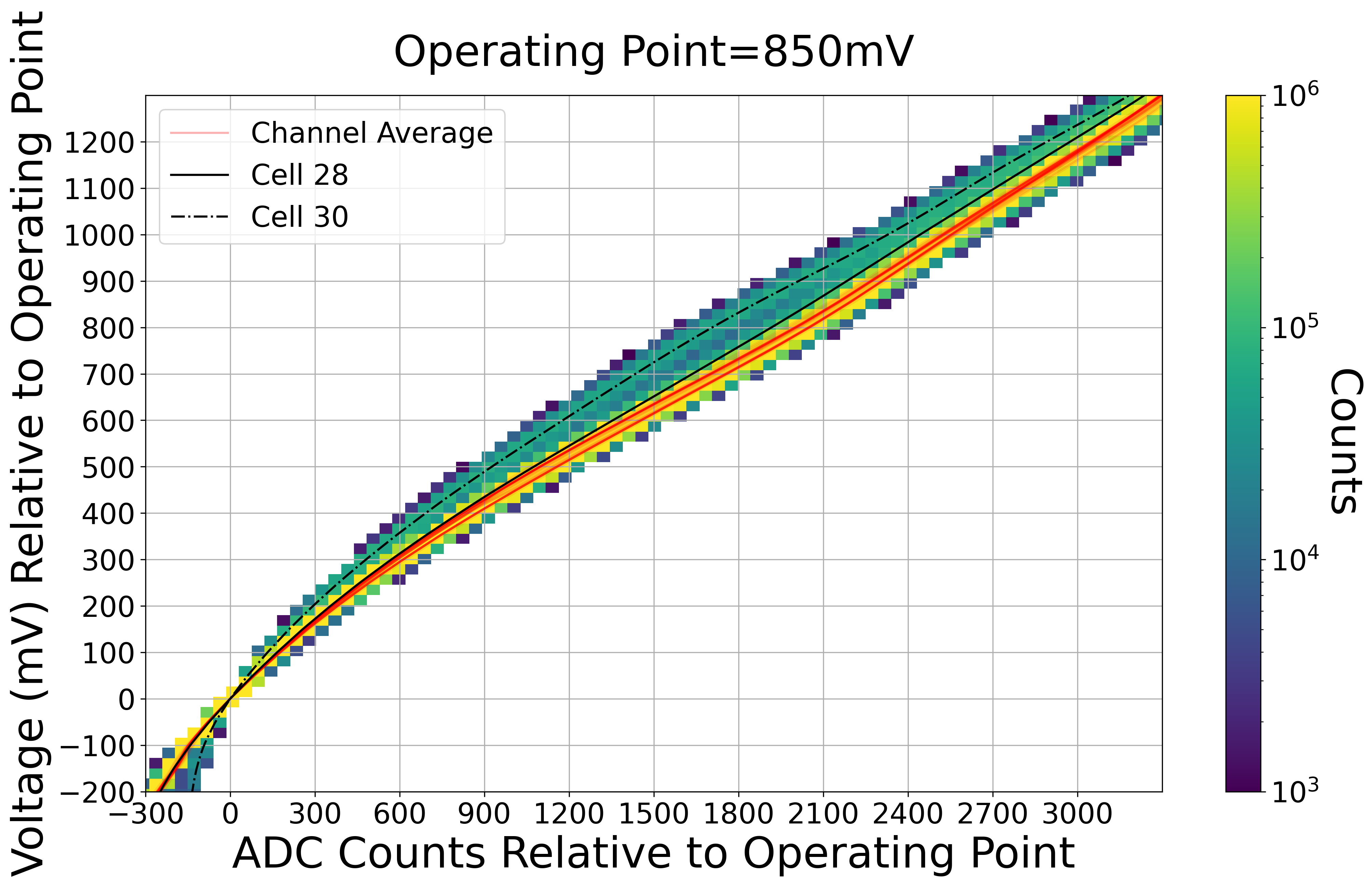}
    \caption{Plot of the TARGET DC transfer function which relates the measured ADC counts to physical voltage. The channel averages are plotted in red over a 2-dimensional histogram of all values. The histogram contains over 400 millions entries so bins with less than 1,000 values are not plotted to remove outliers. Only the useful range of the transfer function is plotted. The differing behavior of cell 28 and cell 30 of each block, as described in Ref.~\citenum{2024NIMPA106969841S}, is averaged and plotted in black.}
    \label{fig:Target_DC_TF}
\end{figure}

To construct the TARGET DC transfer function, data were recorded at 45 pedestal voltage values from \qty{150}{\mV} to \qty{2350}{\mV} in steps of \qty{50}{\mV}. The pedestal voltage was set as a DAC value which was converted from mV using the \texttt{Vped} DAC transfer function. At each pedestal voltage value, approximately 300,000 events were recorded using a \qty{1}{\kHz} external trigger. The trigger rate resulted in a period that did not align with the \qty[group-separator = {,}]{16384}{\ns} period of the storage array and the consecutive events therefore walked through the storage array sufficiently randomly in order to cover all 16,384 storage cells with an adequate number of events: only one readout cell has less than 10 pedestal events for any blocks. Therefore, when considering the distribution of standard errors on the mean of the pedestal value at the operational pedestal, the mean value of the distribution across channels, storage blocks, and readout cells is \qty{0.049}{ADC} counts. A transfer function was calculated for each set of 47 channels and 16,384 storage cells with the channel average plotted in Figure \ref{fig:Target_DC_TF}. The transfer function shows \nor{mostly} linear behavior within the operating range of \qty{850}{\mV} to \qty{2150}{\mV} that extends slightly below and above the range. The \qty{1.3}{\V} of dynamic range extends over a range of approximately 3300 ADC, allowing for a resolution of better than \qty{0.40}{\mV}.

\begin{figure}[htp]
    \centering
    \includegraphics[width=0.9\textwidth]{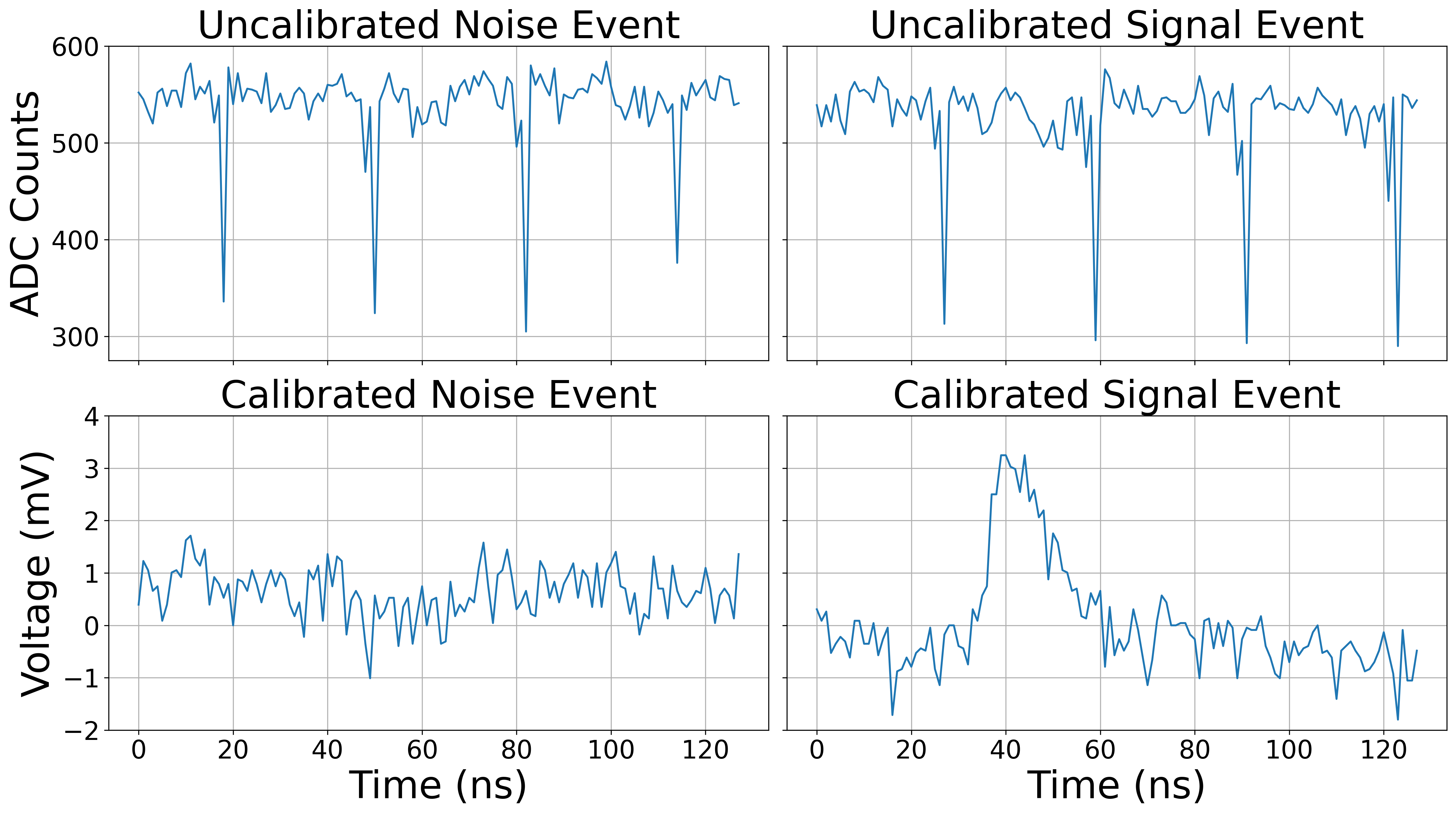}
    \caption{Sample waveform plots for channel 0; the left column of plots is a single waveform with no signal present and the right column is an event with an $\approx$1-2 photoelectron signal. The top row contains the waveforms plotted in raw ADC counts before calibrations. The bottom row contains the calibrated waveforms after pedestal subtraction and transfer function application.}
    \label{fig:Example_Waveforms}
\end{figure}

The effect of using pedestal subtraction and applying the TARGET DC transfer function is shown in Figure \ref{fig:Example_Waveforms}. A low amplitude pulse becomes clearly distinguishable from a noise event and the ADC counts are calibrated to a voltage. \nor{The large drops in ADC counts in the uncalibrated waveforms are a known quality of the \ctc{} ASIC that occur every 32 samples (discussed in Ref. \citenum{2024NIMPA106969841S}) and are removed by pedestal subtraction. For module testing, 4-block readout (128 samples) was used, but for future calibration and data taking, 2-block readout (64 samples) will be used.} 

\section{Performance} \label{Performance}

The tests in Sections \ref{Trigger Performance}, \ref{Baseline Noise}, \ref{Electrical Pulse Injection}, \ref{Optical Pulse Injection}, \ref{Charge Resolution}, and \ref{sect:gain} were performed following the calibrations in Section \ref{Calibration} with the same module and at \qty{35}{\degreeCelsius}, unless otherwise specified. The tests in Sections \ref{Peltier Control}, \ref{Crosstalk}, and \ref{Throughput and Power Consumption} were performed with different preproduction modules. The crosstalk tests in Section \ref{Crosstalk} were performed at room temperature.

\subsection{Testing Setup} \label{Testing Setup}

During testing, the module is placed in an HD-202 environmental chamber which is kept at a stable set temperature by an F4T controller. The module boards are cooled by a QG030-198/12 tangential fan. A Keithley 2200-20-5 DC power supply delivers \qty{12}{\V} to power the module during testing. The module is connected to a custom single module adapter board that provides power, communications, HV, and triggering. A fiber cable runs from the module to a media converter which connects to a lab computer. A Keysight 33612A function generator with two signal and one trigger channel is used to trigger the module, trigger an LED flasher, or directly inject pulses into the module via a custom breakout board. A custom built LED flasher is used for FPM testing. These elements were used in the tests specified in this section except subsections \ref{HV and Current Monitoring} and \ref{Crosstalk}.

For the purpose of trigger testing, \qty{1}{\ohm} resistors were soldered to the test module primary board, connecting the trigger lines, and allowing the module FPGA to count triggers. The production modules will not include the resistors because the camera backplane FPGAs will count the triggers. The custom breakout board is used to inject pulses into the module. For mass calibration, a special trigger adapter board will be used to count triggers via a logic analyzer, as opposed to soldering resistors to the module as the resistors would need to be removed before the module is used with a camera backplane.

\subsection{Trigger Performance} \label{Trigger Performance}

The trigger performance was evaluated for a range of trigger settings at three different ambient temperatures, \nor{\qty{25}{\degreeCelsius}, \qty{35}{\degreeCelsius}, and \qty{45}{\degreeCelsius}}. Electrical pulses from the Keysight 33612A function generator were routed to channel 44 of the module using the custom breakout board. The \qty{10}{\ns} wide pulses with \qty{2.9}{ns} rise and fall times were produced at a rate of \qty{1}{kHz} for \qty{100}{ms} at each injected pulse amplitude and then the number of triggers was read out from a counter on the module FPGA.

The trigger settings \texttt{Thresh} and \texttt{PMTref4}  control the physical trigger threshold. At each temperature, the \texttt{Thresh} was set to a DAC value of 2000 and a search in \texttt{PMTref4} , which is directly related to the physical trigger threshold, was conducted for the noise floor (the lowest \texttt{PMTref4}  setting with any triggers). At each temperature, a total of 60 \texttt{PMTref4} values were scanned in 5 DAC count steps, starting with the \texttt{PMTref4}  setting at the noise floor. At each \texttt{PMTref4}  setting, the injected pulse amplitudes were scanned over in \qty{1}{\mV} increments until rates from 0 to 100 percent of the injected electrical pulse rate were measured. The trigger rates at each amplitude were used to calculate the trigger efficiency - the number of triggers recorded divided by the number of injected pulses.

\begin{figure}[htp]
    \centering
    \includegraphics[width=0.9\textwidth]{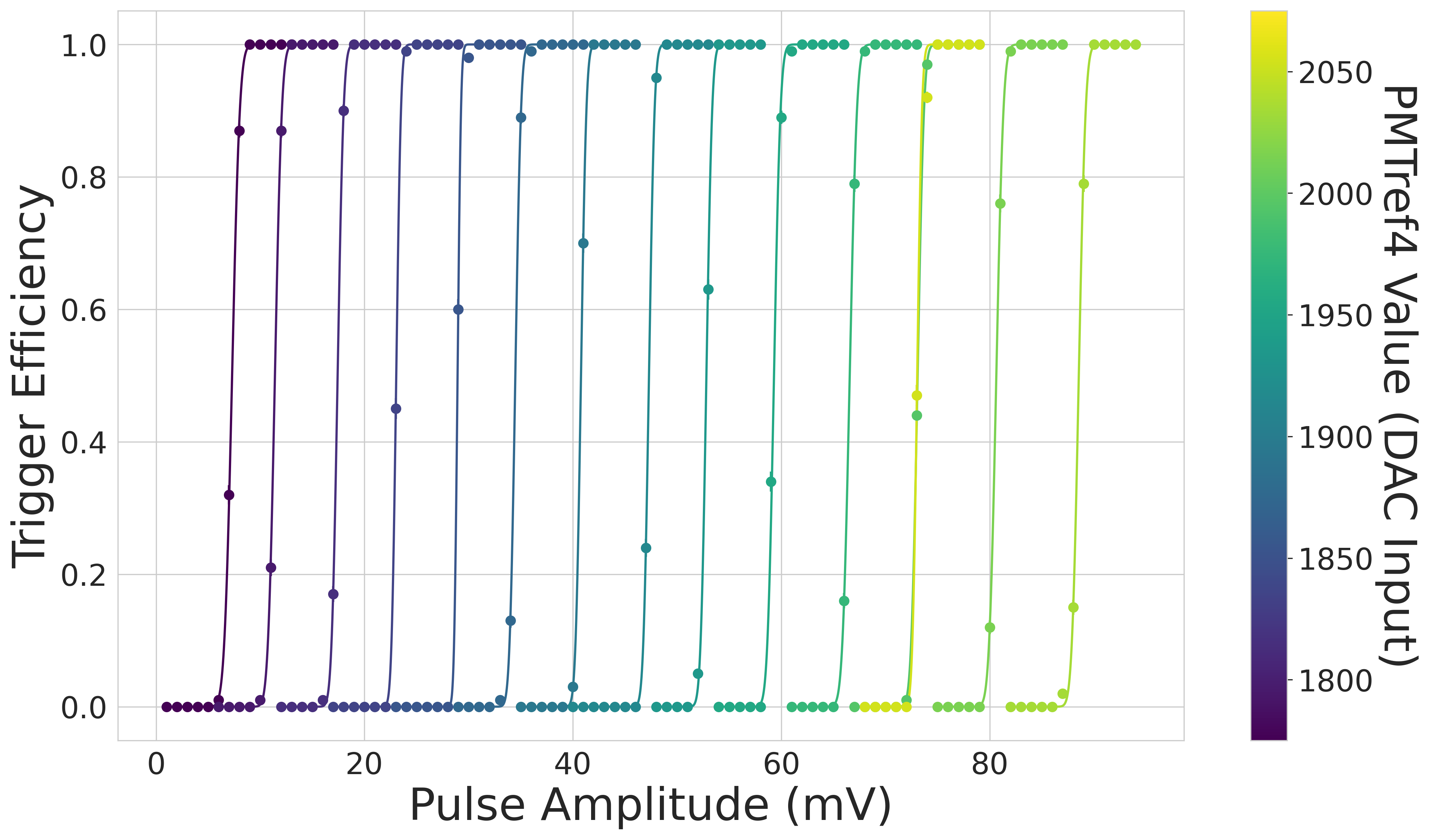}
    \caption{Trigger efficiency of channel 44 plotted as a function of input pulse amplitude for 15 \texttt{PMTref4} DAC values between 1775 and 2075 at \qty{35}{\degreeCelsius}. Only one out of every four \texttt{PMTref4} values were plotted and fit for visual clarity. \nor{The data for some PMTref4 values overlap due to the drop in threshold caused by the 2047-2048 DAC seam.}}
    \label{fig:Trig_Eff_Summary}
\end{figure}

For each \texttt{PMTref4}  setting and temperature pair, the trigger efficiency can be plotted as a function of injected pulse amplitude and fit with equation 1 from Ref.~\citenum{2024NIMPA106969841S}. The $\mu$ and $\sigma$ extracted from the fit are the trigger threshold and the trigger noise. The data points and fit are plotted in Figure \ref{fig:Trig_Eff_Summary} with a color bar for the \texttt{PMTref4}. \nor{The plot includes the large 2047-2048 DAC seam so the threshold drops by $\approx$\qty{15}{mV} for the last set of data and overlaps with a lower PMTref4 set of data.} The trigger was tested up to an amplitude of over \qty{1.5}{\V}, although the data is not presented here.

\begin{figure}[htp]
    \centering
    \includegraphics[width=0.9\textwidth]{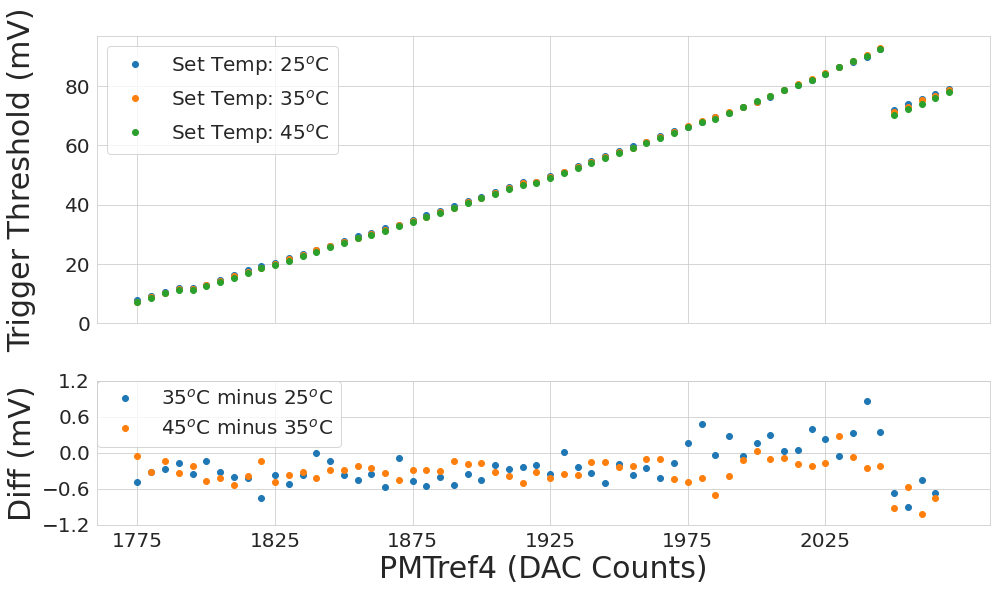}
    \caption{Top: The trigger threshold values of channel 44 plotted as function of \texttt{PMTref4} DAC value for data taken at \qty{25}{\degreeCelsius}, \qty{35}{\degreeCelsius}, and \qty{45}{\degreeCelsius}. A linear fit to the data up to the R-2R seam at a \texttt{PMTref4} DAC value of 2048 gives slopes of \qty{0.30}{\mV} / DAC, \qty{0.31}{\mV} / DAC, and \qty{0.31}{\mV} / DAC for the data at \qty{25}{\degreeCelsius}, \qty{35}{\degreeCelsius}, and \qty{45}{\degreeCelsius}, respectively. Bottom: The difference in trigger threshold voltage between \qty{35}{\degreeCelsius} and \qty{25}{\degreeCelsius} and between \qty{45}{\degreeCelsius} and \qty{35}{\degreeCelsius} as a function of \texttt{PMTref4}  DAC value for channel 44.}
    \label{fig:Trig_Thresh_and_Temp_Diff}
\end{figure}

\begin{figure}[htp]
    \centering
    \includegraphics[width=0.9\textwidth]{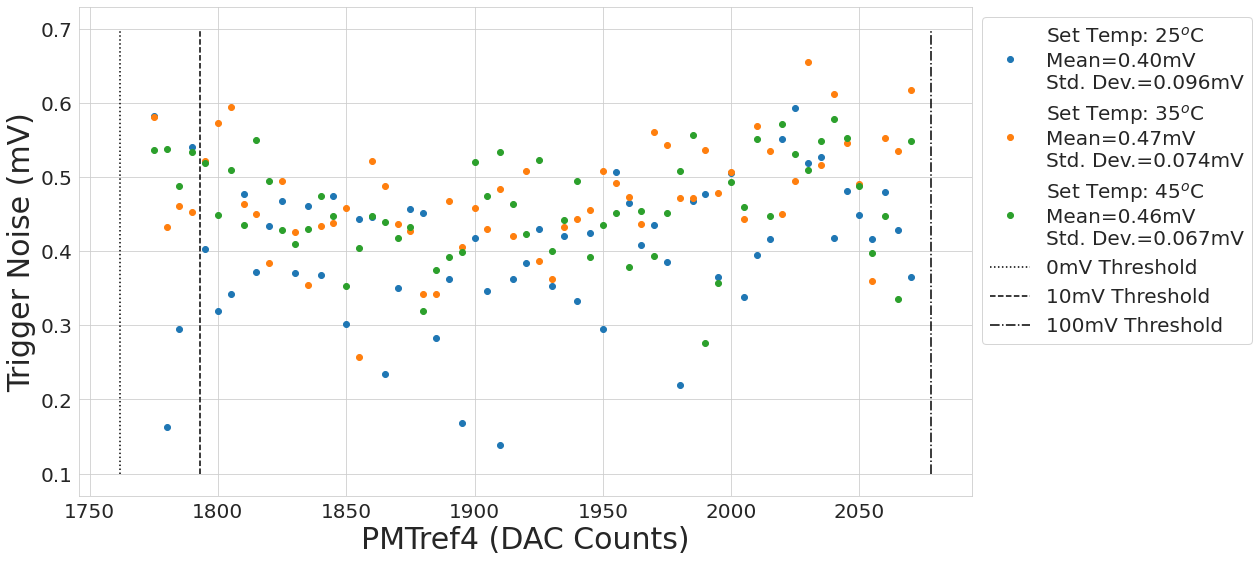}
    \caption{The trigger noise of channel 44 as a function of \texttt{PMTref4}  DAC value for data taken at \qty{25}{\degreeCelsius}, \qty{35}{\degreeCelsius}, and \qty{45}{\degreeCelsius}. \nor{The mean and standard deviation of the distributions are included in the legend. The \qty{0}{mV}, \qty{10}{mV}, and \qty{100}{mV} thresholds for \qty{35}{\degreeCelsius} are plotted as vertical lines for reference.}}
    \label{fig:Trig_Thresh_Sig}
\end{figure}

The trigger threshold can be plotted as a function of \texttt{PMTref4} for each temperature as in Figure \ref{fig:Trig_Thresh_and_Temp_Diff}. The data are mostly linear aside from the clearly visible R-2R DAC seams of the \texttt{PMTref4} DAC. The difference in the trigger threshold values between different temperatures is calculated for each \texttt{PMTref4} and plotted in Figure \ref{fig:Trig_Thresh_and_Temp_Diff} as well. The differences between temperatures are relatively consistent across PTMref4 values and are in general below \qty{0.5}{\mV} of absolute value for a \qty{10}{\degreeCelsius} change. The trigger noise is similar at each temperature and consistent across \texttt{PMTref4}  values, ranging between \qtyrange{0.3}{0.6}{\mV}. It is plotted in Figure \ref{fig:Trig_Thresh_Sig}.

\subsection{SMART Slow monitoring calibration} 
\label{HV and Current Monitoring}

The SMART ASIC is equipped with a low-pass amplifier which allows the measurement of the SiPM mean current with a resolution of \qty{20}{\nA}. This amplifier is based on a differential stage with respect to a reference voltage. This voltage is provided by an internal circuit with a adjustable value. In order to properly exploit the ADC range with each SiPM channel, this internal value needs to be tuned independently for each channel. A procedure is implemented to measure the ADC values in absence of a SiPM signal as a function of the reference voltage, named DAC-17. A linear curve is obtained and the optimal reference value of DAC-17 is set at the rising of the linear increase.

\subsection{SiPM Temperature Control} \label{Peltier Control}

The SiPM temperatures are adjusted with a thermoelectric generator coupled to the SiPM matrices via a copper post. The bias of the thermoelectric generator is pulse-width-modulated by a PID controller as explained in Section \ref{SiPM Temperature Monitoring and Control}. The purpose of the temperature control is to maintain a stable temperature throughout the night, with ambient temperatures expected to fluctuate on the time scale of minutes or hours. The expected camera operating temperature is discussed in Section \ref{sect:design}. The raw thermistor values are converted to absolute temperature by calibrating the raw values at known ambient temperatures. For this purpose, a camera module is placed in a temperature-controlled chamber, and the temperature of the chamber is set to different values. Once the camera module has adjusted to the ambient temperature, it is turned on, and the four thermistor ADCs are recorded instantly. The procedure is repeated at \qty{25}{\degreeCelsius}, \qty{35}{\degreeCelsius}, and \qty{45}{\degreeCelsius}, and the ADC values are then linearly fit to the ambient temperatures to extract the calibration values from the fit function. The calibrated measured thermistor temperatures do not exactly match the ambient temperatures. The most likely explanation is that the use of a linear function to calibrate the ADC count to temperature relationship does not reflect some non-linear behavior of the thermistors or ADCs. 

\begin{figure}[htp]
    \centering
    \includegraphics[width=0.7\textwidth]{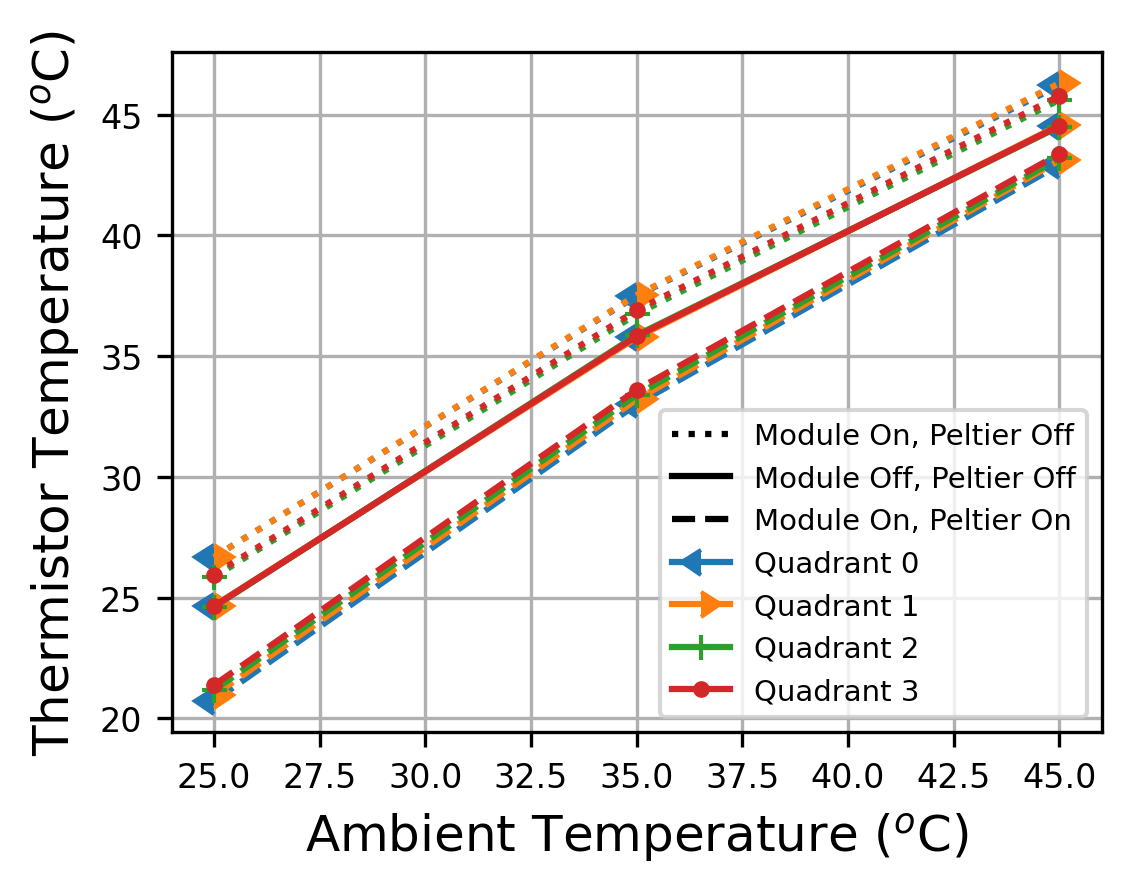}
    \caption{The calibrated thermistor temperature for each module quadrant as a function of ambient temperature. The solid lines indicate the data taken for thermistor calibration purposes, with the module and Peltier element turned off. The dashed and dotted lines indicate the data taken with the module triggering at the nominal data taking rate of \qty{1}{\kHz} with HV on and enabled. The dotted and dashed lines correspond to the Peltier being off and on, respectively. The measured temperatures for each quadrant are very close and overlap on the plot.}
    \label{fig:thermistor_SMART}
\end{figure}

To test the basic functionality of the temperature control, the module was allowed to equilibrate while triggering at a nominal data-taking rate of \qty{1}{kHz} with HV on and enabled. The calibrated thermistor temperatures were then reported for two cases: when the thermoelectric generator was on and when it was off. The measurements indicate that the thermoelectric generator is capable of maintaining a SiPM temperature at a few degrees below the ambient temperature, providing a sufficient cooling margin to maintain a constant temperature during data acquisition. The calibrated thermistor temperatures are shown as a function of the ambient temperature in Figure \ref{fig:thermistor_SMART} for the thermistor calibration run (with the module off) and two different data taking runs where the module was switched on. The temperature control circuit maintained the focal plane at a constant temperature when the control loop was activated and confirmed to be functioning properly. More thorough tests of the control loop in realistic camera conditions are performed in Ref.~\citenum{OTTE201585}.

\subsection{Baseline Noise} \label{Baseline Noise}

\begin{figure}[htp]
    \centering
    \includegraphics[width=0.8\textwidth]{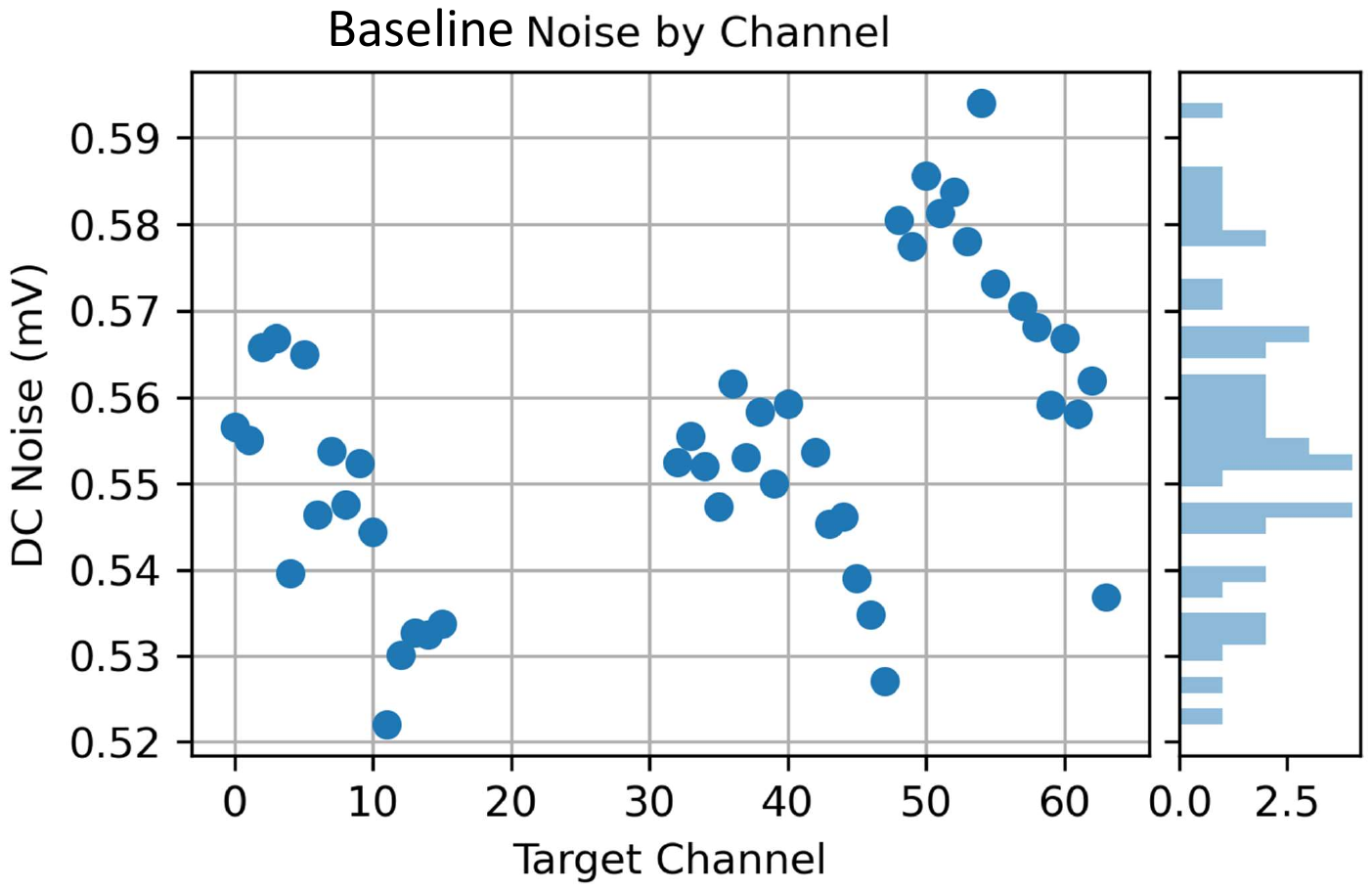}
    \caption{Baseline noise in mV across 46 channels, calculated as the standard deviation across 302,071 pedestal events. A histogram of the values is displayed on the right. Clustering is visible in groups of 16 channels since they share a quadrant of ASICs.}
    \label{fig:DC_Noise}
\end{figure}

\begin{figure}[htp]
    \centering
    \includegraphics[width=0.7\textwidth]{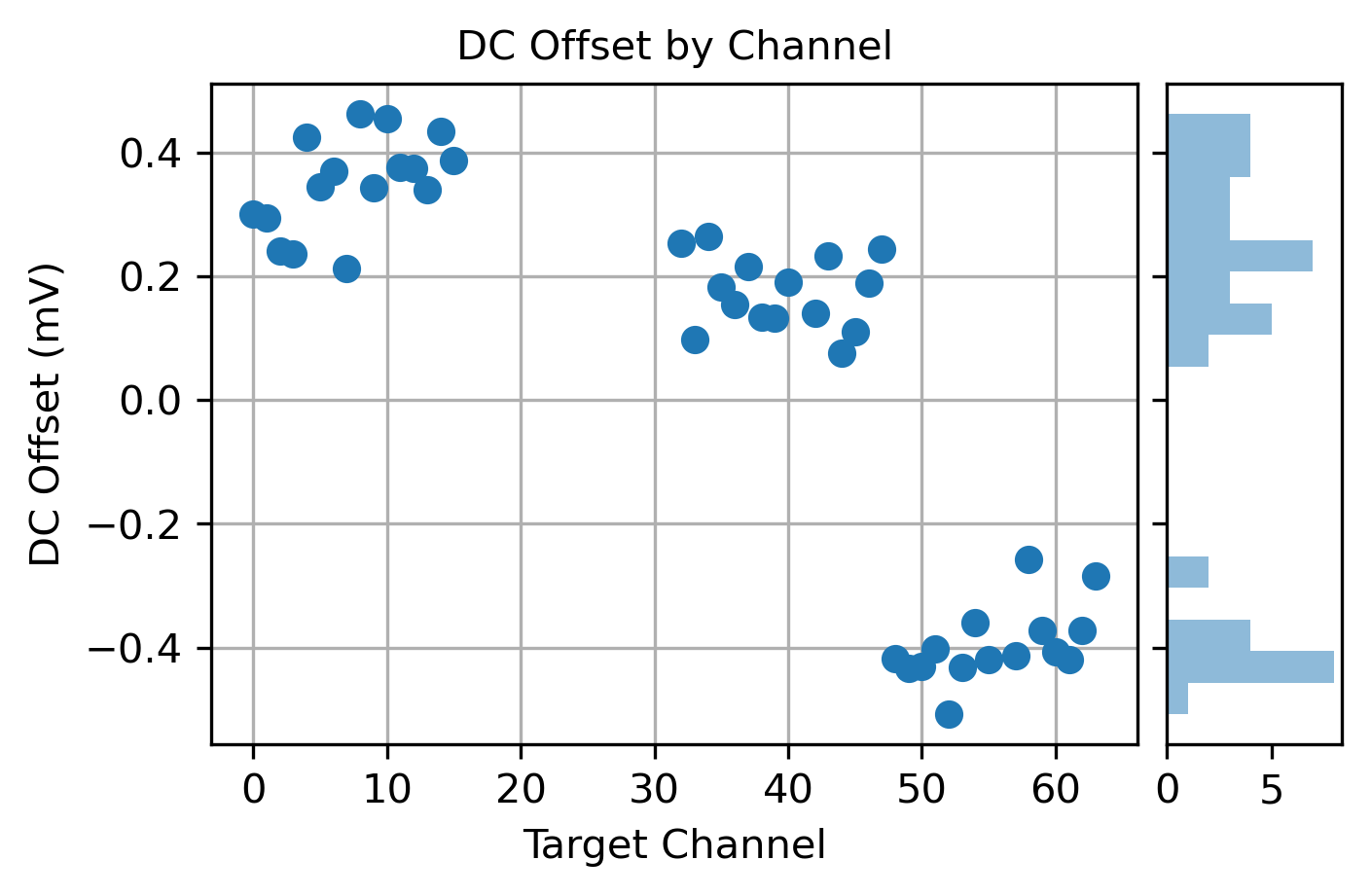}
    \caption{DC offset in mV across 46 channels, calculated as the mean across 302,071 pedestal events. A histogram of the values is displayed on the right. Clustering is visible in groups of 16 channels since they share a quadrant of ASICs.}
    \label{fig:DC_Offset}
\end{figure}

The baseline noise and DC offset were measured by taking externally triggered pedestal events with no signal present. The baseline noise and offset are calculated by taking the standard deviation and mean of waveform sample values, respectively. The baseline noise measured across approximately 300,000 128-sample waveforms is shown in Figure \ref{fig:DC_Noise}. The baseline noise is in the range of \qtyrange{0.5}{0.6}{\mV} for all 46 channels, which corresponds to approximately 1/3 p.e. using nominal gains from Section \ref{Optical Pulse Injection}. The DC offset for the same run is displayed in Figure \ref{fig:DC_Offset} and varies between \qtyrange{-0.5}{0.5}{\mV} for the plotted run. The DC offset is larger and more inconsistent than expected due to the high temperature dependence of a capacitor connected to the Wilkinson ramp. To increase the pedestal voltage stability, the capacitor will be replaced with a low temperature dependence variant on the production modules. Additionally, regular pedestal calibration before or after data runs will minimize any remaining pedestal voltage offset.

\subsection{Electrical Pulse Injection} \label{Electrical Pulse Injection}

\begin{figure}[htp]
    \centering
    \includegraphics[width=0.9\textwidth]{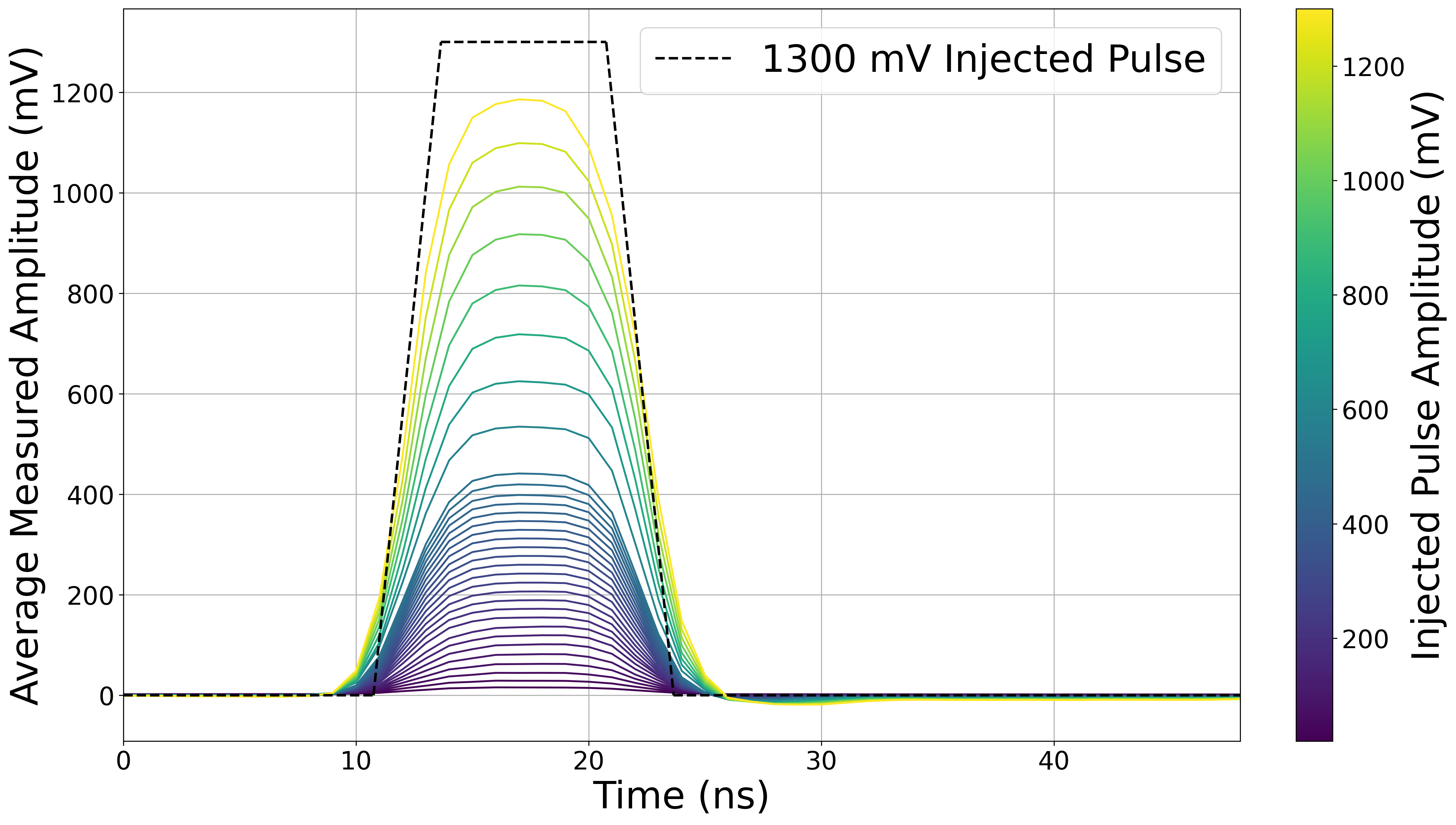}
    \caption{Average of 50,000 waveforms recorded by channel 0 for 33 different electrical pulse input amplitudes that cover the entire dynamic range of the module. The shape of the pulses is chosen to approximately match the shape of pulses from the FPM. The pulse width is \qty{10}{\ns} at FWHM with \qty{2.9}{\ns} rise/fall time. A \qty{1300}{\mV} injected pulse is shown for reference.}
    \label{fig:Elec_Waveforms}
\end{figure}

\begin{figure}[htp]
    \centering
    \includegraphics[width=0.9\textwidth]{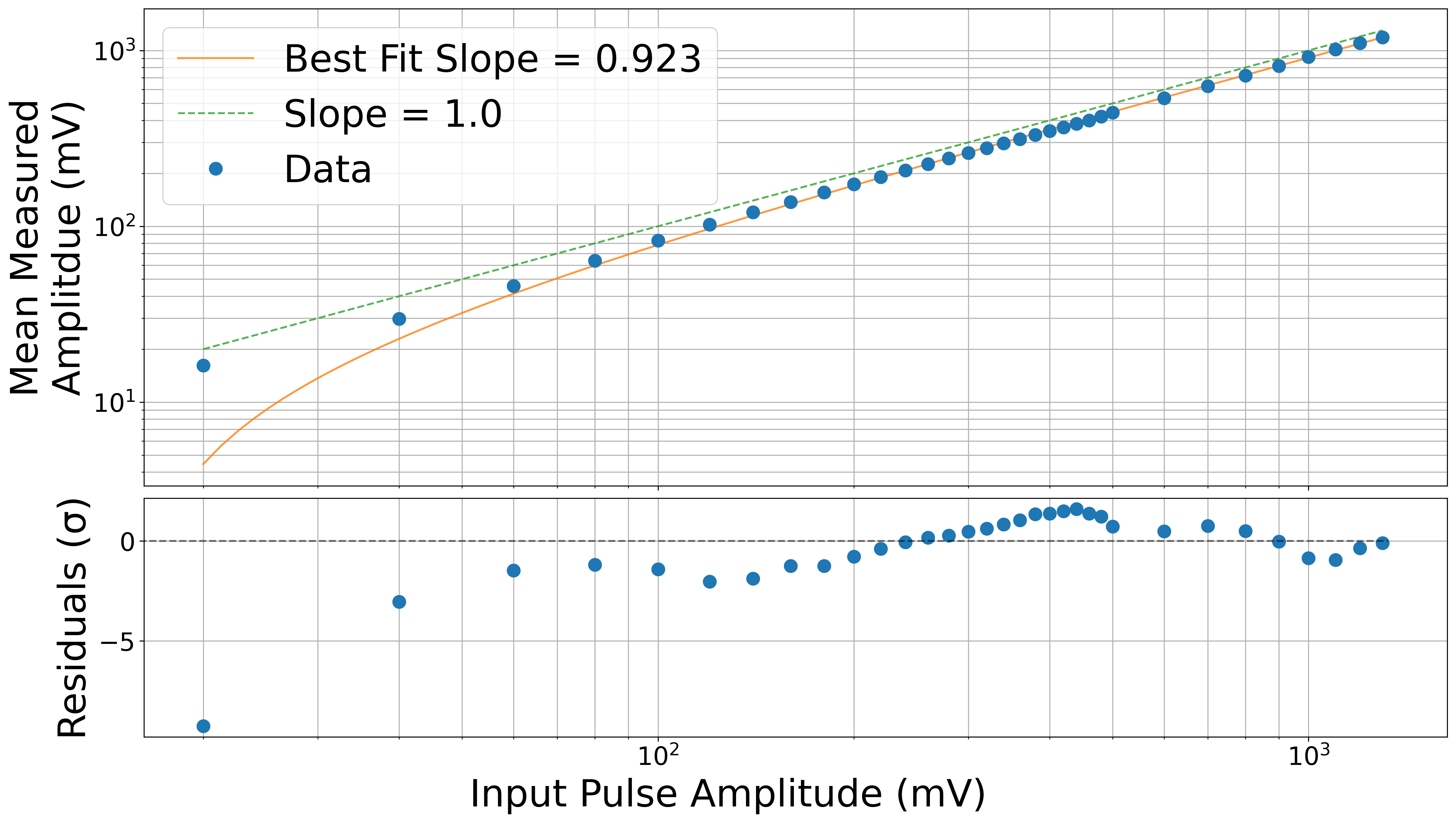}
    \caption{Top: The mean peak amplitude measured by channel 0 as a function of the input amplitude of electrical pulses. Also plotted are the one standard deviation from the mean values, though the error bars are too small to see. A linear fit to the data is included in orange. A line with a slope of 1 is indicated with green dashes. Bottom: Residuals in units of standard deviation between the data and the linear fit.}
    \label{fig:Elec_Lin_Ampl}
\end{figure}

To measure the response of the CTC readout electronics ASIC directly, we inject pulses directly into the FEE using the electric pulse injection setup. Pulses were injected at different amplitudes: \qtyrange{20}{500}{\mV} in steps of \qty{20}{\mV} and \qtyrange{500}{1300}{\mV} in steps of \qty{100}{\mV}. The average of 50,000 waveforms for each amplitude of electrical pulses is shown in Figure \ref{fig:Elec_Waveforms}. As we know the amplitude of the input pulse, we can compare that directly with the measured pulse amplitude (calculated as the maximum \nor{of a three-sample rolling average}). For every event we find the measured pulse amplitude and compare the average of those measured amplitudes to the known input amplitude. This comparison is shown in Figure \ref{fig:Elec_Lin_Ampl}. The amplitude measurement shows a \nor{close to} linear response across all amplitudes, as shown with the red line. The slope of the best fit line is slightly less than one, which is consistent with the measured bandwidth response found in Ref.~\citenum{2024NIMPA106969841S}. The custom breakout board may also have contributed to the signal losses.

\begin{figure}[htp]
    \centering
    \includegraphics[width=0.9\textwidth]{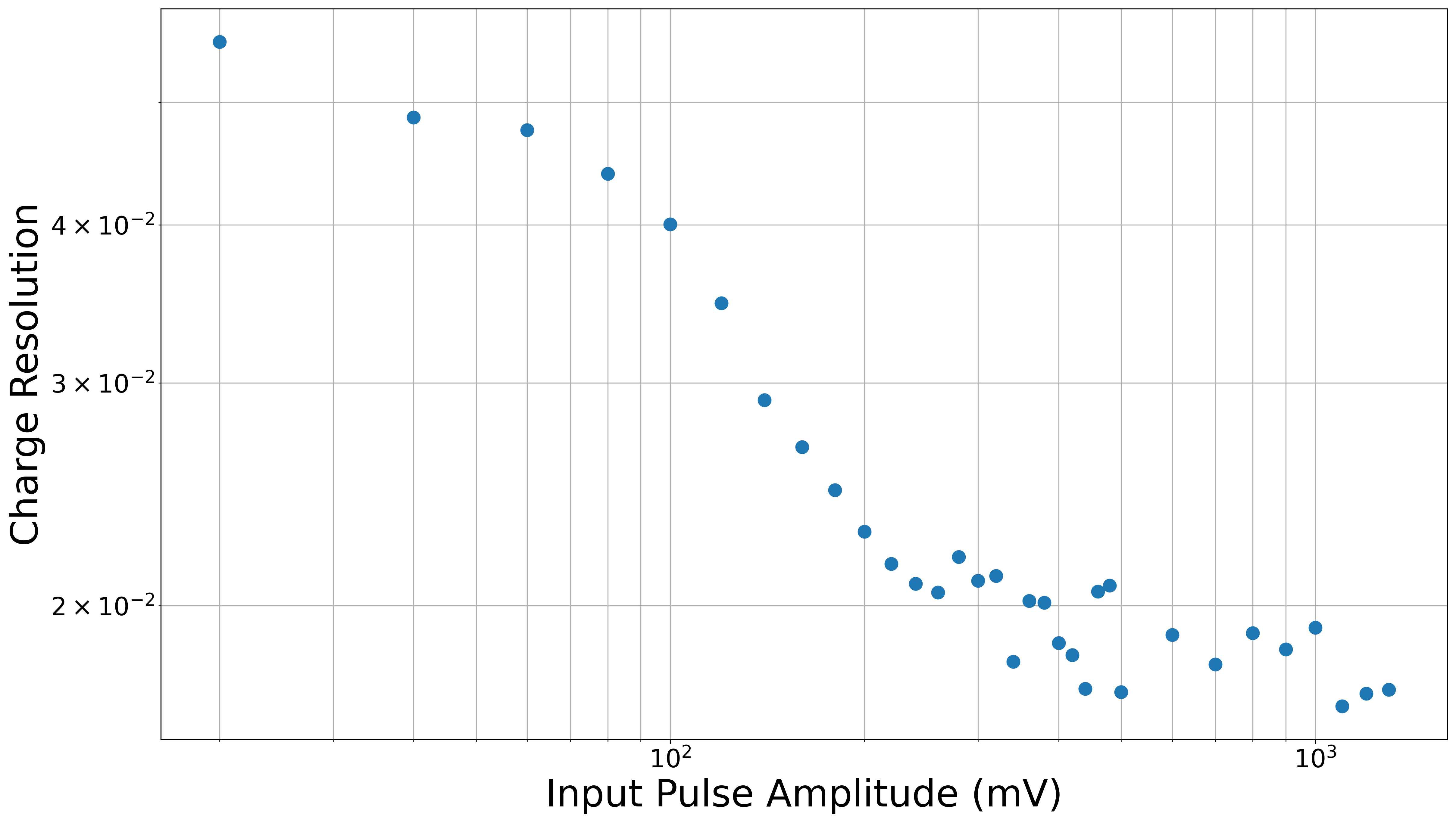}
    \caption{Charge resolution as a function of electrical pulse input amplitudes for channel 0.}
    \label{fig:Elec_Charge_Res}
\end{figure}

We also expect the CTC ASIC to have good resolution to measure pulses. We quantify this through the fractional resolution of the ASIC to resolve pulses across different input amplitudes. This fractional resolution of the integrated charge, called the charge resolution, is shown in Figure \ref{fig:Elec_Charge_Res}. The charge resolution for electrical pulse injection is limited by the electronics noise of the FEE but does not contain the FPM contribution. At the lowest amplitudes the charge resolution of the FEE is approximately 0.05 and it quickly drops down to 0.02. \nor{The plateau around \qtyrange{20}{80}{mV} is a known effect and is seen in Ref. \citenum{2024NIMPA106969841S}. It is caused by a change in slope of the AC response of \ctc{} ASIC and the exact cause of the change in slope is not well understood. The flattening of the curve above \qty{250}{mV} is expected as the irreducible constant electronics noise begins to dominate. The scattering at high amplitudes is likely due to variations in pulse amplitude from the function generator which injects the pulses.}

\subsection{Optical Pulse Injection} \label{Optical Pulse Injection}

\begin{figure}[htp]
    \centering
    \includegraphics[width=0.9\textwidth]{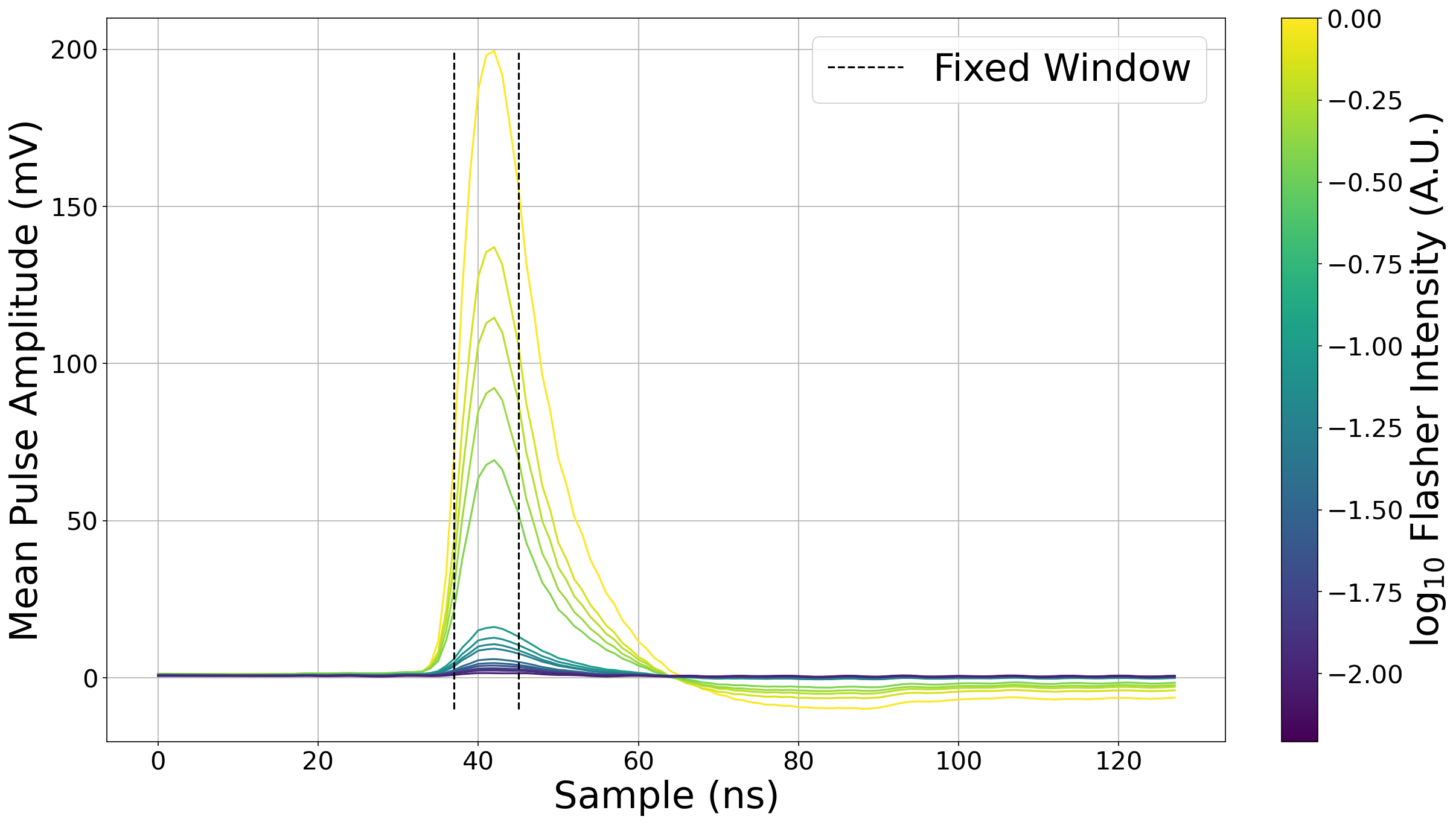}
    \caption{Average of 50,000 waveforms recorded at 16 different LED flasher amplitudes for pixel 0. The amplitude was modified using neutral density (ND) filters. Shown with black dashed lines is the fixed window used for charge integration.}
    \label{fig:LED_Waveforms}
\end{figure}

\begin{figure}[htp]
    \centering
    \includegraphics[width=0.9\textwidth]{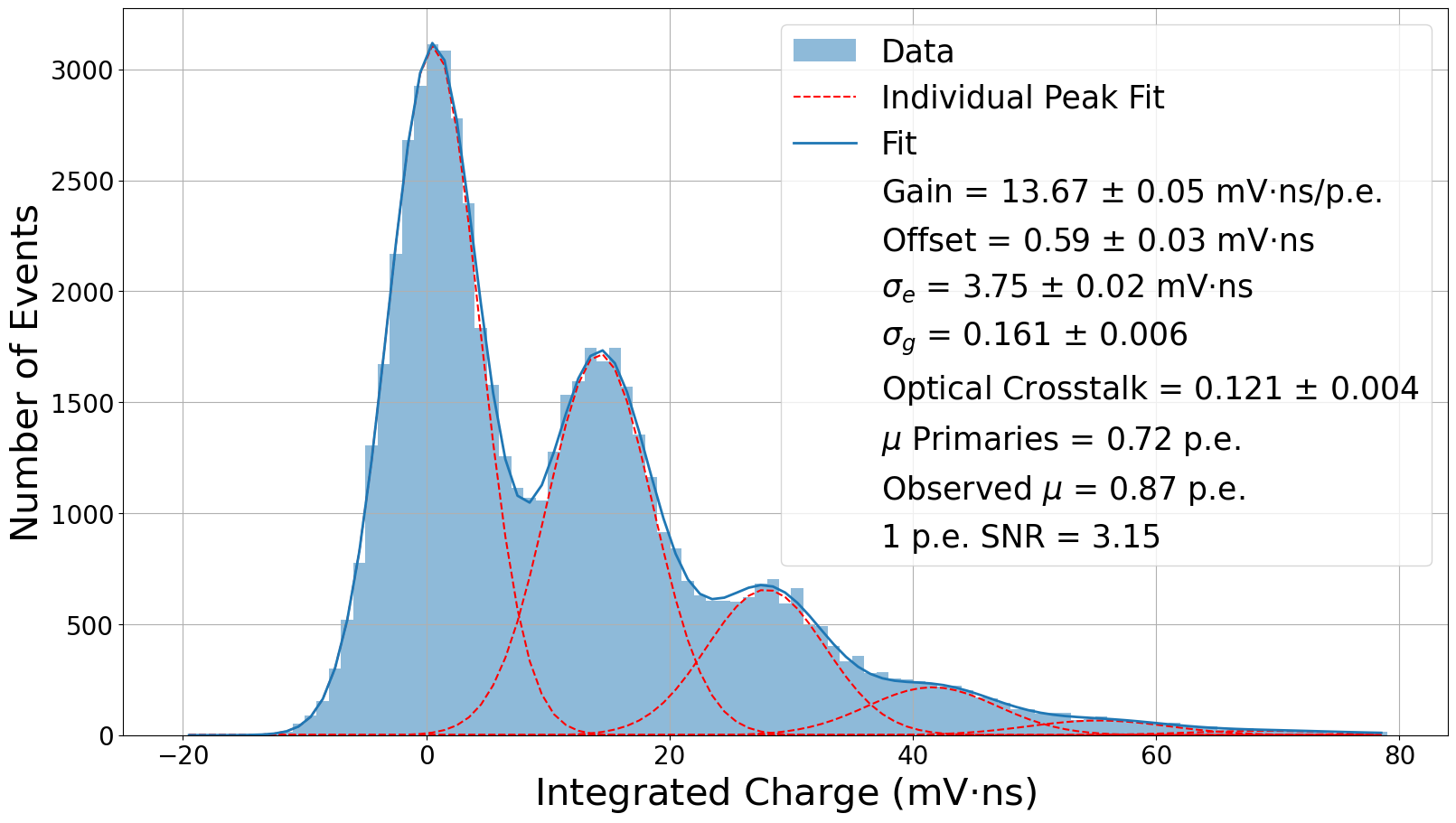}
    \caption{Charge histogram for a low intensity LED flasher run. Each peak corresponds to an integer number of photoelectrons. A Gaussian fit is performed on each photoelectron peak with parameters constrained by the fitted values of the system. The individual Gaussians are displayed with red dashes and the sum of the fit is indicated with the solid blue line.}
    \label{fig:Finger_Plot}
\end{figure}

To measure the module’s response to light flashes we use the LED flasher setup described in \ref{Testing Setup}. The FWHM of the LED flasher pulse is \qty{13.5}{\ns} and the intensity of the LED flasher was controlled by neutral-density (ND) filters attached to the flasher. The nominal filter values used ranged from ND1.5 to ND3.0 in steps of ND0.1, testing the entire linear range of the SMART ASIC, which begins to behave non-linearly above \num{\sim 100} p.e. \nor{The filters were also calibrated using the same LED and an optical power meter, as the measured ND values varied from nominal by up to 25\%, which is expected at near-UV wavelengths.} To reduce any bias from electrical crosstalk, signal was only recorded for pixel 0. The average of 50,000 waveforms from pixel 0 for each different intensity run of the flasher is shown in Figure \ref{fig:LED_Waveforms}. For each intensity of the flasher, we construct charge histograms by integrating the amount of charge in each event. As the setup has a known delay, the charge is calculated by integrating each waveform in a fixed window. This window is set to maximize the signal-to-noise ratio. The window is shown in Figure \ref{fig:LED_Waveforms}. From these charge values we construct a histogram. In the low light intensity regime, the Poisson values for each integer number of p.e. are visible in the charge histogram. We refer to these distinct Poisson peaks as fingers. Figure \ref{fig:Finger_Plot} shows an example finger plot for the module.

To extract relevant information from the finger plot, we perform a multi-Gaussian fit where the parameters of each Gaussian are constrained by the physical parameters of the data. By Poisson statistics, the mean of each Gaussian should be an integer multiple of a specific value called the gain. The means will all be offset by a static value produced by a baseline offset in the waveforms. The width of each Gaussian is the combination of the electronics noise and the gain noise. The amplitude of each Gaussian is taken from the expected number of counts due to the generalized Poisson distribution (see Ref.~\citenum{VINOGRADOV2012247}), the Poisson distribution accounting for optical crosstalk. Putting these all together, the equation for each finger's Gaussian fit $G_n$, where $n$ is the number of p.e. for that finger, (and parameters) are shown in Equations \ref{Gaussian}, \ref{Sigma}, \ref{Mu}, and \ref{Amplitude}.

\begin{equation} \label{Gaussian}
    G_n = \frac{A_n}{\sqrt{2\pi\sigma_n}}e^{\frac{-(x-\mu_n)}{2\sigma_n^2}}
\end{equation}

\begin{equation} \label{Sigma}
    \sigma_n = \sqrt{\sigma_e^2 + (\sqrt{n} \cdot g \cdot \sigma_g)^2}
\end{equation}

\begin{equation} \label{Mu}
    \mu_n = n\cdot g+a
\end{equation}

\begin{equation} \label{Amplitude}
    A_n = \frac{\lambda_n(\lambda_n+p\cdot n)^{n-1}e^{-\lambda_n-p\cdot n}}{\Gamma(n+1)}
\end{equation}

\nor{
\begin{equation} \label{Lambda}
    \lambda_n = \frac{\langle C \rangle - a}{g}(1-p)
\end{equation}
}

Here $g$ is the fitted gain, $a$ is the fitted offset, $\sigma_e$ is the fitted electronics noise, $\sigma_g$ is the fitted gain noise, \nor{$\lambda_n$ is the mean number of primary p.e. (the mean number seen if optical crosstalk was zero),} \nor{$\langle C \rangle$ is the average of the event charges,} and $p$ is the fitted optical crosstalk probability. \nor{Equation \ref{Amplitude} is the functional form of the generalized Poisson distribution from Ref.~\citenum{VINOGRADOV2012247} with the $n!$ replaced with $\Gamma(n+1)$.} The gain noise accounts for differences of a single discharge between micro-cells of the SiPM pixel. As each micro-cell is independent from all others, the uncertainty sums in quadrature, hence the factor of $\sqrt{n}$. The fitting of these values to each finger plot is performed separately for each intensity in the regime where fingers are visible. An example of the fit, and the resulting values, is shown in Figure \ref{fig:Finger_Plot}.

From the \nor{multi-Gaussian fit} we also calculate the 1 p.e. signal-to-noise ratio (SNR) of the pixel. The 1 p.e. SNR is calculated as the mean of the 1 p.e. fit Gaussian divided by the 1$\sigma$ width of the 1 p.e. Gaussian. The 1 p.e. SNR value for the example finger plot is 3.15 and this value of SNR is consistent across all \nor{multi-Gaussian fits}. The optical crosstalk probability of the SiPMs from Ref.~\citenum{2019Senso..19..308G} is $\approx10\%$. The fitted optical crosstalk probability found from the \nor{multi-Gaussian fits} is consistent across the different fits and also consistent with this literature value.

\begin{figure}[htp]
    \centering
    \includegraphics[width=0.8\textwidth]{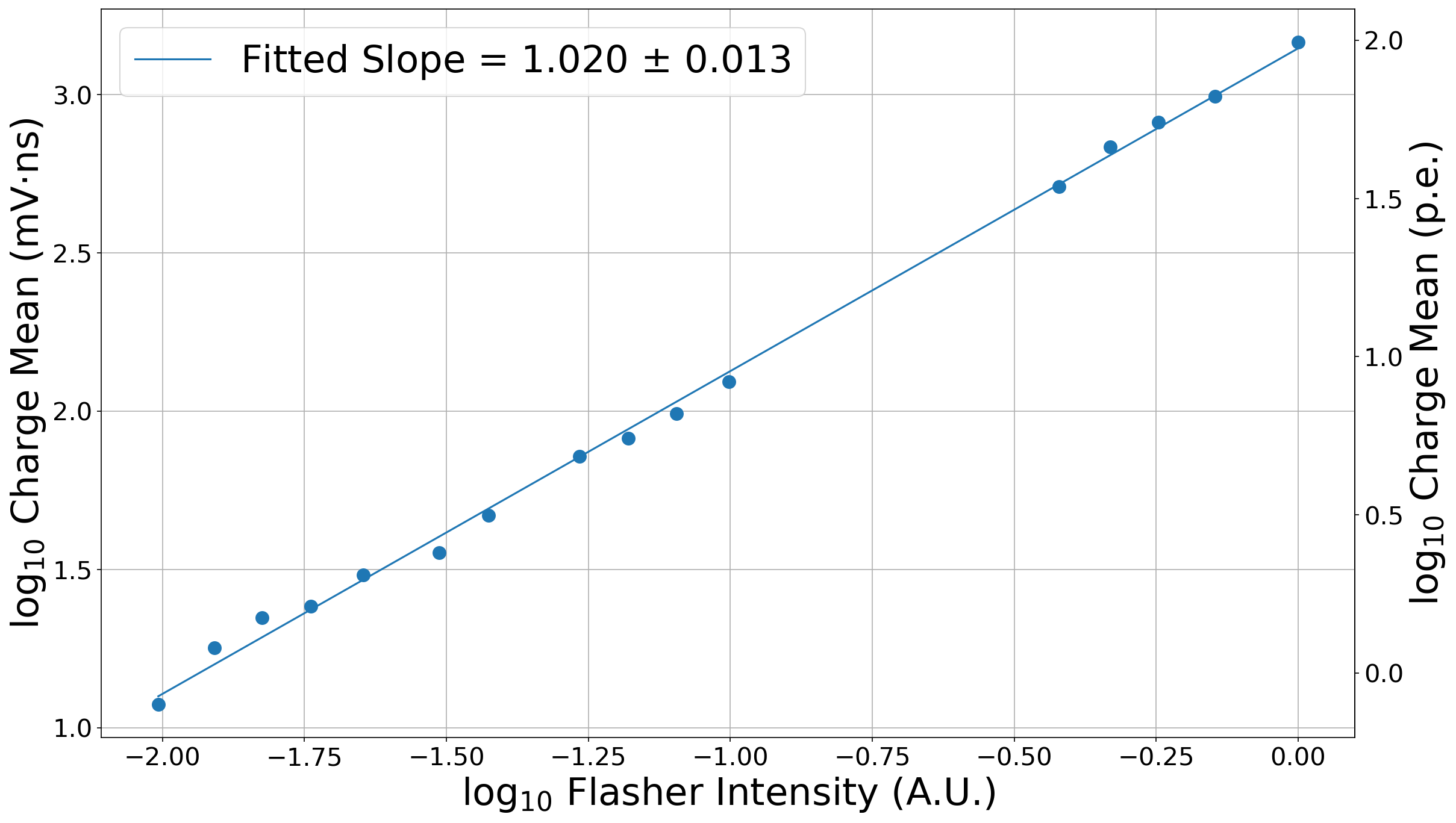}
    \caption{\nor{The recorded mean charge as a function of flasher intensity, both in log$_{10}$ space. From linearity, the mean charge should change by the same order of magnitude as the flasher intensity. This is shown in the linear fit which has a slope close to 1. On the right is shown the charge mean converted to p.e. using the gain found by the multi-Gaussian fits. This dataset show a linear range up to 100 p.e.}}
    \label{fig:Chg_Lin}
\end{figure}

\nor{
At medium light intensities, fingers are no longer distinguishable and less information is available. To verify linearity to higher intensities, the measured mean charge is compared to the flasher intensity from the calibrated ND filter values and is shown in Figure \ref{fig:Chg_Lin}. This relation is linear within the measured intensity range. Using the average gain from the multi-Gaussian fits, it is shown that this linear range extends to 100 p.e. The slope of the fit is close to 1, as expected by linearity.
}

\begin{figure}[htp]
    \centering
    \includegraphics[width=0.8\textwidth]{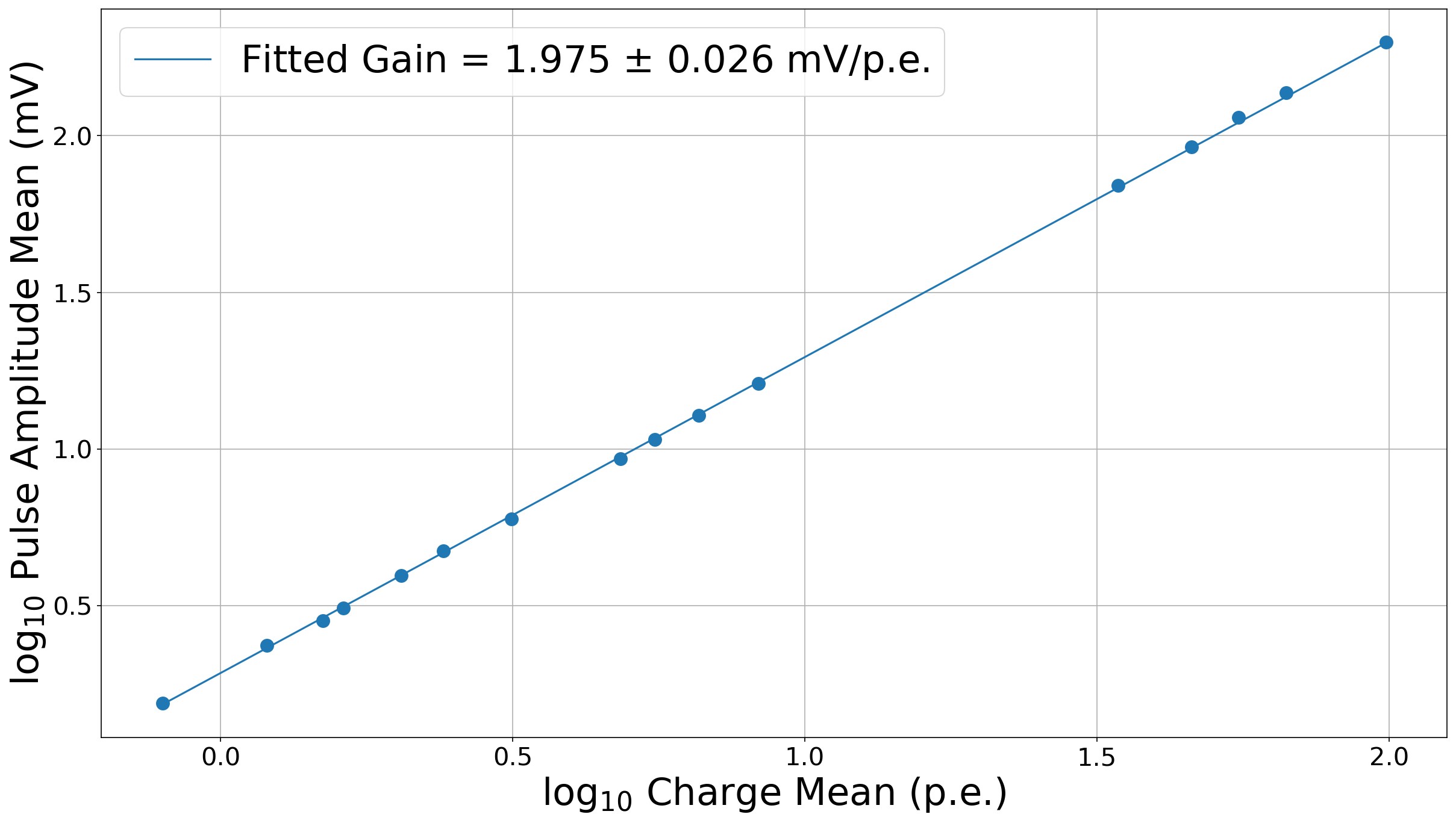}
    \caption{\nor{The average waveform amplitude as a function of charge mean for each intensity, both in log$_{10}$ space. The charge mean is converted to p.e. using the average gain from the finger plot multi-Gaussian fits. The fitted line is shown in blue, which corresponds to the amplitude gain for this module. This confirms the linear range up to 100 p.e.}}
    \label{fig:Amp_Lin}
\end{figure}

\begin{figure}[htp]
    \centering
    \includegraphics[width=0.8\textwidth]{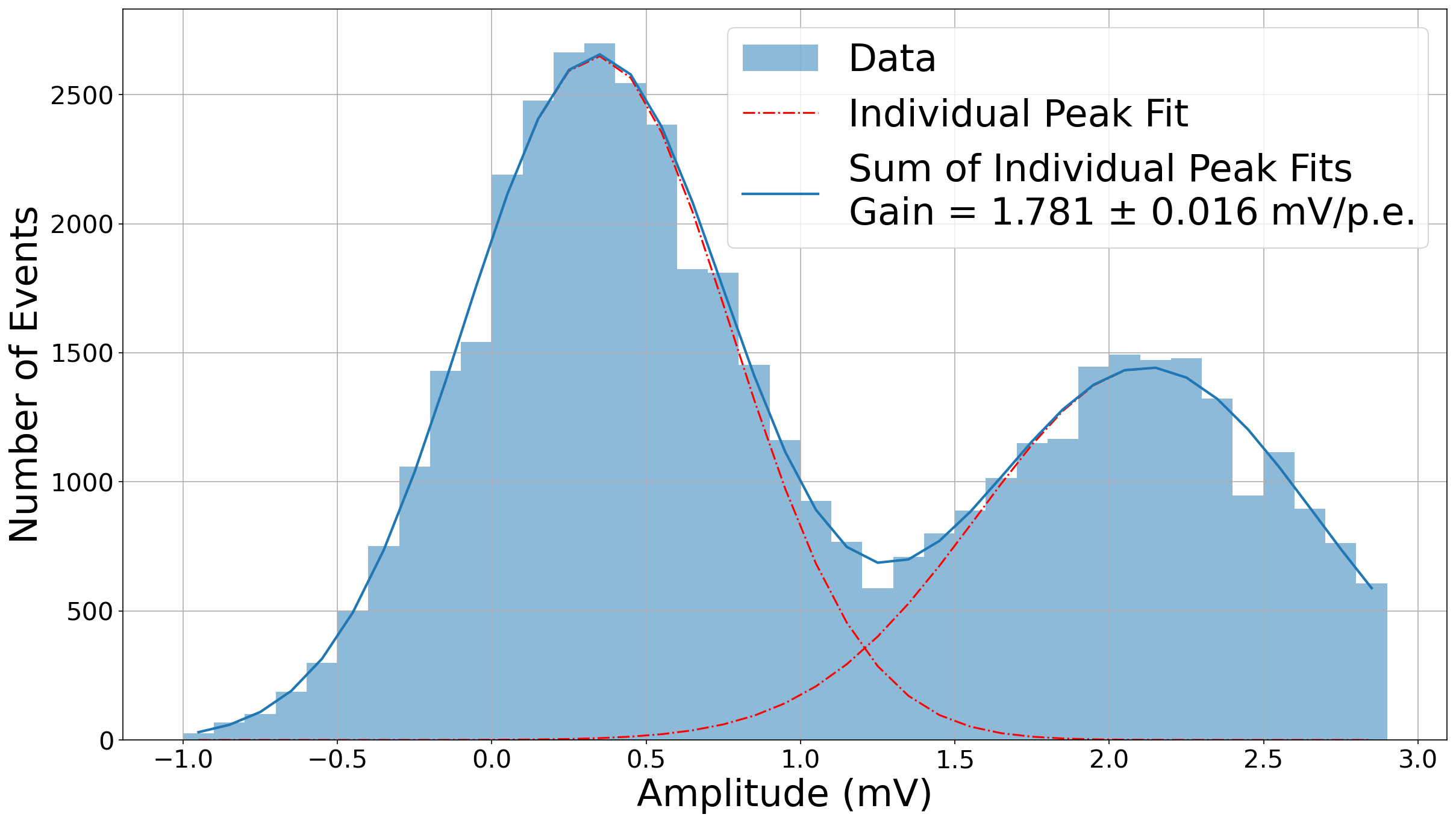}
    \caption{\nor{Amplitude histogram for a low intensity LED flasher run. The amplitude is found by averaging a 3 sample sliding window across the waveform to reduce fluctuations due to noise. Each peak corresponds to an integer number of photoelectrons. The range is cut above the 1 p.e. peak and Gaussians are fit to the 0 and 1 p.e. peaks. The individual Gaussians are displayed with red dashes and the sum of the fit is indicated with the solid blue line. The gain taken from the difference of the Gaussian means is shown in the legend.}}
    \label{fig:Amp_Fin}
\end{figure}

\nor{
Another metric for linearity across the recorded range is to compare the average waveform amplitude, the amplitude of each waveform shown in Figure \ref{fig:LED_Waveforms}, to the charge mean found for each intensity. This is shown in Figure \ref{fig:Amp_Lin}. This relation is linear, as expected, across the measured range. The slope of this relation is the amplitude gain for the module. The amplitude gain can also be calculated from finger plots of recorded amplitude instead of charge. The amplitude is found by averaging a 3-sample sliding window across the waveform to reduce fluctuations due to noise. An example of this is shown in Figure \ref{fig:Amp_Fin}. A simpler multi-Gaussian fit, with each Gaussian parameter fit independently, is used. This fit is done only on the 0 and 1 p.e. fingers to find the amplitude gain. The gain found in Figure \ref{fig:Amp_Lin} is similar to the gains found from the amplitude finger plots like is shown in Figure \ref{fig:Amp_Fin}. These checks confirm both the linearity of the module and the robustness of these fits.
}

\subsection{Charge Resolution} \label{Charge Resolution}

\begin{figure}[htp]
    \centering
    \includegraphics[width=0.9\textwidth]{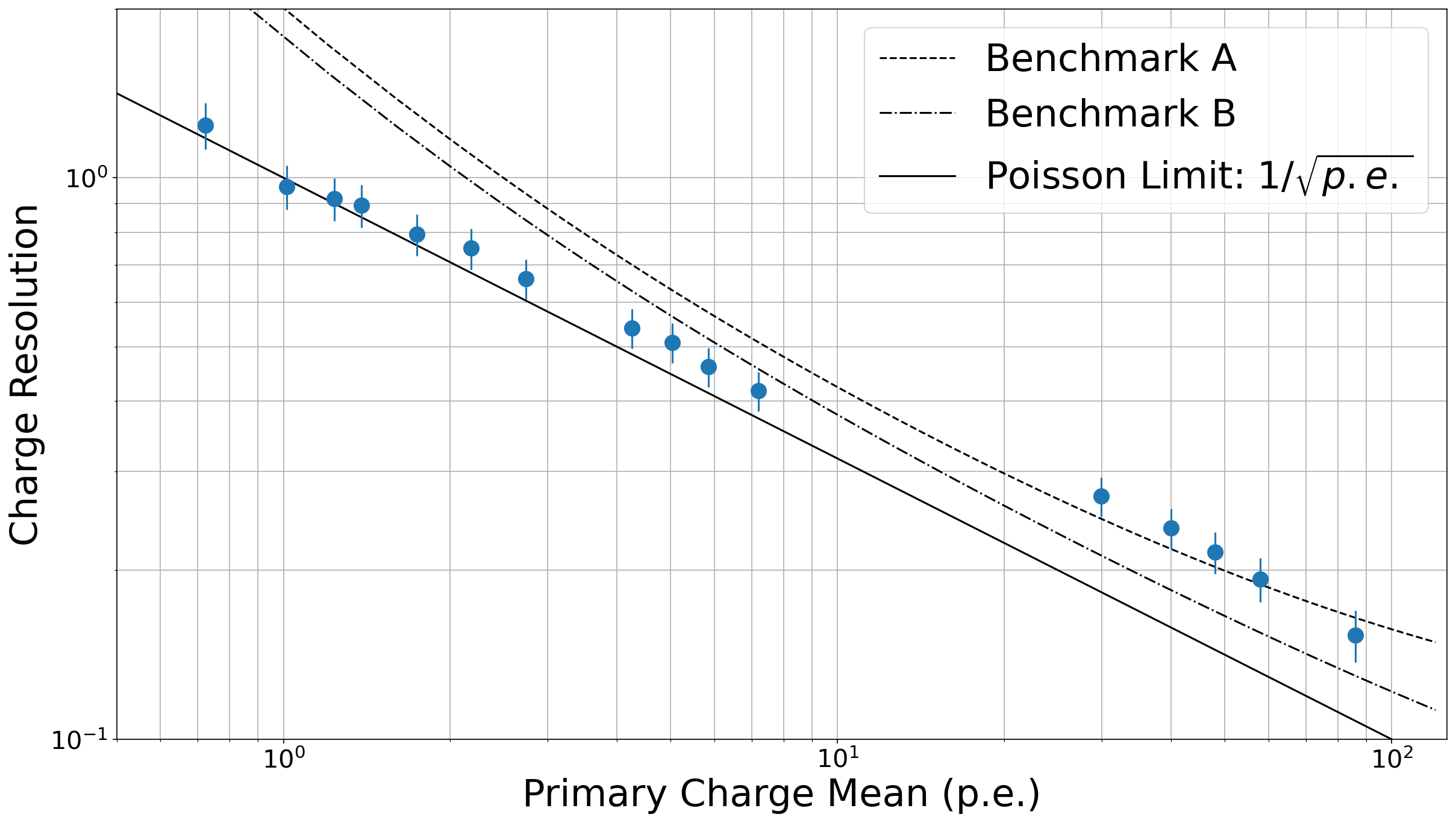}
    \caption{\nor{Charge resolution as a function of primary number of photoelectrons for pixel 0. To calculate the primary number of photoelectrons, the optical crosstalk probability and gain is found for each finger fit, and then the average of those values is used for the higher intensity points where fingers are not visible. Internal charge resolution benchmarks and the Poisson limit are plotted for reference and described in the text.}}
    \label{fig:LED_Charge_Res}
\end{figure}

We look at the charge resolution of the full module which includes the noise contributions from the SMART ASIC and the SiPMs, as compared to the CTC ASIC only. The charge resolution across the \nor{measured} range is shown in Figure \ref{fig:LED_Charge_Res} and compared to internal charge resolution benchmarks, as well as the Poisson limit. \nor{These benchmarks are based on previous CTAO metrics and a future goal is to refine our analysis to compare to the current CTAO intensity resolution requirement.} The benchmarks are defined as:
\begin{equation}\label{Benchmark A}
    \sqrt{0.1^2 + \left(\frac{1.6}{n}\right)^2 + \left(\frac{1.2}{\sqrt{n}}\right)^2}
\end{equation}
\begin{equation}\label{Benchmark B}
    \sqrt{0.05^2 + \left(\frac{1.4}{n}\right)^2 + \left(\frac{1.1}{\sqrt{n}}\right)^2}
\end{equation}
where $n$ is the number of p.e. In each equation, the first term represents the electronics noise, the second the gain uncertainty, and the third a scaling of the Poisson limit. \nor{These benchmarks do not account for optical crosstalk, so the x-axis of Figure \ref{fig:LED_Charge_Res} is the primary number of photoelectrons incident on the pixel.} The charge resolutions of the low intensity data are all better than the stricter Equation \ref{Benchmark B} (Benchmark B). \nor{The multi-Gaussian fits begin to perform worse as the intensity increases since the individual fingers begin to overlap and are harder to distinguish; this likely contributes to the increasing distance between the charge resolution points and the Poisson limit around a few p.e. At medium intensities the charge resolution is not quite meeting Equation \ref{Benchmark A} (Benchmark A). Additionally, the change in the slope of the AC response of the \ctc{} ASIC (discussed in Section \ref{Electrical Pulse Injection}), causes a bump in the charge resolution in the \qtyrange{10}{100}{mV} (\qtyrange{5}{50}{p.e.}) range, so it is not clear whether the poorer charge resolution relative to the benchmarks in that range is due to the \ctc{} ASIC response or systematic uncertainties on the measurement.} Low charge pixels are most important for reconstructing Cherenkov showers so optimizing charge resolution at low intensities is the primary goal.

\subsection{Gain Equalization}
\label{sect:gain}
A flat-fielding procedure was implemented to equalize the pixel-to-pixel gain variations, which can depend not only on pixel-to-pixel differences in gain versus over-voltage (OV) dependence, but also on slightly different pixel-to-pixel breakdown voltages, which are expected to be of the orders of tens of mV \cite{AMBROSI2022167359}. We remark that the module employed for this test is a different preproduction module, on which one of the ASICs was not working properly. For this reason, we performed the flat-fielding procedure only on 48 channels. 

To implement the gain flat-fielding procedure we exploited the SMART DAC register. SMART allows to change the bias at the low voltage side of the SiPM within \qty{1.15}{\V} with 8 bit resolution. As the gain is directly proportional to the OV, we individually adjusted the OV value of each pixel by offsetting the provided common bias $V_\textrm{ext} = 33\,\textrm{V}$. The value of the DAC register for each pixel was calculated by requiring that all pixels operate at the OV value that would return the same gain as the other pixels. As explained in Section \ref{sect:fpm}, the SMART bias DAC values range from 0 to 255, therefore the bias voltage for each SiPM can be changed from $V_\textrm{bias} = V_\textrm{ext} - 0.75\,\textrm{V}$, when the DAC is set to 0, to $V_\textrm{bias} = V_\textrm{ext} - 1.9\,\textrm{V}$ when the DAC is set to 255.

\begin{figure}[htp]
    \centering
    \includegraphics[width=0.7\textwidth]{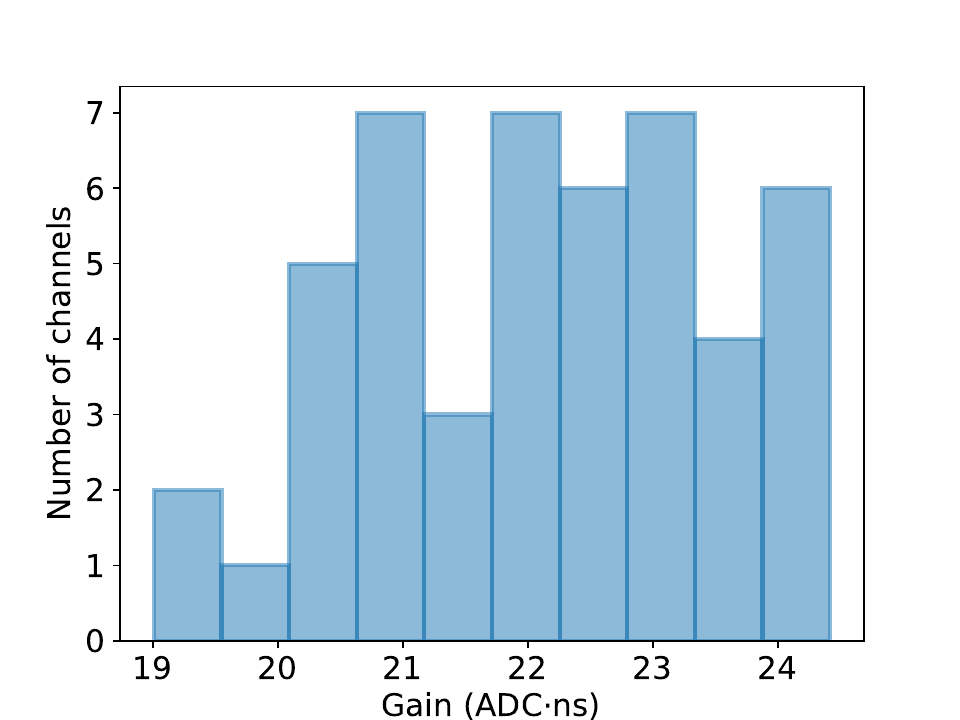}
    \caption{Gain distribution applying the same bias to all channels, using a DAC value of 100. The mean value is \qty{22.1}{ADC \cdot \ns} with a standard deviation of \qty{1.4}{ADC \cdot \ns}.}
    \label{fig:gain_before_ff}
\end{figure}

First, we uniformly illuminated all pixels by placing a diffusing lens between the the laser head and the FPM, and acquired the gain distribution of the 48 channels through fits of finger plots. We used a common bias DAC value of 100, i.e. with the same bias voltage on all SiPMs:
\begin{equation}
V_\textrm{bias} = V_\textrm{ext} - 0.75\, \textrm{V} - (1.15/255)\cdot 100\,\textrm{V} = 31.8\,\textrm{V}.    
\end{equation}
The acquired gain distribution, shown in Figure \ref{fig:gain_before_ff}, ranges from \qty{\sim 19}{ADC \cdot \ns} to \qty{\sim 24}{ADC \cdot \ns}. As can be seen from the plot, gain values are scattered among the 48 channels.

\begin{figure}[htp]
    \centering
    \includegraphics[width=0.85\textwidth]{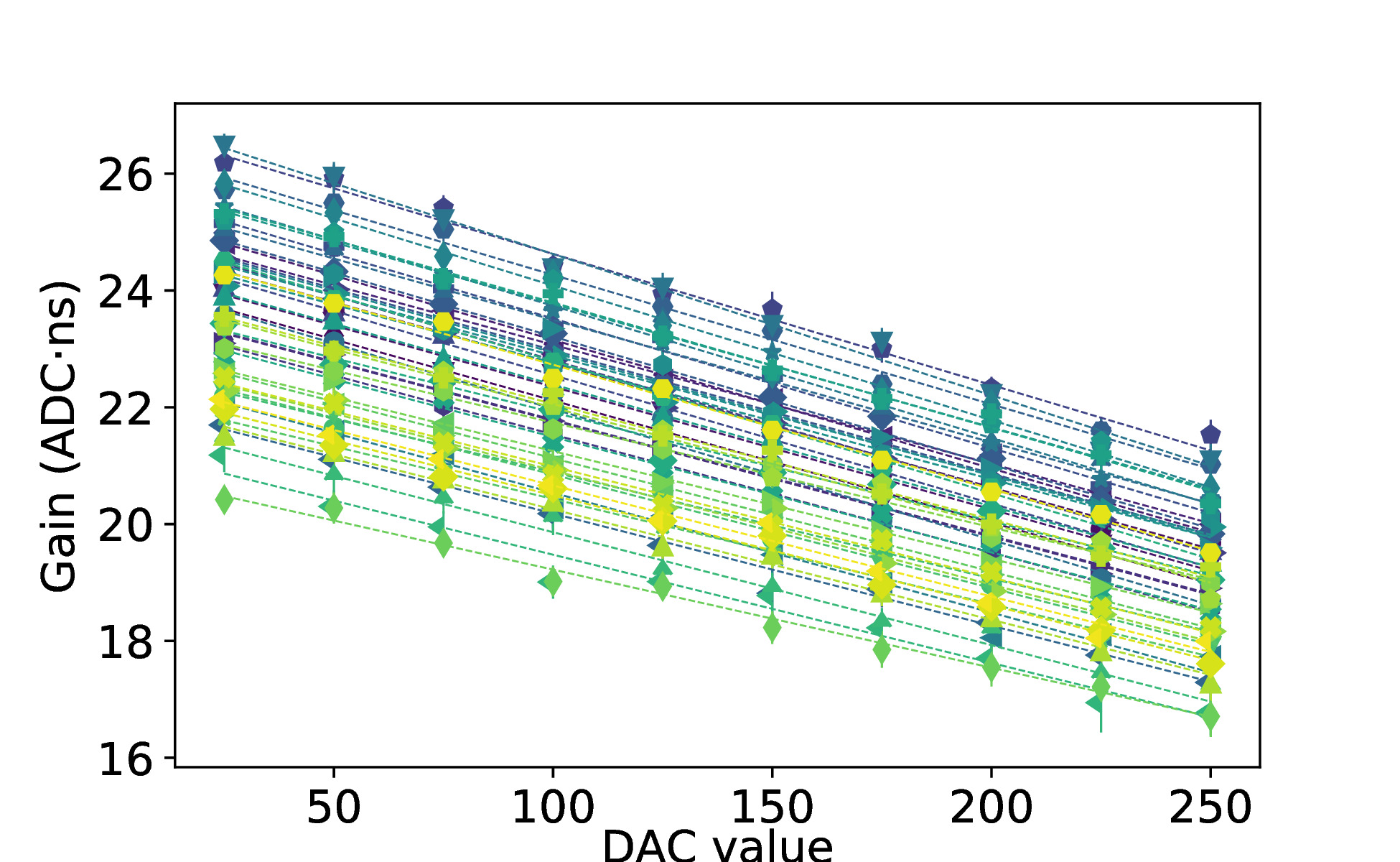}
    \caption{Integrated gain for all the 48 tested channels.}
    \label{fig:gain_vs_dac}
\end{figure}

We repeated the common DAC acquisition with different DAC values, ranging from 25 to 250 in steps of 25, in order to estimate the relation between the bias DAC and the gain for every channel. The curves of gain as a function of the DAC were fitted with a linear function, as shown in Figure \ref{fig:gain_vs_dac}. The inverse correlation with the DAC value is expected as the $V_\textrm{OV}$ decreases with the DAC value.

\begin{figure}[htp]
    \centering
    \includegraphics[width=0.7\textwidth]{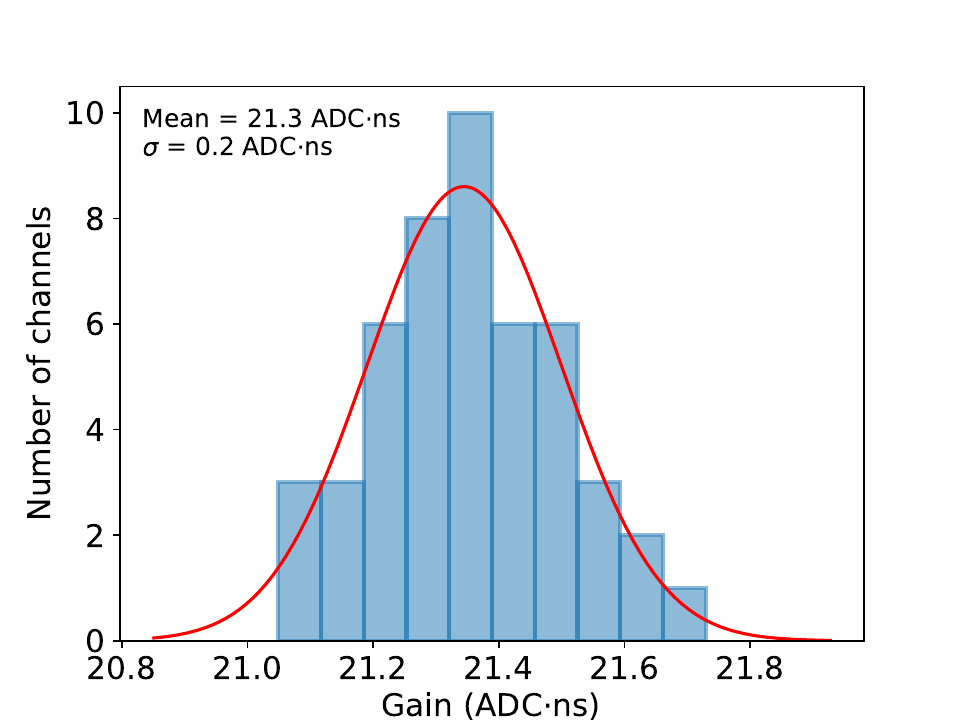}
    \caption{Gain distribution applying the optimized DAC values in order to equalize the gain. A gaussian fit is shown superimposed on data. The mean and standard deviation of the Gaussian fit are also reported in the text. }
    \label{fig:gain_after_ff}
\end{figure}

As can be seen from Figure \ref{fig:gain_vs_dac} the gain ranges from \qty{\sim 17}{ADC \cdot \ns} to \qty{\sim 27}{ADC \cdot \ns}. We set the desired gain value to \qty{21}{ADC \cdot \ns} and used the estimated best-fit lines to calculate the pixel-to-pixel integer DAC values that would correspond to the desired gain value. We then set the calculated DAC value in the corresponding SMART register and performed a further signal acquisition. The corresponding gain distribution is shown in Figure \ref{fig:gain_after_ff}. The mean value of the distribution and its standard deviation, obtained by performing a gaussian fit on the histogram, are found to be \qty{21.3}{ADC \cdot \ns} and \qty{0.2}{ADC \cdot \ns}, respectively.

The gain flat-fielding procedure's results demonstrate a clear improvement compared to the initial situation, proving the possibility to achieve a good uniformity in the responses of the SiPMs to light signals. However, it has to be pointed out that data acquired some time after the flat-fielding procedure resulted in a different gain distribution, possibly due to a temperature or operational conditions dependence. As described in Section \ref{sect:design}, the camera temperature will be highly stable; therefore, with the addition of the thermoelectric element control loop, which was not active during the lab tests, the gain equalization is expected to be much more robust on the telescope. \nor{If necessary, the flat-fielding procedure can be repeated on the telescope to account for any significant temperature changes the SiPMs experience.}

\subsection{Electronics Crosstalk} \label{Crosstalk}
In order to estimate the crosstalk among electronics channels belonging to the same quadrant, we illuminated each pixel individually and measured the amplitude of the signal on the other 15 channels. We developed a procedure to test the full signal processing path crosstalk (i.e. SiPM tiles, SMART ASIC, and CTC ASIC). 

As these tests were conducted on a different preproduction module with respect to other sections in this paper, on which no ASICs nor channels had been damaged, we could perform the tests on all the 64 channels.

\begin{figure}[htp]
    \centering
    \includegraphics[width=0.9\textwidth]{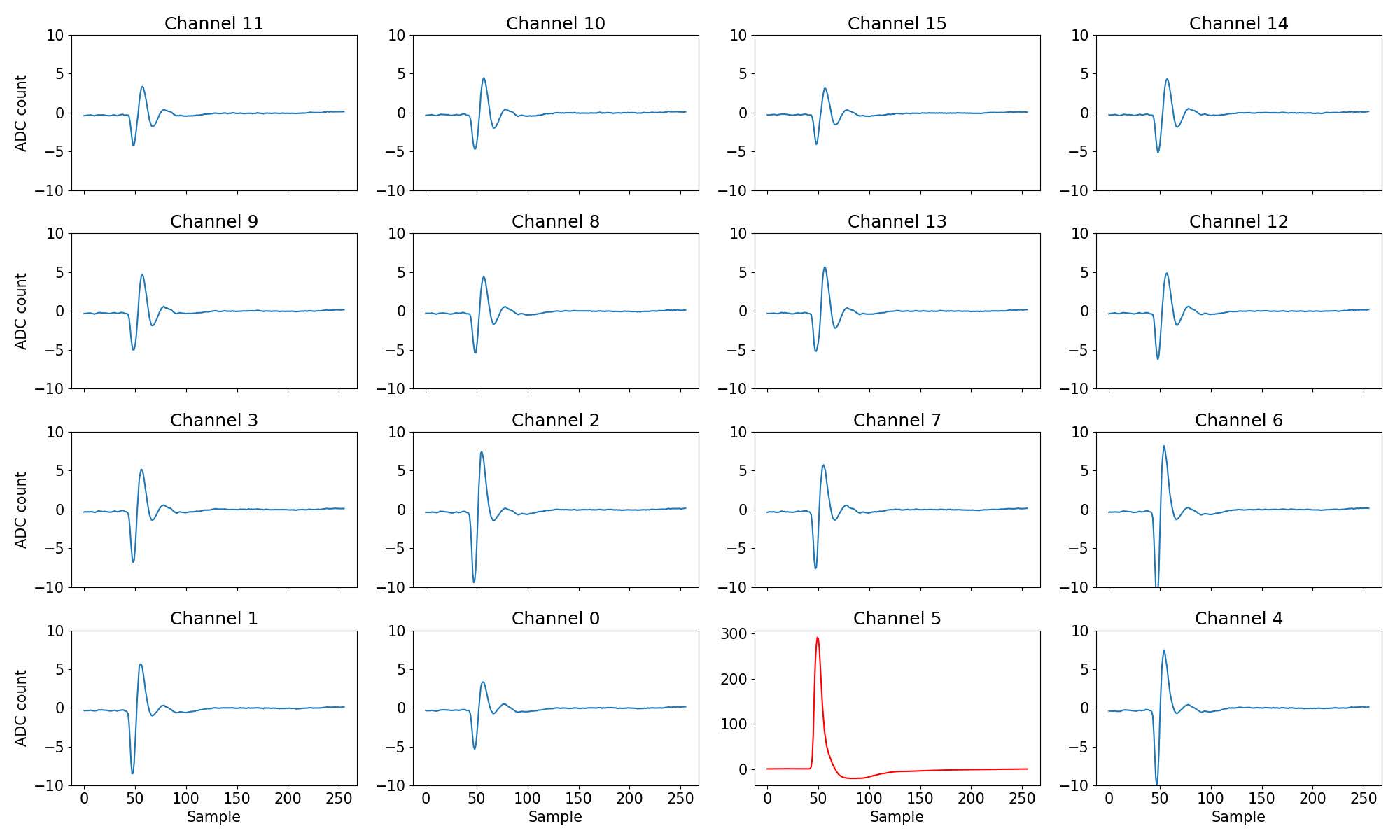}
    \caption{Average calibrated waveform from quadrant 0 channels. Channel 5 is illuminated, while the others only show crosstalk noise. The position of each channel is the physical position of the pixel as seen from the front.}
    \label{fig:avg_wfs_1pixel_on}
\end{figure}

We illuminated each pixel with a \qty{380}{\nm} laser, focusing the light on single pixels by means of a adjustable holder. We changed the position of the laser spot in steps of \qty{6}{\mm} along the focal plane in both the X and the Y direction to change the illuminated pixel. We acquired \num{\sim 2000} events for each of the 64 channels. An example of the average waveforms after the pedestal subtraction for each of the 16 channels belonging to the quadrant 0 is reported in Figure \ref{fig:avg_wfs_1pixel_on}. The illuminated pixel (channel 5 in the Figure) is highlighted in red. Nearby pixels belonging to the same quadrant show a crosstalk signal with different peak-to-peak amplitudes. The crosstalk is likely due to electromagnetic coupling between analog signals traveling through adjacent traces on the printed circuit board. The mapping of the channels changes between detection and readout so the proximity of each channel along the signal path is not constant; therefore, we note that the amplitude of the crosstalk is not perfectly related to each channel's proximity to the illuminated channel in the figure.

\begin{figure}[hbtp]
    \centering
    
    \includegraphics[width=0.9\textwidth]{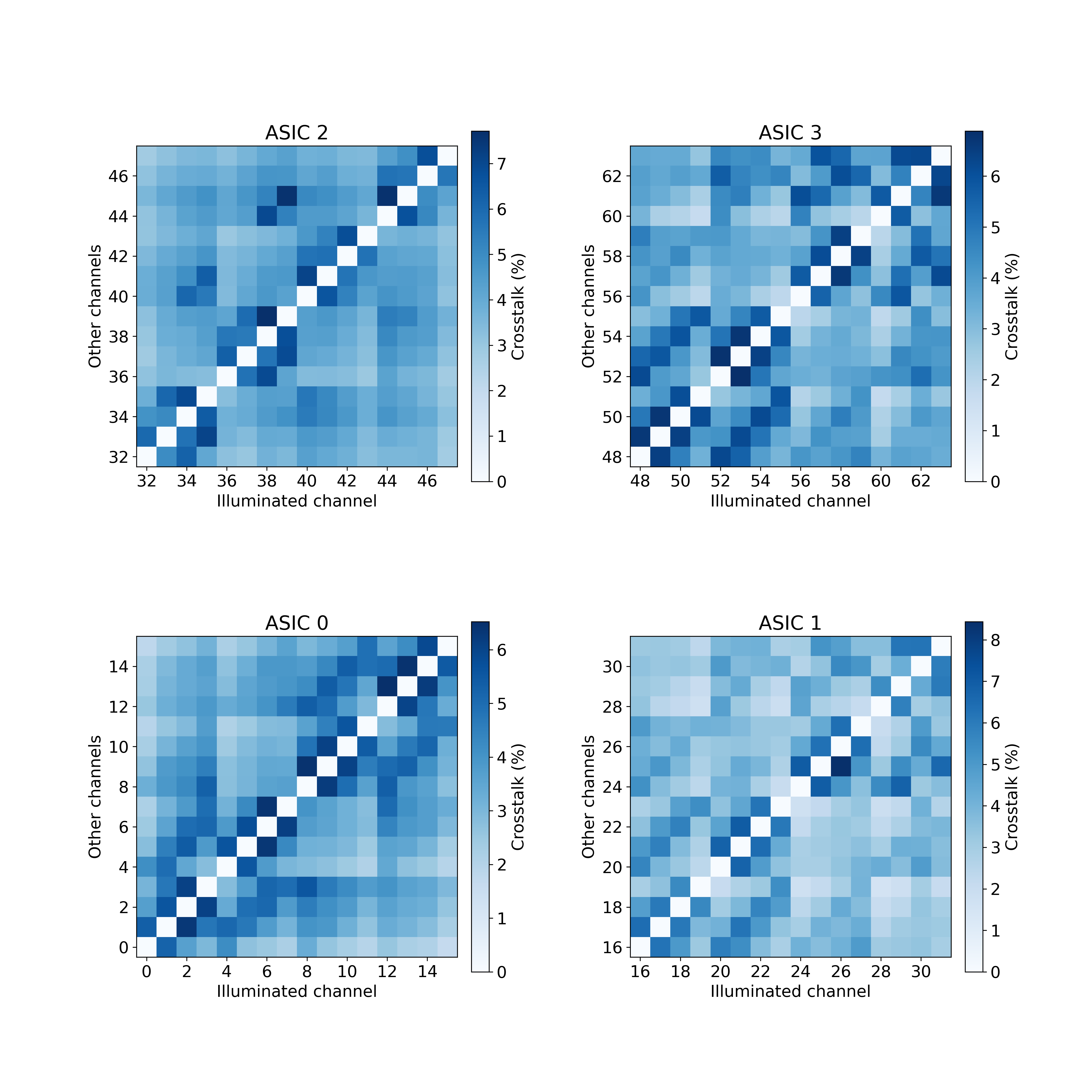}
    \caption{Crosstalk contribution maps. For an easier reading of the map, the self crosstalk contribution of each channel was fixed to \qty{0}{\percent} instead of its value of \qty{100}{\percent}.}
    \label{fig:crosstalk_map}
\end{figure}

The crosstalk contribution to signals was estimated by measuring the peak-to-peak amplitude of average waveforms from non-illuminated SiPMs and dividing this number by a factor 2 to take the undershoot into account. The resulting crosstalk contribution is reported as a fraction of the average waveform amplitude of the illuminated channel, evaluated as the amplitude of the average waveform. The four crosstalk matrices between channels of the same ASICs are reported in Figure \ref{fig:crosstalk_map}.

As can be seen from the crosstalk matrices, the crosstalk contribution on the channels belonging to the same quadrant is below \qty{7}{\percent} of the light-induced signal, and it is stronger within channels belonging to the same trigger pixel as the illuminated channel. We also evaluated the crosstalk contribution between different quadrants. This results in an upper limit on the peak-to-peak crosstalk amplitude of \qty{0.5}{\percent} for quadrants on the same board (i.e. between quadrant 0 and 1, and quadrant 2 and 3), while it is consistent with zero crosstalk between channels on different boards. \nor{Further testing has revealed that the majority (and possibly all) of the crosstalk occurs on the FEE and not on the FPM.}

In addition to amplitude crosstalk, we also evaluated the contribution of crosstalk to the charge integrated signal by integrating all the waveforms in a given time-window of \qty{20}{\ns} around the laser-induced pulse. As expected, given the bipolar shape of the crosstalk pulse, the charge crosstalk results in a  contribution below \qty{0.1}{\percent} for channels within the same quadrant.

It is worth mentioning, however, that the effect of crosstalk primarily affects the waveform amplitude rather than the integrated charge. The crosstalk contribution on a non-illuminated channel (due to multiple input channels collectively inducing a significant signal on the non-illuminated channel, as may happen in a realistic employment of the camera) is linear with respect to the waveform amplitude. However, this linearity does not directly carry over to the integrated charge, since charge integration is performed only after the waveform has already been modified by crosstalk. As a result, evaluating the effect solely in terms of charge, especially when only a single channel is illuminated as in the experimental case here described, can lead to an underestimate of the true impact and the significance of the effect.

We remark that we cannot exclude the possibility that the signal on non-illuminated channels is due to a non-optimal focusing of the laser.
To rule out any possible light contamination, the tests should ideally have been repeated with the illuminated pixel unpowered (i.e., with the HV set to zero) while biasing the remaining pixels. Unfortunately, this was not done at the time of testing and cannot be implemented at the time of writing.

However, in case of laser contamination the shape of the crosstalk pulse would be similar to the laser-induced signal on illuminated pixels, while the signal actually observed show the typical bipolar shape. We conclude that light contamination might partially reduce the amplitude of the negative excursion in the crosstalk waveform, but its effect would be minor, remaining subdominant to the crosstalk itself.

Moreover, since signals are also observed on image pixels physically distant from the illuminated one, as is shown in Figure \ref{fig:avg_wfs_1pixel_on}, the most
likely hypothesis for the origin of this noise is the crosstalk among neighboring channels on the PCB. 

\subsection{Throughput and Power Consumption} \label{Throughput and Power Consumption}

The stated maximum data throughput of the module is \qty{1}{\giga bits \per \s}. \nor{Each sample in a waveform is a 12-bit number. With 64 samples per waveform and 64 waveforms per event, the size of each event is 49,152 bits, excluding the header. With a packet header size of 17,568 bits, the size of each event from the module is 66,720 bits. This corresponds} to a maximum throughput of 14,988 events per second. The most extreme situation would require the module to trigger at approximately 10,000 events per second. The maximum rate of triggers is currently limited by the number of \qty{16}{\ns} clock ticks to digitize each block. This is set to 1312 clock cycles. Each block then takes 1312 $\cdot$ \qty{16}{\ns} + \qty{6}{\us} for digitization (\qty{6}{\us} for data transfer after digitization). With three blocks digitized in every event under expected operating conditions, this gives a duration of \qty{80.976}{\us} to digitize each event. The maximum trigger rate from that duration is 12,350 events per second. Testing has shown the module to take pedestal data at a constant rate of approximately 12,300 events per second with negligible packet losses, reaching the expected limit. \nor{The full camera trigger rate is not expected to exceed \qty{10}{kHz} even in future iterations, so the individual module readout rate meets SCT requirements.} Throughput testing was performed with a \nor{periodic trigger}, not a Poisson trigger. \nor{Future tests are planned with a Poisson source of triggers.} 

To calculate the single module power consumption, the current drawn was measured at the expected operating temperature of \qty{35}{\degreeCelsius} with the module taking data at the expected nominal rate of \qty{1}{kHz}. Testing with different trigger rates showed negligible changes in current draw. The FEE power supply is set to \qty{12}{\V} while the HV supply for the SiPMs is set to \qty{35}{\V}. With the FPM unplugged, the FEE drew \qty{1.07}{\A} or \qty{12.9}{\W}. When the FPM was attached with the HV powered on and enabled, the FEE drew \qty{1.23}{\A} or \qty{14.7}{W} and the SiPMs drew \qty{2.65}{\mA} or \qty{0.093}{\W} in darkness. Exposed to the ambient light of the lab room (an unrealistically bright condition), the SiPMs drew up to \qty{0.3}{\A} or \qty{10.5}{\W}. Additionally, the module power supply drew approximately \qty{1.45}{\A} or \qty{17.4}{\W} when the Peltier was actively cooling the FPM. The power budget allotted for each FEE is \qty{43}{\W} and \qty{9}{\W} for each set of SiPMs. The maximum power consumption for each FEE is well within the budget and the SiPM power consumption is also within the budget under expected operating conditions. These values are summarized in Table \nor{\ref{table:Power Consumption}}.

\begin{table}
    \centering
    \begin{tabular}{||c||c||c|c|c|c||}
    \hline
    \multirow{2}{*}{Measurement} & \multirow{2}{*}{Budget} & \multirow{2}{*}{FPM Detached} & \multicolumn{3}{|c|}{FPM Attached, HV Enabled} \\
    \cline{4-6}
    & & & Darkness & Ambient Light & Peltier Enabled \\ [0.5ex] 
    \hline\hline
    FEE Current & \qty{3.59}{\A} & \qty{1.07}{\A} & \qty{1.23}{\A} & N/A & \qty{1.45}{\A} \\ 
    \hline
    FEE Power (at \qty{12}{V}) & \qty{43.08}{\W} & \qty{12.9}{\W} & \qty{14.7}{\W} & N/A & \qty{17.4}{\W} \\
    \hline
    FPM Current & \qty{0.25}{\A} & N/A & \qty{2.65}{\mA} & \qty{0.3}{\A} & \qty{2.65}{\mA} \\
    \hline
    FPM Power (at \qty{35}{V}) & \qty{8.89}{\W} & N/A & \qty{0.093}{\W} & \qty{10.5}{\W} & \qty{0.093}{\W} \\
    \hline
    \end{tabular}
    \caption{FEE and FPM current and power draw under various conditions.}
    \label{table:Power Consumption}
\end{table}

\section{Conclusion}
\label{sect:conclusion}
The upgraded pSCT camera module has been shown to meet the design goals. Calibrations of all necessary module parameters have been developed and performed without issue. After calibration, performance testing revealed that the module demonstrated stable behavior and produced good quality data. Characterization of the results showed that the module has low noise on the trigger path, ranging between \qtyrange{0.3}{0.6}{\mV}, the electronic crosstalk between signal cables is below \qty{7}{\percent}, and the module's charge resolution is near the Poisson limit and meets the internal benchmarks at the lowest light intensities. Further performance metrics are listed in Table \ref{table:Module Performance}. The behavior of the full pSCT upgrade module is now well understood and well characterized. A few changes and improvements will be made for mass calibration of the pSCT production modules, with the largest improvement being the replacement of a temperature dependent capacitor with a low temperature dependence variant (see Section \ref{Baseline Noise}); this is expected to significantly increase the stability of the waveforms' baseline. Other changes for mass calibration include measuring the \texttt{Vped} transfer functions before the ASICs are soldered to the FEE, changing the \texttt{VpedBias} setting to increase the linear range of the \texttt{Vped} transfer function, and implementing a hard sync internal trigger for calibrations, allowing for faster calibration through the selection of specific storage cells to trigger on.

\begin{table}
    \centering
    \begin{tabular}{||c||c||}
    \hline
    Measurement & Value \\
    \hline\hline
    Dynamic Range & \num{1} p.e. to \num{350} p.e.\\
    \hline
    Linear Range & \num{1} p.e. to \num{100} p.e.\\
    \hline
    Electronics Noise & \qty{0.6}{\mV} \\
    \hline
    \nor{Amplitude Gain} & \qty{2}{\mV} / p.e. \\
    \hline
    Charge Gain & \qty{14}{\mV \ns} / p.e. \\
    \hline
    Charge Resolution & \qty{118}{\percent}, (\num{1} p.e.) \qty{36}{\percent} (\num{10} p.e.), \qty{13}{\percent} (\num{100} p.e.) \\
    \hline
    Minimum Trigger Threshold & \num{3.6} p.e. \\
    \hline
    Maximum Trigger Threshold & \num{750} p.e. \\
    \hline
    Trigger Threshold Resolution & \num{0.15} p.e. / DAC \\
    \hline
    \end{tabular}
    \caption{Summary of module performance parameters. Note that these metrics are approximate and were measured with a specific set of module parameters \nor{and apply to a single image or trigger pixel}.}
    \label{table:Module Performance}
\end{table}

All 177 upgrade modules are expected to be produced, tested, calibrated, and installed in the pSCT camera. Commissioning of the upgraded pSCT camera will be followed by observations of known VHE gamma-ray sources to validate the telescope performance. The modules described in this text are well suited for installation in potential future SCTs at CTAO.

\subsection* {Disclosures}
The authors declare there are no financial interests, commercial affiliations, or other potential conflicts of interest that have influenced the objectivity of this research or the writing of this paper.

\subsection* {Code and Data Availability}
The data that support the findings of this article are not publicly available. They can be requested from the author at riitano@wisc.edu.

\subsection* {Acknowledgments}

We gratefully acknowledge the support of the U.S. National Science Foundation (awards PHY-1707945, PHY-1828168, PHY-2013102, PHY-2413037), the Smithsonian Institution, the Istituto Nazionale di Fisica Nucleare (INFN) in Italy, and the Helmholtz Association in Germany.  J.V. was supported in part by a Vilas Associate award from the University of Wisconsin–Madison. We acknowledge Orisvaldo Salviano Neto and Viviana Rickert (both at the University of Wisconsin–Madison) for their contributions to the ND filter calibrations and crosstalk studies in this work. This paper has been reviewed by the CTAO Consortium.

\bibliography{biblio}   
\bibliographystyle{spiejour}   


\end{spacing}
\end{document}